# Cybersecurity: Past, Present and Future

By

## Shahid Alam, PhD

Department of Computer Engineering
Adana Alparslan Turkes Science and Technology University
Adana, Turkey
salam@atu.edu.tr

© Shahid Alam – 2 October 2022

# Contents

# Chapter 1

# Introduction

*If you know the enemy and know yourself, you need not fear the result of a hundred battles.*
(Sun Tzu, The Art of War, 500 BC)

## 1.1 Why Cybersecurity

With the rising of the Internet and the technology that runs and controls it we are transforming into a new *Digital Age*, This transformation is different than any of the previous such transformations. Advancements in the technical areas, such as natural language processing, machine and deep learning, have tremendously improved the processing and use of social media, Internet of things (IoTs), and cloud computing, etc. These advancements have certainly improved the workings of businesses, organizations, governments, the society as a whole, and day to day life of an individual. These improvements have also created some new challenges, and one of the main challenge is security. The security of the new digital age, also known as *Cyberspace*, is called **Cybersecurity**. A formal definition of Cybersecurity provided by Merriam-Webster dictionary [Merriam-Webster, 2021b] is:

*Measures taken to protect a computer or computer system (as on the Internet) against unauthorized access or attack.*

Another formal but a more comprehensive definition given in [Schatz et al., 2017] is:

*The approach and actions associated with security risk management processes followed by organizations and states to protect confidentiality, integrity and availability of data and assets used in cyberspace. The concept includes guidelines, policies and collections of safeguards, technologies, tools and training to provide the best protection for the state of the cyber environment and its users.*

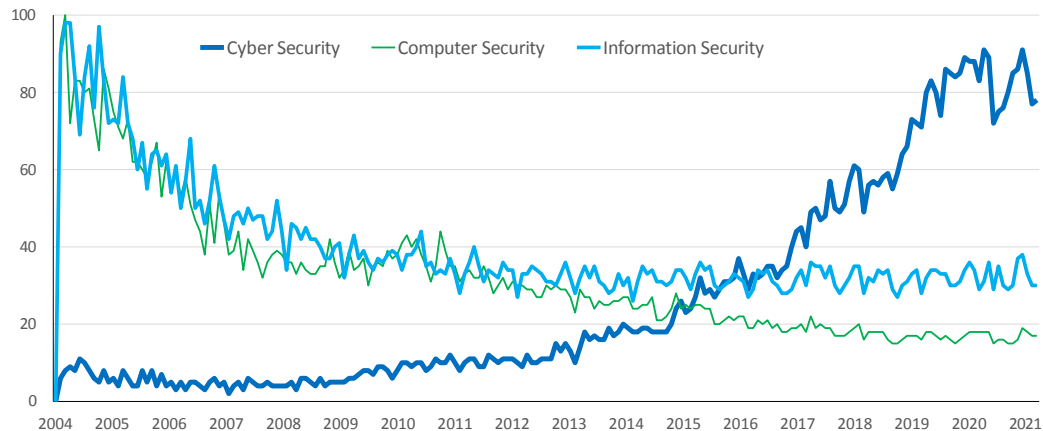

Figure 1.1: Google search trends [Google, 2021] by month, from January 2004 to January 2021, for the terms *Cyber Security*, *Computer Security*, and *Information Security*. The y-axis depicts the relative search frequency for the term. A value of 100 is the peak popularity for the term. A value of 50 means that the term is half as popular. A score of 0 means there was not enough data for this term.

Different terms are being used in the context of security. Some of the popular terms are *Computer Security*, or *Information Security*. Now a days, to include all kind of computer systems and cover everything that is part of the cyberspace, such as smart devices (e.g., smartphones and televisions), cloud computing, IoTs and several others, the new term *Cyber Security* has become increasingly popular. The graphs shown in Figure 1.1 depict the popularity by month of these three terms (mentioned above) when used during Google searches performed from January 2004 to January 2021. This development in the use of terminology based on the current most popular search engine is useful and of value to identify trends. The term *Cyber Security* starts gaining popularity from the year 2017 and achieves peak popularity in the years 2019 and 2020. Whereas, the other two terms show a steady decline in popularity.

To keep pace with the advancements in the new digital technologies like IoTs and cloud, there is a need to expand research and develop novel cybersecurity methods and tools to secure these domains and environments. We need to highlight some of the challenges facing cybersecurity and discuss some innovative solutions to resolve or mitigate these challenges. To know these challenges and provide solutions we have to examine the past, scrutinize the present and ameliorate the future.

## 1.2 A Short History of Cybersecurity

The first comprehensive published work *Security Controls for Computer Systems* [Ware, 1970] which became the foundation in the field of cybersecurity was a technical report commonly called the *Ware Report*. This report, published in 1970, was the result of

```
BBN-TENEX 1.25, BBN EXEC 1.30
@FULL
@LOGIN RT
JOB 3 ON TTY12 08-APR-72
YOU HAVE A MESSAGE
@SYSTAT
UP 85:33:19   3 JOBS
LOAD AV   3.87   2.95   2.14
JOB TTY USER      SUBSYS
 1   DET  SYSTEM   NETSER
 2   DET  SYSTEM   TIPSER
 3   12   RT       EXEC
@
I'M THE CREEPER : CATCH ME IF YOU CAN
```

Figure 1.2: An example of the message produced by the Creeper: source [corewar.co.uk, 2021]

a study carried out by a task force on computer security organized by the department of defense of United States of America. The report concluded, that a comprehensive security of a computer system requires a combination of hardware, software, communications, physical, personnel and administrative controls.

The first computer program (also the first non-self-replicating benign virus in history) that could move across a network was written in 1971, leaving a message trail (I am the creeper, catch me if you can) behind wherever it went. The program was called the *Creeper*. An example of the message produced by the Creeper is shown in Figure 1.2. The first example of an antivirus program (it performed the same functions that an antivirus does today) was written in the year 1973 called the *Reaper*, which chased and deleted the Creeper. Reaper was also the first self-replicating program and hence making it the first benign computer worm (a self-replicating computer program that moves across the network).

In 1977 the *CIA triad*, of *Confidentiality*, *Integrity*, and *Availability* shown in Figure 1.3, was introduced [Ruthberg and McKenzie, 1977]. These are the three basic principles for cybersecurity and are still widely used benchmarks to evaluate the effectiveness of a cybersecurity system.

**Confidentiality** — The confidential information and data should be prevented from reaching the wrong hands. Confidentiality deals with the access, operation, and disclosure of system elements.

**Integrity** — The information and data should not be corrupted or edited by a third party without authorization. Integrity deals with the modification, manipulation, and destruction of system elements.

**Availability** — The information and data should be available all the time and adaptive recovery mechanisms should be established to restore the system and the services provided by the system. Availability deals with the presence, accessibility, readiness, and continuity of service of system elements.

The CIA triad gives the appearance of a holistic security model, but is not in any way a checklist for security systems. It is rather a limited security model and acts as a starting point for many elaborative security systems engineering frameworks, processes and policies [Ross et al., 2016].

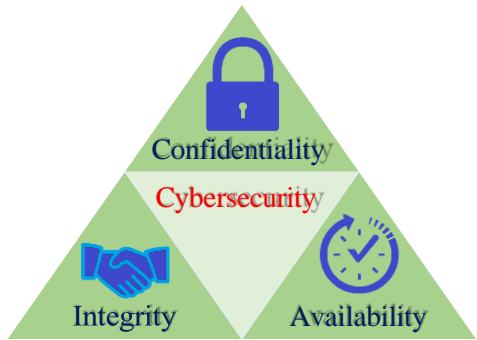

Figure 1.3: The CIA Triad

During the 1980s, the ARPANET (the first wide-area network) became the Internet. As computers started to become more connected computer viruses became more advanced. The Morris Worm (Figure 1.4) developed in 1988 became the starting point of the creation of more effective worms and viruses. This rise led to the development of antivirus solutions as a means for countering the worm and virus attacks.

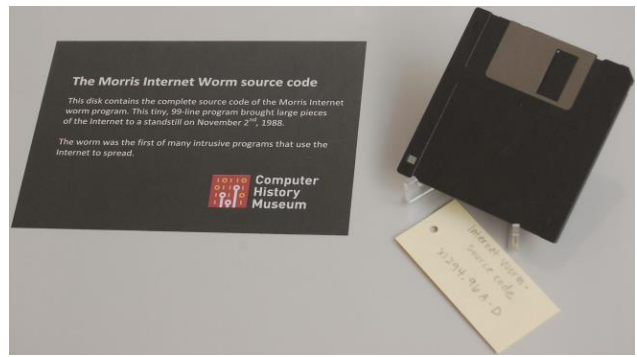

Figure 1.4: The source code of Morris worm. source: [Intel-Newsroom, 2013]

In the 1990s viruses such as Melissa (Figure 1.5) causes the failure of the email systems by infecting tens of millions of computers. These attacks have mostly financial and strategic objectives. 1990s also saw a sudden growth of antivirus companies. These antivirus products suffered with using a large amount of available resources and producing a large number of false positives. Some of the cybersecurity solutions today also suffered similar problems.

As the computers started gaining computing power, the 2000s saw sophisticated malicious software, such as polymorphic and metamorphic malicious programs. These two types can change there shape and structure with a variety of complex modules to execute malicious activities. These and other stealthy malicious programs have also infiltrated the new platforms, such as smartphones and IoTs (Internet of things), etc. Stuxnet (Figure 1.6) found in the year 2010 was the first weaponized malware program, and one of the first instances where cyberattacks were used in espionage. Stuxnet spread using an infected removable drive such as a USB flash drive. This drive contain Windows shortcut files to initiate the malicious executable code.

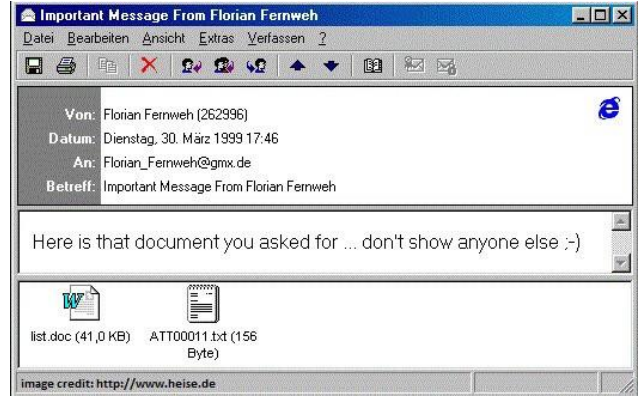

Figure 1.5: The Melissa virus sent through an email. source: [Panda-Security, 2013]

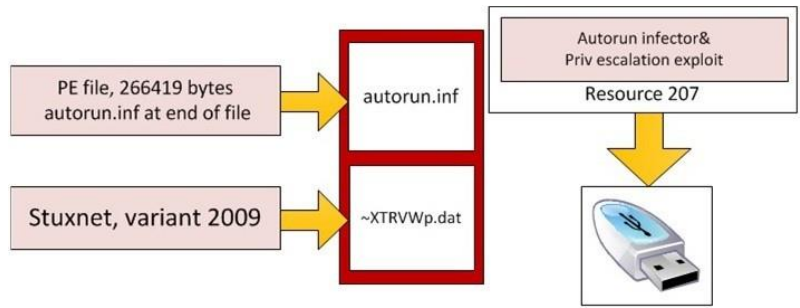

Figure 1.6: The Stuxnet virus spreading through a USB drive. It first injects a malicious module into Windows autorun.inf file. After this, the malicious module executes automatically and loads the Stuxnet main body (~XTRVWp.dat) from the USB drive and injects it into the system process. source: [Gostev, 2012]

Recently, ransomware (malicious software that threatens the victim to publish or block access to victim's data unless a ransom is paid) attacks are on the rise, e.g., WannaCry (Figure 1.7), Clop Ransomware, and Mount Locker. Because of the ongoing digitization and hardware acceleration, attackers also started targeting hardware, with Backdoors, Trojans and side-channel attacks.

With the rise of new technologies, such as artificial intelligence (AI), researchers and industry have also started applying these new technologies to analyse and understand the cyberattacks and improve defense against them. At the same time, attackers and adversaries are using similar techniques to advance their cyberattacks.

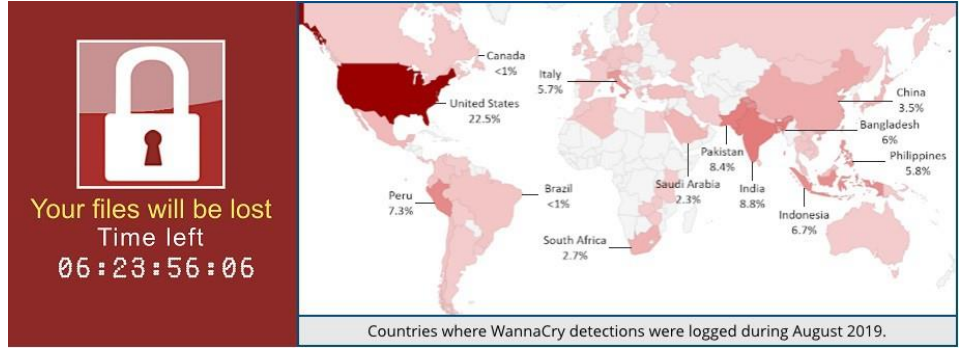

Figure 1.7: The WannaCry ransomware. source: [Naked-Security, 2019]

The best cybersecurity defense for a system is to know and reduce the current and future cybersecurity risks to the system. For example, a novel technique, *Generative Adversarial Networks* (GANs) [Goodfellow et al., 2014], is being applied to give insights into specific behaviors of adversaries and also provide guidance and direction to reduce the cybersecurity risks to a system. GANs can also be used by adversaries to learn specific behaviors of the defenders, and improve their attack strategies to defeat the defense against their attacks. This is the modern warfare (a.k.a **Cyberwarfare**), where friends and foes improve upon their defending and attacking cyber technologies and tools, respectively, to defeat each other.

## 1.3 A Glance into Challenges and Solutions

There are several challenges facing cybersecurity. Some of them are:

1. The increased model complexity of AI makes it difficult for its application and adoption in cybersecurity.

2. Successful development and implementation of human-AI collaboration to improve cybersecurity.

3. Development and improvement of simple user authentication mechanisms suitable for the resource constrained devices, also called wearable and IoT biometrics.

4. Personal privacy is still a big concern when it comes to the design, development, and deployment of biometrics.

5. Cyber forensics of IoT devices and Cloud computing is still immature and faces lot of challenges, such as complexity and diversity of IoT devices, and evidence collection in Cloud environment because the data is distributed between countries and within datacenters.

6. For the successful analysis and detection of the shear number of malicious programs we need to completely automate this process.

7. Design, development, and implementation of in-cloud analysis of malicious programs .

8. Applying machine learning to hardware security faces several challenges. One of the major limitations of the application of machine learning to hardware reverse engineering is the lack of extracted features that can be generalized to other chips.

9. Code obfuscation, smart fuzzing, and secure programming languages are some of the challenges faced by software security.

Next generation cybersecurity started using innovative technologies to improve detection and prevention of new threats and building a safe and secure society. Some of the examples of these technologies are:

1. Multi-factor authentication, e.g., using a combination of personal identification number (PIN), biometrics (fingerprints, and face, etc.) or passwords.

2. Behavioral analysis of malicious programs to detect anomalies.

3. Sandboxing, i.e., creating an isolated and safe environment for testing malicious programs or links.

4. Forensics, the use of scientific methods and expertise to gather and analyze evidences that can be used in criminal or other investigations in a court of law.

5. Applying artificial intelligence (AI) to improve and solve challenging problems in cybersecurity.

Here we have just enumerated some of the challenges and solutions. In rest of the chapters, we are going to discuss in detail these and some other challenges and also some new innovations and solutions. We also present some future directions to resolve or mitigate the challenges faced by cybersecurity.

## 1.4 Organization of the Book

This chapter has provided a peek into the past, present and future of cybersecurity. Rest of the chapters provide detail discussion about the past, present, and future of different major sub-domains and specializations of cybersecurity: including software, hardware, malware, biometrics, artificial intelligence, and forensics.

The rest of the book is organized as follows:

*Chapter 2* first defines what is **Software Security**, and then discusses some of the major software vulnerabilities and techniques to mitigate these vulnerabilities. How to protect and design secure software. It also discusses about software security testing. At the end, it points out some future works and directions in the field of software security.

*Chapter 3* first defines what is **Hardware Security**, and then discusses some of the major hardware vulnerabilities and techniques to mitigate these vulnerabilities. It also discusses about hardware security testing. At the end, it points out some future works and directions in the field of hardware security.

Malware is a major threat to cybersecurity, therefore we have dedicated a separate *Chapter 4* that discusses the **Evolution of Malware**, including classification, analysis, reverse engineering, cyberattacks. It also discusses some of the issues in malware analysis and detection, such as obfuscation, automation, and morphing, and their solutions. At the end, it points out some future works and directions in the field of malware analysis and detection.

*Chapter 5* first covers the basics of **Biometrics**, such as properties, classification, phases, testing, and methods used. It also discussed about multi and soft biometrics, and different biometric attacks and how to protect against them. At the end, it points out some future works and directions in the field of biometrics.

*Chapter 6* first covers the basics of AI, including the definition, different approaches to AI, and explains the processes involved in **Cyber Intelligence**. Some of the challenges plaguing cybersecurity where AI can help are also discussed. It also discusses some practical applications of AI to cybersecurity, such as adversarial machine learning, game theory, and generative adversarial networks. At the end, it points out some future works and directions in the field of cyber intelligence.

*Chapter 7* first introduces the field of **Cyber Forensics**, including basic terminologies used and the general process of cyber forensics. Then it talks about some anti-forensics technique used to make forensic analysis difficult. At the end, it points out some future works and directions in the field of cyber forensics.

# Chapter 2

# Software Security

*Force the enemy to reveal himself, so as to find out his vulnerable spots.*
(Sun Tzu, The Art of War, 500 BC)

## 2.1   What is the Problem

NASA Mars Polar Lander, a 165 million US dollars machine, crashed on December 3, 1999 into Mars due to software errors, generating false signals that led the onboard computer to shut off the engine too soon [Leary, 2000]. This disaster could have been avoided and fixed by a software change in the onboard computer. The investigating reports concluded that due to the pressure of the NASA's credo of *faster, cheaper, better* on project managers and engineers working on the NASA Mars Polar Lander ended up compromising the project. To meet the new constraints, the project managers neglected and sacrificed needed testing and realistic assessments of the risks of failure.

Northeast blackout of 2003 was caused due to a software bug, known as a *race condition*, in the alarm system at the control room. This blackout was a widespread power outage of upto two days of the Northeastern and Midwestern United States, and the Canadian province Ontario. Almost 55 million people were affected by this power outage. The blackout's cause was stalling of the alarm system at the control room, causing the failure of both audio and video alerts of the system state.

Buggy smart meters open door to power grid botnet (interconnected devices each running one or more malicious programs) [Goodin, 2009]. An experimental worm developed at IOActive (a cybersecurity company) self-propagated across these meters, and was able to take over large number of them. The vulnerabilities present in the software on these smart meters, because of the use of insecure C functions `memcpy()` and `strcpy()`, were exploited for this attack.

The above three real life examples demonstrate and reveal that software bugs are responsible for producing security vulnerabilities in a program that can be exploited by attackers to fail the software system. Complexity is the enemy of security, i.e., a complex program will contain more bugs. Not all the bugs lead to security vulnerabilities. A security analyst is only interested in bugs that have security implications and are called security bugs/flaws. If we eliminate all the bugs from a software system that will also eliminate all the security bugs from the system. This proposition becomes a premise of the relationship between software engineering and software security. Improving the quality of a software system would result in an increase in security of the system. However, many security bugs go undetected, because traditional software development processes does not cater for security. For example, during quality assurance of a software system only reasonable range of user inputs are tested, whereas the attackers are rarely reasonable and will devise inputs that break the system.

**DEFINITION**: A software *vulnerability* is a set of conditions, bugs, flaws, or weaknesses in the software that can be exploited by an attacker. A software *exploit* is a piece of malicious program that takes advantage of a vulnerability in a software system to cause damage to the system. A *mitigation* is a method, technique, or process that prevents or limits exploits. **Software Security** is a process/practice to either mitigate or design and develop software without vulnerabilities to escape exploitations.

Figure 2.1 shows the growth in the number of software vulnerabilities from 1988 − 2020 in the National Vulnerability Database maintained by NIST (National Institute of Standards and Technology) USA. The number of vulnerabilities grew from just 2 in the year 1988 to 16958 in the year 2020. In the last four years there is a significant increase in these numbers. Following are some of the rationals for this sudden increase in software vulnerabilities over the last four years. There are several, but we only list some of the major causes.

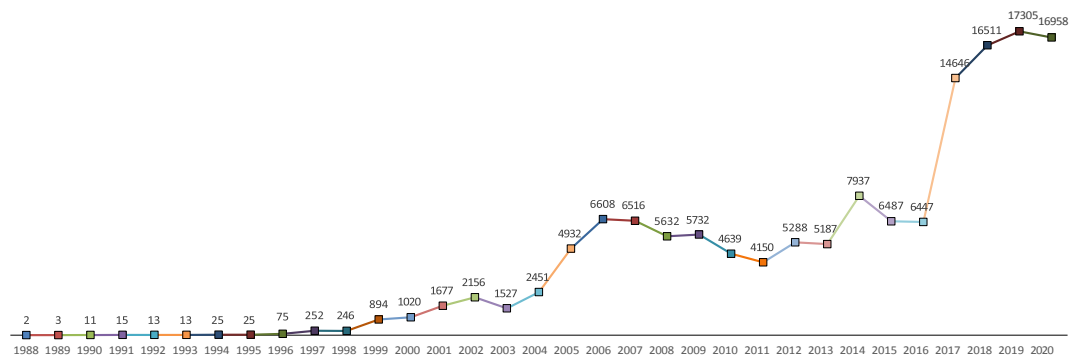

Figure 2.1: Statistics on software vulnerabilities reported from 1988 − 2020 Source: National Vulnerability Database maintained by NIST, USA [NIST, 2020]

1. The competition and pressure from businesses and organizations demand a shorter time to market of the software products from the vendors. This in turn forces the vendors to release products faster and reduce the resources and time for quality assurance and testing.

2. With the continuous evolving technology, mobile devices and their applications,

cloud computing, social media, and Internet of things (IoT) have become ubiquitous. This has enabled the deployment of new products and services and have opened opportunities for innovative businesses, though with relatively low maturity in software development and lacking skills in secure coding.

3. Complexity is the enemy of security. Software applications' are increasingly becoming complex because of the new wave of technology. Modern applications are often built using open source components which become a source of major vulnerabilities in these applications. The complexity problem is worsen by the use of unsafe programming languages (e.g., C and C++).

4. Legacy systems, that for different reasons are still in use, are also a major source of vulnerabilities.

5. Cybersecurity is now on top of the agenda on every nation state. Therefore, they are actively engaged in vulnerability research to support lawful interventions in software systems. With cyberattack becoming a profitable activity and business, cybercriminals are also heavily involved in vulnerability research to support their attacks.

6. A vulnerability is introduced by design, such as back doors, for various purposes (software updates, security patches, and hidden commercial agendas, etc) by the software vendors.

## 2.2 Impact of Software Vulnerabilities

The impact of software vulnerabilities on consumers and businesses is in the form of, loss of data, privacy, other resources, such as time and money, etc. Following are some of the key impacts because of a vulnerability in a software system/application.

**Denial of Service**: An attacker partially or completely disrupts the normal function of a computer service (e.g., an online service such as a website) and prevents other users from accessing the service. This can cost a business or an organization both time and money while their resources are unavailable.

**Information Leakage**: An attacker gets access to confidential information present in a software system/application, such as technical and proprietary details about the application, developer comments, or user specific data. This data can be used by the attacker to further exploit the target application, its hosting network/cloud, or its users. Depending on the type of software, this attack can cause serious financial damages to the business, or sabotage the company or organization, owning the software.

**Privilege Elevation**: An attacker attempts to gain more permissions or access with an existing account that the attacker has compromised. The result of this is that the attacker with more privileges can perform unauthorized and malicious actions, such as deleting important accounts, files or other such resources on a system. In addition to other damages, these actions can compromise the trust and privacy of users and cause a loss of assets of the system.

Software is everywhere. It runs your car, controls your cell phone and household items. It is the backbone of banks and nations' power grids. As businesses, organizations, individuals and nations depend heavily on software, we have to make it better and more secure. Software security is no longer a luxury but a necessity.

## 2.3 Software Vulnerabilities and Mitigation

Here we discuss some of the major software vulnerabilities that can be exploited by an attacker (a person or program) to inflict damage on the system. We also present and discuss some of the strategies/techniques to mitigate them.

### 2.3.1 Buffer Overflow

A *buffer overflow* occurs when a program writes data outside the buffer, i.e., the allocated memory. Buffer overflow vulnerabilities are usually exploited by an attacker to overwrite values in memory to the advantage of the attacker. The first exploit of buffer overflow was carried out in 1988 by Morris Worm (a malicious program). The worm exploited a buffer overflow vulnerability in some of the UNIX applications, sendmail, finger, etc., infecting 10% of the Internet at that time. After over 3 decades we would expect that buffer overflow would no longer pose a significant threat. But, recently in 2019 a buffer overflow vulnerability in WhatsApp VOIP on smartphones was exploited that allows a remote attacker to execute arbitrary code on the target system. WhatsApp was broken with just a phone call, the user did not even have to pick the phone [Newman, 2019]. This indicates how dangerous is this vulnerability and justify the importance of preventing buffer overflow.

Here is a simple buggy program overflow() that is susceptible to buffer overflow attack.

```
void overflow() {
    char buff[32];
    gets(buff);
}
```

`gets()` continues to read input until the end of line character, an attacker can overflow the buffer, `buff`, with malicious data. In the current environment the above code would be quickly labeled as unsafe, because `gets()` is almost universally understood to be an unsafe function. The presence of the above code in legacy applications still makes this a real threat today, especially on older platforms. For a better understanding of a classic buffer overflow exploit, consider Figure 2.2

There are three different versions of execution of function `overflow()`. In the first version, before execution, space is reserved for the buffer and then the return address of the caller. In the normal execution, the string `"Hello World"` is read from the input and stored in the buffer. Then the function jumps to the return address, i.e., returns to the caller. In the third execution, during the buffer overflow attack the attacker has already filled the stack with a series of `nop` instructions, the exploit code, and the address of the beginning of the buffer, hence the function instead of returning to the caller executes the exploit code. There are also other types of buffer overflow, such as heap-based, integer overflow, etc. Interested readers are referred to [Hoglund and McGraw, 2004] for more details and in-depth coverage of buffer overflow exploits.

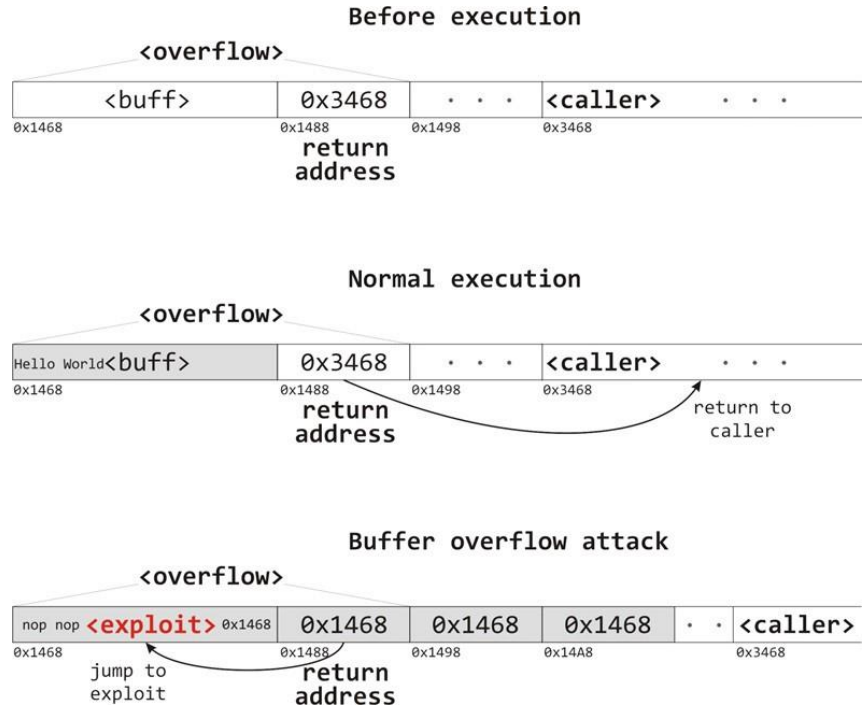

Figure 2.2: Three different varieties of a simplified stack frame for the function `overflow()`: the top one before execution, the middle one after normal execution, and the bottom one after a buffer overflow attack

**Mitigation**

☐ *Tracking Buffer Size*: The most common mistake that leads to buffer overflow exploit is not keeping track of the buffer size. Memory safe languages, such as Java and Python, keep track of the buffer size and use this size during different operations performed on the buffer. But, a very large number including the legacy programs are written in the two most popular languages C and C++, which are memory unsafe languages. A general solution to explicitly track size of the buffer is to keep/store length of the buffer as a separate value and update whenever the buffer size changes. In C, this can be accomplished with the following data structure.

```
typedef struct {
        char* ptr;
```

```
        int length;
    } buff;
```

☐ *Use of Safe Functions and Languages*: As shown above in the code snippet for buffer overflow example, the function `gets()` is an unsafe (dangerous to use) function. There are several other such functions that are also dangerous, such as `scanf()`, `strcpy()`, `strcat()`, and so on. All of them can be replaced by the safe functions, that check the array bonds, such as `gets_s()`, `scanf-s()`, `strcpy-s()`, `strcat-s()`, and so on. There is a safe dialect of C and C++ programming languages and contain all the safer functions. The best way to avoid buffer overflow is to either use safe programming languages, such as Java and C#, or use safe dialect of a programming language. The problem is not yet solved, because there already exists a large number of legacy programs that contain these unsafe functions. Many of these legacy programs are in native code, i.e., the source code is not present. That means it is difficult/impossible to replace the unsafe functions in these programs. What to do about these programs. There are different solutions to get around this problem, and here we discuss two of them.

☐ *Address Space Layout Randomization*: During a buffer overflow attack the main purpose of the attacker is to execute its own malicious code. The first solution, address space layout randomization (ASLR) randomly arranges the address space of data areas of a process, including the stack, heap and libraries. In this case the attacker has to first successfully guess the address of the injected malicious code which is a non-trivial process. Recently researchers have shown that ASLR can be bypassed by using a type of side-channel attack [Evtyushkin et al., 2016]. A side-channel attack is any attack based on the information gained from the internal knowledge, such as timing information, power consumption, etc., of a computer system, rather than the weaknesses, such as software bugs, etc.

☐ *Data Execution Prevention*: The second solution, data execution prevention (DEP) attempts to ensure that the contents of only those memory locations can be executed that are pre-defined for such purpose, and data/code in all other memory locations become non-executable. DEP helps prevent buffer overflow attacks that inject and execute code and rely on some part of the memory, usually the stack, both being writable and executable. DEP can be implemented in either hardware or software. A combination of DEP and ASLR becomes a successful defense against buffer overflow and other such attacks. All the major operating systems, including Linux, Android, iOS, Windows, and Solaris implement both ASLR and DEP technique to avoid buffer overflow and other such attacks.

### 2.3.2 Never Trust Input

One of the most talked about mantra of security professionals/gurus is to *Never Trust Input*. But, when you ask a group of programmers about this mantra, you will notice a mystified look on their faces. If you can not trust input, how can your program do anything useful? Example 2.3.1 illustrates the gravity of this mantra.

**Example 2.3.1.** How a hacker might get access to all the usernames and passwords in a database. If the input is not checked and a user is allowed to enter anything, then a malicious user can enter the following input in the user id field of a web database application:

UserID:  1003 OR 'a'='a'

The SQL statement generated is:

SELECT * FROM Users WHERE UserID = 1003 OR 'a'='a';

This is a valid SQL statement, since `'a'='a'` is always true. If the `Users` table also contains names and passwords then the above SQL statement becomes:

SELECT UserID, Name, Password FROM Users WHERE UserID = 1003
OR 'a'='a';

This returns all the ids, usernames and passwords from the `Users` table in the database.

The above example elucidates a SQL injection attack because of the poor/no input validation. This example also shows how *metacharacters* are used to exploit a vulnerability in the SQL. Metacharacters are characters that have special meaning in the input language, such as single quote (') in SQL, double period (..) in file system paths, and semicolon (;) and double ampersand (&&) in command shells. Using these metacharacters, several other such attacks are possible because of inadequate input validation. A program have to accept input, but it must not trust it, then what it should do? A program must implement adequate input validation, such as performing sanity check, corroborating the input, and controlling and limiting only the acceptable values, etc.

**Mitigation**

□ *Validate All Input*: This is the basic mitigation, to check all the inputs from all the sources. For example, if your application consists of more than one process, validate the input for each of the process. Validate input even if it comes from a secure or trusted connection. Every validation done denies the attacker an opportunity and provides an added degree of assurance. There are two types of validation: *Syntax Validation* – checks the format of the input, and *Semantic Validation* – checks the input based on the logic and function of the application. Perform input validation not only on user input, but also on data from any source outside the program/application, such as command line parameters, configuration files, data from database, and network services, etc.

□ *Parameterized Requests*: Example 2.3.1 shows a metacharacter vulnerability exploit. Many such and other metacharacter vulnerabilities can be eliminated by keeping data and code information separate. The main cause of SQL injection is the mixing of data and code. Trusted code is sent via a code channel and untrusted data provided by user is sent via a data channel. The database clearly knows the boundary between code and data. The code hidden in data by an attacker manipulating metacharacters (example 2.3.1) will never be treated as code, so it will never be executed.

☐ *Whitelist Validation*: In this type of validation only known good inputs are accepted. The known good inputs are based on the expected type, length or size, numeric range, or other format standards. For example a credit card number should only contain numbers between 16 digits long and passes the business logic check (validity of a number based on the last digit of the card). Regular expressions are generally used to implement whitelist validation. Example 2.3.2 lists a Java program using regular expressions to validate US phone numbers.

☐ *Blacklist Validation*: In this type of validation known bad inputs are rejected. The known bad inputs are based on the presence of the number of known bad characters, strings, or patterns in the contents of the input. Because, in general the list of potentially bad characters is extremely large, this validation is weaker than the whitelist validation. Like whitelisting, a common way of implementing a balcklist is also to use regular expressions.

In practice, whitelist validation is first used and then if needed it is supplemented with blacklist validation. There are certain validations where it becomes difficult to get a full list of possible inputs and use the whitelist validation. For example, applications with certain inputs that are localized in languages with large character sets, e.g., Chinese and Japanese. In these cases to validate the application inputs whitelist is supplemented with blacklist validation.

**Example 2.3.2.** Example, taken from [Chess and West, 2007], of a Java program using regular expressions to validate US phone numbers.

```
// A separator is some whitespace surrounding a . or -.
String sep = "\\s*[.-]?\\s*";
// A US phone number is an optional 1 followed by
// a three digit area code, a three digit prefix,
// and a four digit number.
String regex = "(1"+sep+")?\\d{3}"+sep+"\\d{3}"+sep+"\\d{4}";
```

### 2.3.3 Access Control

Access control is one of the most important security features when designing and developing an application. It sets permissions and protects the application's resources from unauthorized access. Applications may manage their own resources or use the services provided by the operating system (OS). Different OSs have their own access control model. Here we briefly discuss the traditional access control model of the UNIX OS (Android OS is based on Linux, which is a variant of UNIX).

Users of a UNIX system have a user name, which is identified by a user ID (UID). Every user belongs to a group and consequently has a group ID (GID). The information required to map a user name to a UID is maintained in the file `/etc/passwd`. The login program authenticates a user by examining the `/etc/passwd` file to validate the UID and password of the user. Each file in a UNIX system has an owner (UID) and a group (GID). Only the owner and root can change permissions for a file. The permissions are `read`, `write`, and `execute`. These permissions can be granted or revoked for each of the following types of users: Owner, group, and others.

Figure 2.3 shows an example of an access control exploit. In this attack, an attacker exploits the race condition bug in the program to gain access to a file which the attacker could not otherwise access. The problem occurs when the access to a file is checked and then the file is opened if access is granted. There are two operations performed here, one is checking access, and the other is opening the file. In the time between these two operations (window of vulnerability) an attacker can change the link of the file that is being operated upon. In this way the attacker can gain access to a file that otherwise is not accessible to the attacker. A shown in Figure 2.3 the attacker changed the link of the file `/home/foo` to point to `/etc/passwd`. Now the program is tricked into opening a file on behalf of the attacker that it should not have.

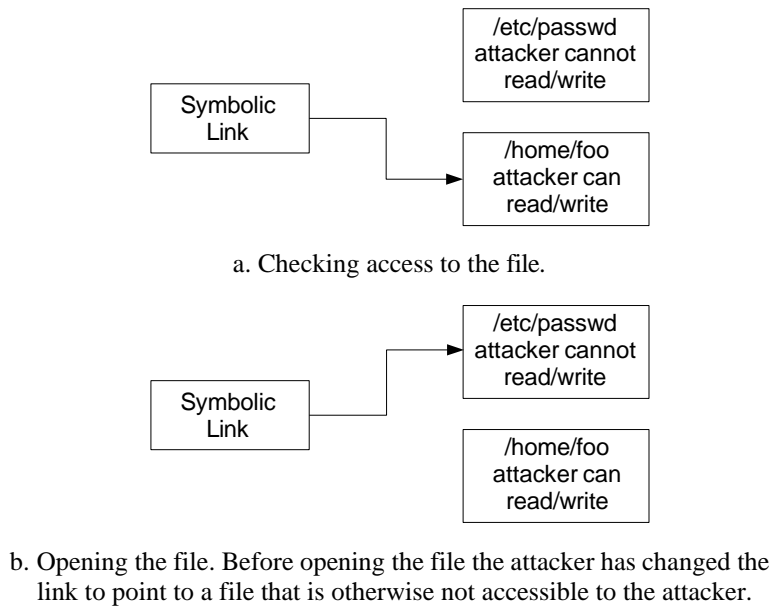

a. Checking access to the file.

b. Opening the file. Before opening the file the attacker has changed the link to point to a file that is otherwise not accessible to the attacker.

Figure 2.3: Example of an access control exploit by utilizing race condition software vulnerability

The race condition vulnerability is more common in multithreaded programs because of the uncontrolled concurrency found in these programs. These type of vulnerabilities are difficult to detect, reproduce, and eliminate. In these cases, similar to the above example, to eliminate the race condition a race window is to be identified. A race window is a piece of code that accesses the race object in a way that opens a window of opportunity for the other processes/threads (e.g., an attacker) can access and alter the race object. A race window is not protected by a lock or any other mechanism.

Generally a web application provides a web interface to manage user accounts and other resources. An administrator logins using this web interface to manage the resources. Example 2.3.3 shows an access control exploit at the application level. In which an attacker gains privileges of an administrator.

**Example 2.3.3.** Some applications determine the user's access rights or role at login depending on the parameters passed. For example the following URL:

`https://insecure.com/login/home.jsp?admin=false`

when changed (admin=false means an unsuccessful login) by an attacker (unauthorized person – not an admin) to:

`https://insecure.com/login/home.jsp?admin=true`

gives the attacker access to carry out administrative functions.

**Mitigation**

☐ *Closing the Window of Vulnerability*: The race condition vulnerabilities exist only during the (race) window of vulnerability. The obvious solution is to eliminate this window. We discuss here some of the techniques to eliminate these windows. (1) Checking for symbolic links and only give access to a file if it is not a symbolic link. This way even if the attacker changes the symbolic link will not be able to gain access to the file. (2) Using synchronization primitives available in OSs (locks, semaphores, etc) these windows of vulnerability can be protected as mutually exclusive critical sections. Care must be taken to minimize the size of the critical sections. (3) In a multithreaded application it is not enough to only avoid race conditions within the application's own instructions. The invoked functions could be responsible for race conditions. In this case, use of thread-safe functions avoid race conditions. Thread-safe functions can be called by concurrent threads with the function being responsible for any race condition. (4) Use of atomic operations. These operations can not be interrupted until run to completion. It is this atomic property that makes these functions useful for synchronization and helps eliminate race conditions.

☐ *Principle of Least Privilege*: A process, user, or program must only access the resources that are necessary for their legitimate purpose to complete the job [Saltzer, 1974]. This mechanism mitigates race conditions as well as other vulnerabilities. For example, if a program is running with elevated privileges and accessing files in shared or user directories, there is a chance that the program might be exploited and perform an operation for which the user of the program does not have the appropriate privileges. One of the methods used to implement this mechanism is in the microprocessor hardware. For example, in the Intel x8 architecture there are four modes (ring 0 – ring 3) of running with graduated degrees of access. The most privileged is ring 0. The OS kernel runs in ring 0, device drivers may run in rings 1 or 2 and applications in ring 3. It can also be implemented in software. As in OSs the elevated privileges are dropped, as soon as the process is finished with the required job, before accessing the shared resources.

□ *Managing Permissions*: Managing privileges is a good strategy for controlling the access, but it will not control the initial permissions given to a process, user, or program. This is the responsibility of the administrator and the programmer. We discuss here some of the techniques that can be used to manage permissions. (1) Securing Directories – A secure directory is a directory where no one other than the user, or the administrator, has the ability to manipulate (create, rename, and delete, etc) files. All the directories above a secure directory are owned by either the user or the superuser (administrator), and can not be manipulated by other users. Other users may only read or search the directory. This eliminates the possibility of an attacker masquerading as an other user will not be able to exploit a file system vulnerability in a program. (2) Restrict Initial Access Permissions – A process inherits the permissions from its parent process. A parent process gets the default permissions set by the OS, that can be adjusted by the administrator when installing the OS. These permissions should be set appropriately keeping the principle of least privilege as discussed above. There are certain rules (mostly based on common sense) to follow when managing the initial access permissions. Here we will just list couple of them. For executable files, only permit the file owner to execute the file. For sensitive files, only permit the file owner to read and write.

□ *Multi-Factor Authentication*: Multi-factor authentication (MFA) is a process of granting access to a user or application after successfully presenting two or more factors of authentication. These factors may include: something you have – e.g., a bank card or a key; something you know – e.g., a password or a PIN; and something you are – e.g., fingerprints or eye iris. A typical example of two-factor authentication is the correct combination of a bank card and a PIN to withdraw from an ATM. This kind of authentication reduces identity thefts and online frauds as the victim's password would not be enough to access the information.

### 2.3.4 Denial of Service

Denial of service (DOS) attack is one of the easiest and most popular attacks, that it needs a separate mention here. We have already discussed some of the vulnerabilities that can cause DOS, such as buffer overflow and access control vulnerabilities. Here we explain DOS with an example and then discuss some of the mitigation techniques. Resource starvation attacks consumes all the available resources to prevent normal clients from obtaining access to the system. In a service-based system, if your application is single threaded, it may not be able to process the requests as quickly as they are received, i.e., your application will only support a simple service with one thread. To support more number of clients, a pool of threads is used. If the size of the pool is fixed, an attacker can simply send requests faster than an application can serve. If the pool of thread is not fixed, then the application can absorb lot of DOS attacks without dropping the service entirely. But, then at some point the application is going to consume system resources without bound. Therefore, instead of just breaking the service provided by an application a DOS attack can bring down the entire system. An extension of DOS attack is a DDOS (Distributed DOS) attack. A DDOS attack is carried out from multiple machines to starve the resources of the target system. Figure 2.4 shows a DDOS attack carried out by an IoT (Internet of things) botnet. Several Internet devices (including computers, smartphones, and IoTs) connected and running one or more bots is called a botnet (the network of bots). A botnet is controlled through a command and control (C&C) software. A bot is a software application that rns automated tasks, such as web crawling, etc., over the Internet.

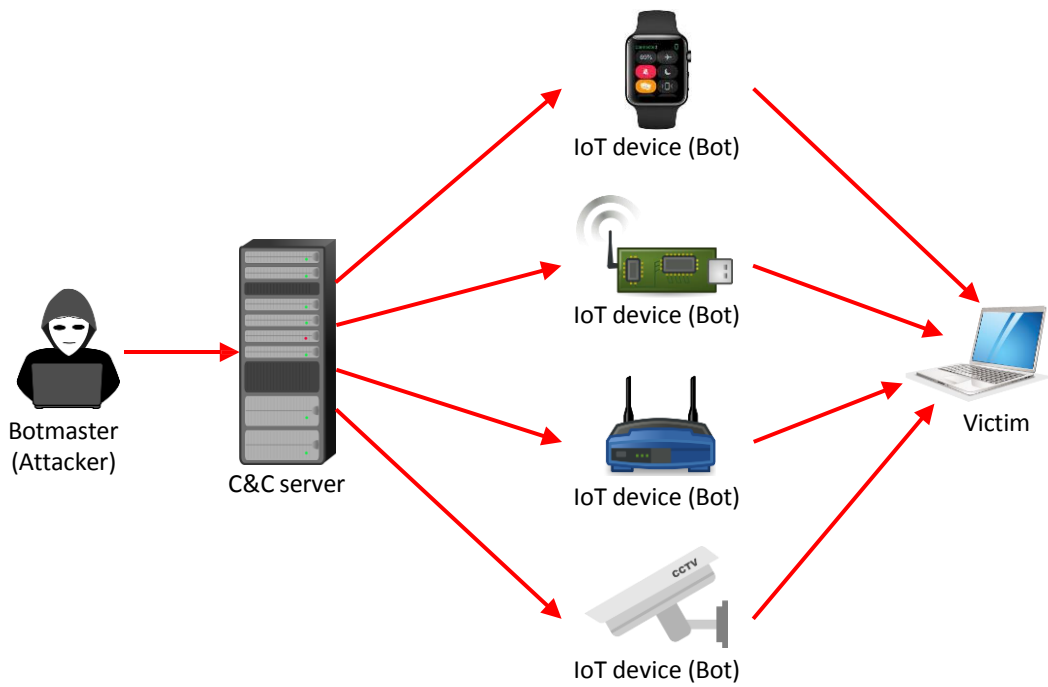

Figure 2.4: A DDOS attack carried out by an IoT botnet

**Mitigation**

☐ *Managing Clients*: A common mistake in writing multithreaded application is to create a new thread every time a new task is initiated, such as a new connection from a client. An attacker can quickly flood (leading to a DOS attack) the server with false or incomplete connections. To avoid this attack, place limits on the number of connections from each client. If a client tries to consume more than the allowed number, then these additional requests from the client are simply refused. Also, make the limit of connections configurable. This will save the server from DOS attack, and the attacker has to launch a DDOS attack to deny service to anyone other than themselves. To mitigate DDOS attack is to implement timeouts. If a client does not send requests in a timely manner, the connection with the client can be closed, making the worker thread available for a new connection. This response can also be based on the utilization of the worker threads. If an attack is perceived, the connections can be dropped more quickly. Do not right away allocate memory/resources just to respond to a client who wants to connect to your application. First validate and authenticate a client before starting any complicated process or allocating much memory for the client. If each client sends 10MB of data, then it will not take many clients to run your machine out of memory. Set limits on the amount of data the application will accept. Make this limit configurable, and drop clients that exceeds this limit.

☐ *System Resource Management*: System resources, such as memory, CPU time, disk space, and open files, etc., are best managed by system quotas. A running application does not have knowledge if these quotas are enabled or not. Therefore, it is better to be defensive and write code that is more aware of the system resource management. This is also part of good programming practice. Some of these good programming practice rules are listed as follows. (1) Minimize the number of file reads and writes. (2) Try to avoid CPU intensive code or loops. (3) Do not use stack for allocating large size of buffers. (4) When possible, avoid using system calls.

☐ *Intrusion Prevention System*: Intrusion prevention system (IPS) is a relatively recent development in cybersecurity. The main purpose of such systems is to monitor network or system activities for malicious activity. Whenever a DOS attack takes place these systems can log information about the attack, report and attempt to block or stop the attack. There are two methods that these systems use to detect attacks. (1) Signature-Based – This method monitors packets in the network traffic with known attack patterns as signatures. (9) Anomaly-Based – This method monitors network traffic and compare it against an established baseline. The baseline determines what is normal and what is anomaly.

### 2.3.5 Poor Error Handling

In general programmers pay less attention to abnormal situations in the code than they do to the normal cases. This makes errors (abnormal cases) a natural path for the attackers to follow. These abnormal cases can be, the network is down, database connection failed, page not found on the server, or memory not allocated, etc. Example 2.3.4 shows that on Error 404 (HTTP standard error for the requested resource not found) sensitive information is revealed/leaked by a server that can be used by an adversary (opponent – e.g., an attacker) to design and launch an attack.

**Example 2.3.4.** Generation of error message containing sensitive information
Following is the message returned by a server for a non-existent page:

```
Error 404 - Page not found.
The requested page /page.html was not fond on this server.
Apache/2.2.3 (Unix) mod_ssl/2.2.3 OpenSSL/0.9.7g DAV/2
PHP/5.1.2 Server at Port 80
```
The server reveals valuable information, web server version, OS, modules, and code used, that can be used by an attacker to design and launch an attack.

**Mitigation**

☐ *Managing Exceptions*: Exceptions allows code to separate between normal and abnormal cases. There are two types of exceptions, *checked* and *unchecked*. Checked exceptions can not be ignored, whereas unchecked exceptions do not have to be declared or handled. To force the programmer to properly take care of the abnormal case catch all exceptions whether checked or unchecked. To accomplish this, handle exception at the top level of the exceptions' hierarchy. In Java the base (top) class is `Throwable`. Example 2.3.5 shows a snippet of Java code for handling exception at the top level.

**Example 2.3.5.** All the top-level Java methods should catch Throwable:
```
protected void getHost(HttpServletRequest req) {
    try {
        String ip = req.getRemoteAdddr();
        InetAddress addr = InetAddress.getByName(ip);
        out.pritnln("Hostname: " + addr.getHostName());
    }
    catch (Throwable e) {
        out.println("Caught at the top level: " + e);
    }
}
```

Certain other rules to follow for managing exceptions are: (1) To avoid duplicated try/catch blocks handle exceptions in a centralized place, and ensure that all abnormal behaviors are correctly handled. (2) Error messages should be verbose enough to enable proper user response, but must not display or leak critical data. (3) Careful testing and verification of error handling code should be performed.

☐ *Return Values*: Functions or methods return a value to convey their success or failure. In case of a failure, this value is used by the calling code to handle the error. The problem with this is that function that returns the value depends on the caller to check the error. Further, this prevents returning something more interesting, like the function result. Moreover, there is no universal convention for communicating error information, so error returned from each function call is handled differently. Uncheck return values become problems and can lead to buffer overflows, information leakage and other software vulnerabilities. To avoid these problems, the designers of C++ and Java included exceptions as part of the language feature. Use exceptions wherever possible but handle/manage them properly as discussed above.

☐ *Managing Resources*: Resources including database objects, file handles, sockets, and allocated memory etc., should be released properly, otherwise they can cause serious performance and information/data leakage problems. There is a special connection between error handling and resource leaks, that error handling should address resource management properly. There are basically three steps to eliminate a resource leak: (1) Detect the leak. (2) Find the source of the leak. (3) Release the resource at the proper location in the code to avoid the leak. For example, if a call to reading a file fails the buffer allocated for the read operation should be freed. If this is not done properly a resource/memory leak will occur, as shown in example 2.3.6. Example 2.3.7 shows a proper way of performing the error handling and hence avoiding the resource leak.

**Example 2.3.6.** A memory leak takes place if the call to reading the file fails:

```
char *readFile(int handle) {
    char *buffer = (char *) malloc(SIZE);
    if (buffer == NULL); {
        return NULL;
    }
    else if (read(handle, buffer, SIZE) != SIZE) {
        return NULL;
    }
    return buffer;
}
```

**Example 2.3.7.** Handling the error (reading the file fails) properly to avoid the memory leak shown in example 2.3.6:

```
char *readFile(int handle) {
    char *buffer = (char *) malloc(SIZE);
    if (buffer == NULL); {
        return NULL;
    }
    else if (read(handle, buffer, SIZE) != SIZE) {
        free(buffer);
        return NULL;
    }
    return buffer;
}
```

☐ *Debugging*: Debugging is one of the best practices to catch any error/bug in the software. This practice gives insight into a running program and assists in finding errors or abnormal behaviors in the program. Logging is another technique that saves hours of manually debugging a program. The problem is *to log or not to log*. Here are some of the good practices to follow taken from [Chess and West, 2007]: (1) Log every important action, including authentication attempts, attempt to modify the ownership of an object, account creation, password reset requests, purchases, sales, or anything of value, etc. (2) Use centralize logging, to provide one consistent and uniform view, and to facilitate changes. (3) Time stamp log entries to form a timeline of events that greatly helps in forensic investigations. (4) Protect the logs, i.e., prevent attackers from gaining access to the log entries.

Keep debugging aids and back-door access code out of production. Isolate the debugging code within the program so that it does not show up in the production code. Back-door code is a form of debugging code that is written to test different components of the software system and allows the test engineers to access the software in a way that are not intended for the end users/customers. If back-door code ends up in the production code then an attacker can use it to bypass security mechanisms and gain access to sensitive parts of the system. There are different methods, such as using version control tools, preprocessors, etc, to keep the debugging code out of production. Example 2.3.8 shows how to isolate debugging code from the production code using C/C++ preprocessors.

**Example 2.3.8.** Keeping debugging code out of production.

```
#define DEBUG
...
#if defined( DEBUG )
// debugging code
...
#endif
```

## 2.4 Software Protection

Brain, the first IBM PC virus was written (for benign purpose) in 1986 by the authors to protect their software from illegal use and copying. Brain effects the PC by replacing the boot sector of a floppy disk with a copy of the virus, making the disk unusable. The virus came complete with the address and phone numbers of the authors and a message asking the users to call them for inoculation. Similar techniques or concept is used now to protect against software piracy and tampering (reverse engineering). Malware writers often use reverse engineering techniques to find vulnerabilities in a software (usually binary programs) system and write malware to exploit the system. Protecting software from reverse engineering is an often overlooked software engineering topic, and has no easy answers. Section 4.3 in Chapter 4 discusses and explains several reverse engineering techniques. Here before discussing some of the anti-reverse engineering techniques, we first elaborate upon why and how of software protection.

### 2.4.1 Why and How

The first thing an attacker is going to ask before cracking a software is, if the software is popular and in high demand. For example, there are more malware programs written for Android than iOS. The reason is the popularity and demand of Android. This way the exploit written has a greater impact and produces more benefits (financial and others) for the attacker. On the other hand the software company will try to make cracking the software a difficult or impossible task. The second question an attacker is going to ask is, if the software is highly and genuinely valuable, such as special algorithms, or applications controlling and managing sensitive facilities/resources of a state or a company, etc. This way also, the exploit written has a greater impact, and more professional actors, including states, get involved in researching and developing the exploit. In such cases, it is worth spending large amounts of time and money to protect the software.

When writing and developing a software protection mechanism, it should be assumed that the attacker has absolute control over the physical and software components of the system on which the application is running. The basic goal of software protection is to increase the level of competence and time required to breach the software. The achieve this goal stealthy techniques are used to protect the software. For example a strong encryption only serve to lead the attacker directly to the location of protection. Then it is a just a matter of time before the software is breached. The location and effects of protection should be hidden, so that the attacker finds it difficult to trigger the protection.

There is always some kind of cost involved when protecting software. The cost to breach the software is always much less than the cost to develop the protection. There are certain type of costs that one should keep in mind when developing software protection. (1) *Developing Time* – This is the time to develop effective software protection. Skilled and experienced programmers are required to write good software protection. Moreover, the time spent on writing software protection is the time away from the application development. Sometimes, a third-party software protection is used. Well known and commercial software protections are known to the attackers and can be bypassed. So, it is always wise to supplement it with in-house protection mechanisms. (2) *Debugging Time* – While developing software protection new bugs may be induced into an otherwise working application. Moreover, protected software itself is difficult to debug. With more new bugs the time to debug the application is significantly increased. (3) *Maintaining Time* – Software protection makes it difficult to understand the code, which in turn makes it difficult to maintain. After months or years of the protection being implemented, the maintenance engineers may no longer be able to understand the protection or the code it protects.

### 2.4.2 Anti-Reverse Engineering Techniques

#### Checksums

Checksum is a datum (a single piece of information usually numerical) used as a signature of a program to check for any modifications done to the program. A good checksum algorithm will generate a significantly different value even for small changes made to the input. That is why usually a cryptographic hash function (CHF) is used for generating the checksum. CHF is a special hash function with the following properties: (1) *Compression* – Output of a CHF, i.e., the hash value, is fixed size. (2) *Deterministic* – A CHF produces the same hash for the same input, no matter how many times it is used. (3) *One Way* – There is no practical or feasible way to invert the hash. (4) *Collision Resistant* – It is impossible to find any two inputs that hash to the same value.

### Obfuscations

Obfuscation is a technique in which a program is transformed into another program that keeps the behavior of the original program but is hard to understand. This makes it non-trivial to reverse engineer the obfuscated program. The quality of obfuscation depends on these four metrics [Collberg et al., 1998]: (1) Potency – How much more obscure an obfuscated program becomes. (2) Resilience – How much an obfuscated program holds up to an automatic reverse engineering tool. (3) Stealth – How well an obfuscated program blends in with the rest of the program. (4) Cost – How much time and space is required to obfuscate a program. Section 4.5.2 in Chapter 4 presents and discusses some of the obfuscations, including some basic and advance techniques, used by malware (malicious program) to escape detection, and can also be used for software protection. A more complete list of different obfuscation techniques can be found in [Nagra and Collberg, 2009].

### Watermarking

Software watermarking is a defense technique whereby a signature is embedded inside a software application to reliably representing and identifying the owner. The embedded signature should be resilient to code transformations, optimizations and obfuscations,, i.e., the signature should be stealthy and hard to tamper with. Moreover, the signature should be retrievable. There are two types of software watermarking, static and dynamic [Nagra et al., 2002]. *Static* watermark is defined as one which is stored in the application executable itself. *Dynamic* watermark is defined as one which is stored in a program's execution state, rather than in the program code itself. Some important metrics [Nagra et al., 2002] to measure the effectiveness of a software watermarking technique are: (1) The computational cost of developing, embedding, and recognizing the watermark. (2) The running time and memory consumption of embedding a watermark into a program.

### Hardware

There are different methods that the hardware can be used to protect the software. Here we discuss few of these methods. (1) A typical hardware based protection is to use a trusted processor to execute the software. The trusted processor uses specific cryptographic techniques to check and verify a software application before executing the application. This makes it strenuous to execute the application on any other processor, and becomes enigmatic for the pirates to produce the same hardware. (2) Another approach used is to associate the software with a particular machine by secretly sending the serial number associated with the hardware to the software company. (3) A smart card or a dongle – trusted hardware – is used to protect the software, by preventing the software from executing unless some specific information from the trusted hardware is retrieved.

## 2.5 Designing Secure Software

Different software systems have different security requirements depending on type of the system and environments where the system is deployed. For example, software system for a bank requires a higher level of security than for a university. Despite the presence of these differences among systems, there are some general guidelines that are universally applied when designing system security solutions. Here we discuss some of these guidelines. For a full list and discussion readers are referred to [Sommerville, 2016].

- □ *Base Security Decisions on an Explicit Security Policy* — A security policy of an organization explicitly lays out fundamental security conditions for the organization. Software and system designers should use the security policy as a framework for making and evaluating design decisions. For example, a hospital security policy may state that only the authorized clinical staff may modify the patient records. Design of the access control for the hospital should be based on such security policies. If an organization do not have an explicit security policy, which is generally the case, then the designers may have work this out from examples and confirm with the managers of the organization.

- □ *Defense in Depth* — In any critical software system, it is a good design practice to avoid a single point of failure. For software security this means, instead of relying on a single mechanism to ensure security employ several different techniques. This approach is sometimes called *defense in depth*. For example, using multifactor authentication to recognize and authenticate users to a system.

- □ *Fail Securely* — Failing securely means, when a system fails use fallback procedures that are as secure as the system or more secure than the system itself. So that when the system fails an attacker should nt access data that would not normally be allowed. For example, if a server in a hospital fails before deleting certain patient data from the database, that data can be accessed by an attacker. The patient data should be encrypted before storing on the client, so that an unauthorized user cannot read the data.

- □ *Compartmentalize Assets* — Compartmentalizing means organizing the information in a system into compartments. A user should have access to the information that they need for their work, rather than to all of the information in a system. An attack compromising a user account can be contained. This means some information may get damaged or lost, but not all of the information in the system. For example, clinical staff should not be given access to all the patient records, but only to those patient records who have an appointment with them. If an intruder steal their credentials, then the intruder cannot damage all patient records.

□ *Design for Recovery* — A system should always be designed for recovery. That means the designer should always assume that a security failure could occur, and a recovery mechanism should always be designed so that the system can be restored to a secure operational state. For example, if the password authentication subsystem is compromised, then there should be a backup authentication sbsystem that can replace the compromised subsystem.

## 2.6 Secure Programming

Most successful attacks on software rely on software vulnerabilities that were introduced during the program development. We have already discussed some of these vulnerabilities and their mitigation in section 2.3. There are some universally accepted programming practices and principles that reduces the chance of introducing software vulnerabilities into programs. Here we discuss some of these basic principles of secure programming. For a full list and discussion readers are referred to [Sommerville, 2016].

□ *Control the Visibility of Information in a Program* — A principle that is adopted by military organizations is the *need to know* security principle. Individuals (users) are given access to only a particular piece of information that they need to carry out their duties. When programming control access to the variables and data structures that you use. Hide these from the components that do not use them. For example, abstract data types, in which the structure and attributes are not externally visible, and all access to the data is through operations.

□ *Minimize the Use of Error Prone Constructs* — There are some programming approaches/constructs that are more likely to cause and introduce errors in programs than others. These constructs should be avoided as much as possible. Some of these approaches/constructs are: `goto` statements, dynamic memory allocation, aliasing, and unbounded arrays, etc. For example, dynamic memory allocation where storage is explicitly managed. it is very easy to forget to release the storage in the program, that can lead to a memroy leakage and eventually can become a security hazard.

□ *Check Array Bounds* — Not all programming languages automatically check array bounds. For example the two most popular languages C and C++ does not provide any standard construct or library for checking array bounds. For example in C/C++, array[123456] would access the word that is 123456 locations from the beginning of the array, irrespective of whether or not this was part of the array. This lack of bound checking leads to security vulnerabilities, such as buffer overflow discussed in section 2.3. For these and other such languages there is a need to include checks that the array index is within bounds.

□ *Check all Inputs for Validity* — We have already discussed some general techniques about checking and validating inputs in section 2.3. Wrong input values may lead to produce wrong results and may cause system failure. Here we discuss different type of checks specifically used during software coding and programming. (1) *Range Checks*: Check if the input is within the range. For example, a sensor value should be within a particular range, e.g., temperature value between 0 and 100 degrees Celsius. (2) *Size Checks*: Size of the input could be important for various financial transactions, such as 8 characters to represent a bank account, and less than 40 characters to represent name of a person, etc. (3) *Representation Checks*: There are specific formats to represent different inputs, such as name of a person does not include numeric values, and the two parts of an email address are always separated by a @ sign, etc. (4) *Reasonableness Checks*: If the input belongs to a series and there is a relationship between members of the series, then we can check if the input is reasonable or not. For example checking for outliers in the readings of a household electricity smart meter. The current amount of electricity used should be approximately the same as in the previous years. If there is a huge difference (from 3000 kWh $\rightarrow$ 10 kWh) then that means something has gone wrong, e.g., the reading of the smart meter has been manipulated by violating the software.

## 2.7 Security Testing and Assurance

To assure the security of a software system, security testing [Potter and McGraw, 2004, Sommerville, 2016] of the system is carried out. The verification and validation of software system should not only look for the correctness of the system and that the system is fit for purpose, but also focus on security assessment. During security assessment, the ability of the system to resist against different types of attack is tested. One of the other goals of security assessment is to find vulnerabilities that can be used by the attackers to gain access to the system or cause damage to the system or its data. To carry out security assessment of a software system different types of testing is carried out, some of them are discussed below.

□ *Experience-Based Testing* — This testing is based on the experience of the validation team. The team analyses a system against the types of attacks and vulnerabilities that are known to them. For example, testing for buffer overflow errors, validating all inputs, and so on. They also check if the design and programming guidelines have been followed.

□ *Penetration Testing* — This testing is based on the experience of the validation team and as well as experiences drawn from outside the validation team. The main goal of this testing is to break into the system. For this purpose they simulate different type of known attacks, and also develop new (unknown) attacks based on their ingenuity. One of the main advantages of penetration testing is that it simulates the behavior of a real cybercriminal to uncover the critical security issues in a system. This also uncovers the steps carried out to exploit a vulnerability and the steps required to fix the vulnerability, before it is exploited for real.

□ *Tool-Based Testing* — This testing is based on the use of security tools such as static and dynamic analysis tools to find errors and vulnerabilities in a software program. Anomalies revealed during this analysis can be directly fixed or help identify part of a system that requires more testing to find the real possibility of a risk to the system. In this approach experience of security flaws is embedded into the tools used for testing.

The main problem with security testing is the time available for testing. The validation team has a limited time to test the software, because of reasons beyond their control, such as the pressure from the customers to release the software as soon as possible, and other financial constraints. The time available to attackers is almost unlimited. They can spend more time discovering vulnerabilities than the system test engineers. They are also willing to experiment and try things that are outside the normal activities. A better approach to testing is a risk-based [Potter and McGraw, 2004] analysis of a system, that focuses on the most significant risks faced by the system.

## 2.8 Future Work

The number of attacks and threats posed to software are increasing rapidly. Therefore, research in software protection is growing at a fast pace. To encourage further research in this fast-growing area, this section discusses some future research directions including the problems and challenges faced.

### 2.8.1 Code Obfuscation

To protect software we want to make exploitation of vulnerabilities harder. One of the basic techniques used for this purpose is code obfuscation. This makes the code more complicated and harder to comprehend, and hence difficult to find vulnerabilities. Even if adversaries get hold of the code it takes more time and energy, and in turn makes reverse engineering of the code harder and costlier. Some of the future research opportunities or gaps in this area are [Hosseinzadeh et al., 2018]: (1) Lack of a standard metric for reporting the overhead and effectiveness of the obfuscation techniques. More empirical studies are needed to measure and present the performance and space requirements of the different obfuscation techniques. (2) There is a need to do more research on applying the obfuscation techniques to other domains/environments, such as mobile platforms, cloud computing, and IoT. For example Android applications are vulnerable to reverse engineering. Most of the studies on obfuscation in mobile environments only discuss techniques used by malware to evade detection. There is a need to develop new obfuscation techniques for mobile platforms to increase their software security. There is a limited number of studies on the obfuscation techniques for securing cloud computing and IoT environments.

### 2.8.2 Smart Fuzzing

Fuzzing is the most widely used technique in practice for finding vulnerabilities in software [Li et al., 2018, Godefroid, 2020]. It generates and executes tests automatically with the goal of finding vulnerabilities. Compared to static analysis fuzzing produces less number of false positives, but is more compute intensive. Any application that processes untrusted data and input must be fuzzed. The main processes of fuzzing are shown in Figure 2.5. Fuzzing starts with the generation of a set of program inputs (testcases). The input should meet the requirements of the target program input format as far as possible. The generated inputs are fed to the target programs. One of the main challenges of fuzzing is that the inputs should be broken enough to fail the program. Therefore, one of the future works in fuzzing is to automate the generation of input grammars for complex formats, possibly using machine learning or some other techniques. Another future work is the effective fuzzing of large distributed applications such as the entire cloud services. Smart fuzzing is another area to improve, such as collecting as much information about the target program as possible to improve fuzzing.

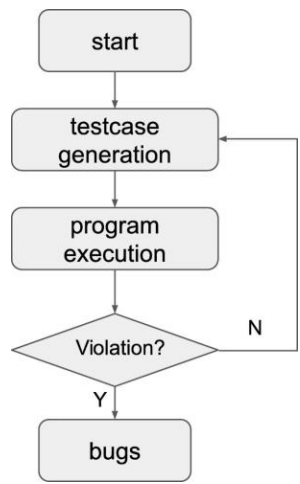

Figure 2.5: Main processes of fuzzing: source [Li et al., 2018]

### 2.8.3 Secure Programming Languages

Programming languages are used to communicate ideas between humans and machines, and design and write computer programs that perform some kind of computation. Recently, because of the importance of designing secure information systems, programming language researchers have made security a prime concern in language design and implementation.

**Safe Languages**

Many attacks exploiting software vulnerabilities in language design and implementations. The most well known example of such an attack is the buffer overflow attack (Section 2.3.1, such as the `strcpy` function in C and C++, in which the overflowing data overwrites unallocated memory e.g., for malicious purposes. Therefore, a safe language is crucial for designing secure systems and avoiding software vulnerabilities, and provides more principled defense against buffer overflow and other exploits [Salka, 2005]. For example, Java employs dynamic checks to enforce safety. Safe languages like PLT, a safe dialect of Lisp, is used to write an advanced web server that, unlike web servers written in C like Apache, is not subject to buffer overflow exploits [Graunke et al., 2001].

**Language-Based Access Control**

A programming language provides access to local resources such as CPU, memory, and files, which can be exploited by the attackers. *Sandboxing* is a technique that is used to access control to resources. Java and other such language platforms can achieve sandboxing by combining lower level safety and security controls. For example, Java prevents untrusted platforms such as Applets to load classes with the same name as the existing library names to protect against spoofing attacks (a program identifying as another).

**Language-Based Information Flow**

We explain the importance of secure information flow with the following example. If `H` is a variable and has a higher security level than the variable `L`, then we would allow flows from `L → L, L → H, H → H`, but not `H → L`. That means:

the following explicit flows
`H = L;` and `L = 1234;` are legal
whereas, the following explicit flow
`L = H;` is illegal
and also illegal is the following implicit flow

```
if ((H % 2) == 0)
    L = 0;
else
    L = 1;
```

This copies (the information leaks) the last bit of `H` to `L`

If we do not check the flow of information from `H → L` then some information can be leaked. To check and catch these insecure and illegal information flows, information flow analysis analyzes a software program before its execution to inspect and propel the security of data that flows through the program. In this, program variables and statements are augmented with labels to specify data use policies. For example, Jif (Java + information flow) [Myers, 2021, Sabelfeld and Myers, 2003], an extension to the Java language, has labels that express security policies to restrict the information flow in a Jif program. A Jif compiler will flag the above implicit flow (the `if` statement) as illegal, because the label associated with it is high (`H`) while there is assignment to the low label variable (`L`).

**Unsafe Languages**

How to make unsafe languages, such as C and C++ safer? The researchers have extended to create safe variants, CCured [Necula et al., 2002], Cyclone [Hicks et al., 2003] and others, of these languages. All of these extensions bear some form of weaknesses, such as an unacceptable runtime overhead, and incomplete detection of memory bound violations etc., and as such these extensions are not widely accepted or used in practice. Because of their popularity, these languages are still in use, and moreover, lot of legacy code is written in these languages. Therefore, there is a need to develop methods and techniques to automatically secure a program written in these languages.

Research in these topics (discussed above) is ongoing, but the current most important task at hand is the successful integration of modern language security technologies. Why developers repeat the same mistake? Instead of relying on programmers we should develop tools that systematize the known security vulnerabilities and integrate them directly into the development process [Evans and Larochelle, 2002].

# Chapter 3

# Hardware Security

*To know your Enemy, you must become your Enemy.*
(Sun Tzu, The Art of War, 500 BC)

## 3.1 What is the Problem

Hardware security is a new and compounded research area. The purpose of this chapter is to introduce briefly the topic of hardware security. We have kept the discussion simple and to the point, and referred further readings to get a thorough understanding of this subject area.

Computer hardware consists of all the physical parts of a computer, including central processing unit (CPU), motherboard, monitor, case, mouse, keyboard, data storage, etc. Most of the time, hardware is directed by the software to execute instructions to perform certain tasks, but sometimes it is just the hardware that accomplishes a task.

Software functions and primitives either running or implemented on a hardware platform assume that the hardware is resilient to attacks. This is not always true, as shown by recent vulnerabilities found and attacks on some of the hardware platforms. The first real world backdoor in a military FPGA (field programmable gate array – a chip that can be reprogrammed, also called dynamically programmable chip) was detected in the year 2012 by [Skorobogatov and Woods, 2012]. Backdoor is a covert method to gain remote access to a computer by bypassing normal authentication processes. Using this vulnerability an attacker is able to perform malicious actions, such as reprogramming crypto and access keys, modifying low level silicon features, reverse engineering the design or permanently damaging the device. The most damaging possibility in this case was that it was not possible to patch the backdoor in the FPGAs already deployed. Organizations using this FPGA either have to live with the fact that they can be compromised or physically replaced after a redesign of the silicon itself. There are other examples [Rostami et al., 2014] of finding such vulnerabilities in smart cards and microcontrollers (small computers embedded inside devices). Counterfeit (imitation of something authentic for malicious purposes) electronics is prevalent in computers, automobile, control, and defense systems [Rostami et al., 2014].

**DEFINITION**: A hardware *vulnerability* is a set of conditions, bugs, flaws, or weaknesses in the hardware that can be exploited by an attacker. A hardware *exploit* is any malicious action that takes advantage of a vulnerability in a hardware system to cause damage to the system. A *mitigation* is a method, technique, or process that prevents or limits exploits. **Hardware Security** is a process/practice to either mitigate or design and develop hardware without vulnerabilities to escape exploitations.

## 3.2 Impact of Hardware Vulnerabilities

The impact of hardware vulnerabilities on consumers and businesses is in the form of, loss of data, privacy, other resources, such as time and money, destruction of hardware, etc. Following are some of the key impacts because of a vulnerability in a hardware device.

**Denial of Service**: In this attack a chip is rendered partially or fully unusable. The performance of the chip degrades and the functional units of the chip, such as registers, data path, logic units, remain in an intentionally induced state of overload or deadlock. Eventually, either the chip is reset or replaced.

**Information Leakage**: An attacker gains access to a chip and leaks sensitive information, such as the design of the chip, technical and proprietary details about the chip, encryption keys, trade secrets, etc. This sensitive information can lead to further exploits and attacks on the chip or on the overall system, including the hardware and software. This can lead to serious financial damages to the companies, and organizations owning or using the chip. The sensitive information leaked can be used to espionage and finally lead to sabotaging the critical infrastructures of states and organizations.

**Reduce Reliability**: Reliability is the quality of a hardware for being trustworthy or performing consistently well without failures. If the reliability of a chip is reduced, the software running on it suffers and may fail. This is one of the root causes of software failures. Moreover, reliability of data storage is paramount for modern hardware and software systems. Unreliable chips can not be installed in a mission critical system, such as nuclear power plants, medical devices, and intelligent transport systems, etc.

Hardware is everywhere, even where there is no software. It runs your car, controls your cell phone, household items, mission critical systems, and saves lives in the form of medical devices. It is the backbone of banks and nations' power grids. As businesses, organizations, individuals and nations depend heavily on hardware, we have to make it better and more secure. Hardware security is no longer a luxury but a necessity. Cybersecurity starts with first securing the hardware.

## 3.3 Hardware Vulnerabilities and Mitigation

Some of the main goals of the adversaries attacking hardware are, leakage of sensitive information, reducing reliability, denial of service, stealing the design, and identifying trade secrets. Following type of attacks [Rostami et al., 2014, Prinetto and Roascio, 2020] on the hardware are possible to achieve these goals.

### 3.3.1 Hardware Trojans

Any malicious modifications introduced during the design and manufacturing of a chip is referred to as *Hardware Trojans* [Chakraborty et al., 2009]. Due to the stealthy nature (precondition for activation is a very rare event) of Hardware Trojans, it is difficult to detect them. Activation of a Trojan depends on a condition, such as certain values from sensors, a particular input pattern, or value of an internal counter. The possibility of vast spectrum and large number of Trojan instances, and large variety of structure and operation mode an adversary can employ also make Trojans difficult to detect.

**Mitigation**

Unlike Software viruses/Trojans Hardware Trojans are difficult to detect and eliminate, and therefore, are more harmful to computer systems. Lot of research have been done on detecting Hardware Trojans, and different detection and mitigation methods have been proposed. These methods can be divided into three main categories [Jin, 2015, Becker et al., 2017]. Here we provide a brief discussion of these three categories. For a more complete list and detail discussion of detection and mitigation methods of Hardware Trojans readers are referred to [Chakraborty et al., 2009].

1. **Enhanced Functional Testing** — This method is based on the fact that Hardware Trojans are triggered by rare events. Therefore, monitor all the rare events during the testing stage of a chip. The problem with this testing is that there is no standard definition of rare events. This makes it difficult to generate all the rare event patterns for testing.

2. **Side-Channel Analysis** — Every electronic device when in operation emits different signals, such as electrical and magnetic fields. These signals can be analyzed to measure and gain information about the state and the data which the device processes. These measurements if different from the Trojan-free device indicates that there may be a Trojan present on the device. For this purpose, using advanced data analysis methods a side-channel fingerprint is generated of the Trojan-free device and is then utilized to detect Trojans. Various side-channel parameters, such as The success of this method depends on the availability of a golden model of the device for comparison.

3. **Hardware Obfuscation** — The first step in counterfeiting or stealing the design of a chip (to introduce a Hardware Trojan) is to reverse engineer the chip. Obfuscation methods are used to defend against reverse engineering the chip. For a detail discussion of these and other methods readers are referred to [Becker et al., 2017]. Here we just list couple of these methods: *Camouflage Gates* – The logic function of the gates used in the design of a chip are camouflaged to make them hard to detect visually. *Obfuscating the Connectivity* – Connectivity of individual gates reveal a lot of information to a reverse engineer. Generally, dummy wires are used to conceal the connectivity.

### 3.3.2 Side-Channel Attacks

When a computer is in operation, it performs certain activities, such as performing computations, reading and writing memory, consuming power, releasing energy in the form of electromagnetic waves, and producing sound, etc. All these activities if monitored properly with other structural information about the computer and algorithms executed may help an attacker gain access to sensitive information. Side-channel attacks are based on the information leaked through a side-channel as listed above (such as power, and sound, etc). We divide side-channel attacks into two different types, *Passive Attacks* and *Active Attacks*.

**Passive Attacks**

These attacks just perform reading (i.e., they are passive) of the sensitive information from the side-channels. General classes [Prinetto and Roascio, 2020] of passive side-channel attacks are:

- □ **Timing Attacks**: These attacks are carried out using the timing information of different operations performed in the hardware. For example, processing of passwords or encryption keys have a correlation between the computation time and the actual values of the input. If an attacker knows this correlation, the sensitive information (the password or the encryption key) can be extracted. Some of the possibilities and examples of timing attacks are presented in [Kocher, 1996, Bernstein, 2005].

- □ **Power Attacks**: Similar to the timing attacks, this attack is carried out using the power consumption information of different operations performed in the hardware. Power consumed during different operations is different, so by measuring this variation sensible data is read from the device. Some of the possibilities and examples of power attacks are presented in [Kocher et al., 1999].

- □ **Acoustic Attacks**: These attacks exploit the sound and vibrations produced by different hardware components during a computation. These acoustics are recorded by a covert device placed close to the computer. Latter these acoustics are analyzed using signal analysis and machine learning algorithms to detect the keys pressed [Koushanfar, 2012, Halevi and Saxena, 2015] or correlate the acoustic emission with the computer activity [Genkin et al., 2014].

- □ **Optical Attacks**: These attacks are carried out using the light emitted by the hardware (e.g., a transistor used as a switch) in the form of photons for a very short period. Light emitted by the LEDs (light emitting diodes) that may be used to indicate the device activity can also be exploited to steal sensitive information from the device. Some of the possibilities and examples of optical attacks are presented in [Ferrigno and Hlaváč, 2008, Loughry and Umphress, 2002].

☐ **Electromagnetic Attacks**: An electromagnetic field is produced around an operational hardware due to the current flowing in the hardware circuit. The radiation caused by the electromagnetic field carries useful information about the original digital source. By capturing this information the original digital signal can be reconstructed [Yilmaz et al., 2019], which can lead to information leakage.

**Active Attacks**

During these attacks, faults are introduced (i.e., they are active) into a chip so that it becomes easy to steal sensitive information from the side-channels present/available in the chip. These attacks are also called *Fault Attacks*. General classes [Prinetto and Roascio, 2020] of active side-channel attacks are:

☐ **Power Supply Attacks**: In this attack power supply of a chip is tapped to wither increase or decrease the power supplied to the chip. In case the power is decreased, the delay of logic gates increases and wrong values may get injected. This implies that one or more faulty bits as injected into he system [Dusart et al., 2003]. In case the power is increased, a serious damage can be inflicted on the chip.

☐ **Glitch Attacks**: Similar to power supply attacks, it is possible to change the clock speed of a chip. For example, by doubling the speed of the clock some of the old data could be used as the new data has not arrived yet [Kömmerling and Kuhn, 1999]. This can cause corruption of some of the data. This process introduces a glitch which can also influence conditional jumps, by either not performing them or performing them at wrong time [Anderson and Kuhn, 1997].

☐ **Heating Attacks**: Raising the temperature of a chip generates random currents, which may lead to bit flipping. This can corrupt the data stored inside memory of the chip. This can also cause aging of the circuit, and in extreme cases when the temperature rises to a certain threshold can also destroy or damage the chip [Hutter and Schmidt, 2013].

☐ **Optical Attacks**: Like heating attacks, a light with specific wavelength focused on a chip can also flip the bits. For this to be successful the light should be able to reach the chip, and any protective layer needs to be removed. An example of such a precise attack is presented in [Skorobogatov and Anderson, 2002].

☐ **Radiation Attacks**: A simple and practical method to induce faults in a chip is to cause strong electromagnetic disturbances near the chip. This induces currents in the circuit which in turn alter the level of a signal, i.e., it gives the attacker control over the bits [Schmidt and Hutter, 2007]. At a certain level of disturbance the components of the chip may stop working or even be physically destroyed [Schmidt and Hutter, 2007].

**Mitigation**

Side-channel attacks rely on the information present in the leaked data. Therefore, to mitigate and counter these attacks, there is a need to eliminate either the information leakage or the correlation between the leakage and the secret data. In the second case it will be difficult to make any sense of the information leaked. There are different methods and techniques [Rostami et al., 2014] used to achieve these goals, here we discuss few of these methods.

1. **Leakage Reduction** — These methods reduces the relationship and dependency between the side-channel leakage and the secret data. For example to decrease the data dependencies of the power consumption, various dynamic and differential logic styles have been proposed [Tiri et al., 2002]. Other techniques used to decrease the data dependencies are: hiding – making the leakage constant; masking – making the leakage dependent on some random value [Goubin and Patarin, 1999]; encryption of buses [Brier et al., 2001]; and time randomization [May et al., 2001].

2. **Noise Jamming** — Jamming the channel with artificial noise reduces the amount of information in the side-channel leakages. Artificial noise can be introduced by dummy circuits or operations [Joye, 2009]. This makes it difficult, i.e., increase the required work, for an attacker to extract the secret data.

3. **Data Scrambling** — This method scrambles the data to make it difficult to accumulate the secret data. The scrambling is not random and uses a predefined sequence of the secret data (e.g., secret keys) and synchronized timings to ensure that the data is consistent for both the communication parties [Kocher, 2003]. Similar to data scrambling, blinding techniques [Chaum, 1983] can be used to prevent side-channel attacks. Blinding can change the algorithm's input into some unpredictable state, hence preventing some or all leakage of sensitive information.

4. **Physical Unclonable Functions (PUFs)** — PUF [Gassend et al., 2002] is a function implemented in a chip that maps challenges to responses that can be used to authenticate the chip. PUF exploits the inherent randomness during manufacturing of a chip and provides a unique *digital fingerprint* to the chip. Due to the effectiveness of side-channel attacks against PUFs, circuit countermeasures should be used as proposed in [Rührmair et al., 2013].

### 3.3.3 Physical Attacks

So far we have discussed non-invasive hardware attacks, that can be carried out without any physical contact with the hardware device. *Physical attacks* are carried out by taking invasive actions against the attacked hardware. These actions could include depackaging or disconnecting internal components of the device, etc. Physical access to the device is required for these type of attacks. In the following paragraphs we discuss some of these attacks.

**Microprobing**

Microprobing is the process of affixing microscopic needles onto the internal wiring of a chip in order to steal information or to carry out fault attacks. A well known method for getting access to the internals (circuit and paths) of a chip is by making a hole through the chip with a Focused Ion Beam (FIB) [Wang et al., 2017]. This hole can be filled latter with metal such that a surface contact is created. This can be used by an attacker to access the internal lines for eavesdropping on a signal.

**Reverse-Engineering**

The main goal of reverse-engineering is to understand the structure and functions of a chip. Several techniques and tools [Torrance and James, 2009, Quadir et al., 2016] have been developed to reverse engineer a chip. In general, all the layers used in chip fabrication are removed one be one to determine the internal structure or functions of the chip. Reverse-engineering can be used both for benign (to verify the chip design, to detect hardware Trojans, etc.) or malicious purposes (stealing or pirating design of the chip, etc.). A successful reverse-engineering attack can help the attacker understand the chip technology, and eventually illegally fabricate the chip [Quadir et al., 2016].

**Data Remanence**

Data Remanence is the remnants of the data that persists even after erasing or destroying the data from the storage media. Computers usually store their secret data in volatile memories. Once the power is down the content of volatile memories are erased. In practice, that is not the case. The information stored in these memories is in the form of a charge. After the power is down the charge takes some time to decay. During this time an attacker can read the contents of the memory. An attacker usually uses liquid nitrogen to increase the decay time of the charge and gains precious time to dump the memory for subsequent analysis. This attack is called a *cold-boot attack* [Halderman et al., 2009]. Recent research has also shown that even some sensible information can be extracted from non-volatile memories [Skorobogatov, 2005], and the hard disks even if the hard disk is fully encrypted [Halderman et al., 2009].

**Mitigation**

1. **Shielding** — In this method a shield is placed on the outer most layer of a chip. This shield carries signals and helps detect holes milled by FIB [Wang et al., 2017]. This method provides defense against microprobing attacks.

2. **Obfuscation and Camouflaging** — Hardware obfuscation and camouflaging techniques discussed in section 3.3.1 can thwart reverse-engineering attacks. The design layout can be obfuscated, and a camouflaged layer can be added on top of obfuscation.

3. **Protect Data** — In this method the secret data is stored outside the memory, such as storing the data using the registers of a microprocessor (the chip) [Müller et al., 2011]. The secret data, such as the secret keys are stored in processor registers throughout the operational (when data is in motion or use) time of the system. This substantially increases the security of the system and especially provide a strong defense against attacks that target the volatile memory, such as cold-boot attacks.

4. **Memory Scrubbing** — To protect data on a hard disk, one of the techniques used is to fully encrypt the hard disk. This capability is currently provided by Windows, Linux, and Mac operating systems. Recently researchers have shown [Halderman et al., 2009] that such systems can be compromised using a cold-boot attack. The basic countermeasure is to avoid storing the keys in the memory. Even if they have to be stored in memory, software should overwrite keys in the memory (scrub) when they are no longer required and store them at a place other than the memory, such as processor registers etc.

5. **Trusted Platform Module** — Trusted platform module (TPM) is a processor that can provide additional security capabilities at the hardware level. A TPM chip is typically found on a computer's motherboard. TPM allows hardware-based cryptographic operations and is commonly associated with ensuring boot integrity and full disk encryption. TPM is also used to protect the keys to encrypt the storage devices. Recently researchers have shown [Halderman et al., 2009] that such systems can be compromised using a cold-boot attack employing a USB bootable drive.

## 3.4 Security/Kill Switch

The purpose of a security or kill switch is to protect a computer (laptops, desktops, smartphones and other similar devices) from unauthorized access and prevent malicious manipulation of the computer [Adee, 2008]. A kill switch is any manipulation of the chip's software or hardware that would instantly destroy (kill) the chip. For example, shutting off a missile launching electronics. Whereas, a security switch or a backdoor provides access to the system through software or hardware to disable or enable a specific function. For example, an attacker can use it to bypass encryption. Chip alteration can be done even after the manufacturing and packaging is done. A skilled technician can use a FIB to edit a circuit similar to using an eraser and a pencil to it. The process is also referred to as microsurgery and can take from hours to several days. This kind of editing is done by companies when they are designing and debugging prototype chips. But, the same process can be performed by an adversary with a substantial amount of experience and skill in digital circuitry.

Generally the purpose for these kind of attacks is to sabotage mission critical and highly classified chips. Recently, companies and other organizations have also started using these techniques to prevent malicious manipulation of their devices. Several devices, mostly in the smart home space, designed by Google, Apple and other tech giants are already using hardware kill switches [Schwab, 2019] to protect the privacy of their users. But, kill switches are still far from commonplace, especially in mobile phones that need them the most.

## 3.5 Security Testing and Assurance

To assure the security of hardware, testing [Majzoobi et al., 2008] of the hardware is carried out. To test if a chip is behaving as expected, testing is usually performed during the design and manufacturing processes. This type of testing is very good for identifying accidental design and manufacturing flaws but may not be suitable for testing against various hardware attacks. To carry out security assessment of a chip different type of testing is carried out, some of them are discussed below.

- □ *Hardware Fault Injection* — During this testing all the known vulnerabilities are injected into the hardware device for testing the resilience of the device against these vulnerabilities. This is mostly based on the experience of the testing and validation team. There are two approaches [Hsueh et al., 1997] used to inject the vulnerabilities into a chip: (1) Hardware Fault Injection: This approach requires an additional hardware to inject faults into the target, and is very useful in injecting low level faults into locations that are inaccessible to software. (2) Software Fault Injection: Software implemented falt injection does not require any extra expensive hardware. Unlike hardware fault injection methods, they can be used to target applications and operating systems.

- □ *Penetration Testing* — Hardware penetration testing is based on the experience as well as ingenuity of the testing and validation team to generate known and also new (unknown) attacks. The main purpose is to exploit a vulnerability present in a device and break into the device. The main advantage is that it simulates the behavior of a cybercriminal to uncover and unfold critical security issues in a device. This gives in detail the steps required to exploit a vulnerability and carry out the attack and also the steps required to fix the vulnerability before it is exploited for real.

- □ For examining the security properties of PUFs four different testing approaches are proposed [Majzoobi et al., 2008]: (1) Predictability – This test identifies the difficulty of predicting the output of a PUF for a given input, i.e., testing how difficult it is to predict the PUF response to a new challenge. (2) Collision – This test checks how often two different PUFs produce the same outputs to the same given inputs, i.e., testing for collision between two different PUFs. (3) Sensitivity – This test checks the ability of a PUF to operate in a stable way even when the components are imperfect, i.e., testing the PUF sensitivity to component imperfections. (4) Reverse Engineering – This test checks the hardness of a PUF components, i.e., testing resilience of the PUF to reverse engineering.

The problem with hardware testing is that an exhaustive testing of a hardware device is never possible. Moreover the time available to a testing team is limited because of reasons beyond their control, such as pressure from the customers to release the product as soon as possible and other financial constraints. Whereas, the time available to the attackers is unlimited. They can spent more time finding vulnerabilities than system test engineers. They are also willing to try things that are outside the normal activities. Similar to software testing, a better approach to hardware testing is a risk-based [Potter and McGraw, 2004] analysis of the system and focusing on the most significant risks faced by the system.

## 3.6 Future Work

The number of attacks and threats posed to hardware are increasing rapidly. To encourage further research in protecting the hardware, this section discusses some future research directions including the problems and challenges faced.

### 3.6.1 Machine Learning

Reverse engineering is normally viewed in a negative aspect, e.g., in the use of cloning hardware designs and leaking sensitive information, but it can also be used to detect malicious alteration of the hardware or finding vulnerabilities in the chips [Botero et al., 2020]. Other techniques used for this purpose are either limited or ineffective. For example, monitoring the hardware at runtime increases the resource requirements, such as power consumption, memory utilization etc. Due to these and other limitations, hardware reverse engineering is gaining popularity and acceptance as an effective approach.

An overview of an automated reverse engineering process [Botero et al., 2020] is shown in Figure 3.1. The two main blocks of this process are image analysis and machine learning. After the image is acquired physical reprocessing is carried out until a clear view of the physical layer is visible. There are also some challenges during image analysis, such as noise and effective tuning of the image parameters. To overcome these challenges machine learning techniques are used. This improves the quality of the acquired image. The feature extraction and analysis steps reduce the noise and improve the tuning of the image parameters.

One of the major limitations of the application of machine learning to reverse engineering is the lack of extracted features that can be generalized to other chips. Another major challenge is the availability of a large number of diverse, high quality images as inputs and their corresponding accurate labels as ground truth for effective learning of machine learning algorithms. To the best of our knowledge, a dataset meeting this criteria is not available. Therefore, currently to apply machine learning successfully to reverse engineering, these two major challenges should be resolved. The other major area for future works is the full successful automation of the reverse engineering process as shown in Figure 3.1. This will enhance the level of assurance when validating or verifying a design. Qualitative evaluation of reverse engineering process at each stage and also the whole process, is another area for future works.

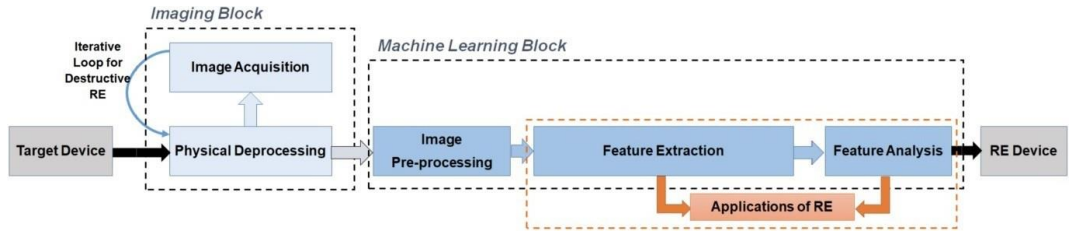

Figure 3.1: Overview of a reverse engineering process: source [Botero et al., 2020]

### 3.6.2 IoT Devices

With the proliferation of IoTs (Internet of things) we have seen a new wave of digital transformation. It is possible to connect everything (IoT devices) through a network and controllable from any remote location. Several domains, such as health care, farming, traffic control, security systems, etc., have started using this technology to improve their functionality. The deployment and lifetime of an IoT device is much longer than other computing devices. Therefore, security during operation and information exchange of these devices becomes a critical task. IoTs face the same security threats that the other hardware faces. Software security is not enough for securing the IoTs. There is a strong need for implementing security in IoTs at the hardware level.

IoTs are resource constrained devices. This limits them to process complex security algorithms. Due to these reasons it is very challenging to implement cryptographic algorithms in IoT devices. Implementing encryption in these devices is a current field of interest for researchers [Mohd and Hayajneh, 2018].

Trusted Platform Module (TPM) is a microprocessor with added cryptographic functionalities, and can be used for device authentication and authorization purposes. The challenge for using TPM for IoTs is their limited space and power, and the increased cost for such resource constrained devices.

Physical Unclonable Function (PUF) is a lightweight security primitive that can be used for device identification, authentication, secret key generation and storage. Resource limitation in IoT devices makes it difficult to develop and embed traditional PUFs. Currently researchers are working on developing and implementing lightweight PUFs (such as silicon PUFs) specifically for IoTs [Zhang and Qu, 2018].

# Chapter 4

# Evolution of Malware

*You can be sure of succeeding in your attacks if you only attack places which are undefended.*
(Sun Tzu, The Art of War, 500 BC)

## 4.1   Malware

According to an earlier definition [McGraw and Morrisett, 2000]: *Malware or malicious code, is any code added, changed, or removed from a software system to intentionally cause harm or subvert the intended function of the system*. Initially malware writers were hobbyists and their purpose was not to cause harm or damage but prove their capabilities to others. As we are stepping into the digital age this phenomenon has also changed. Now there are other incentives, such as financial gains, intelligence gathering, and cyberwarfare, etc., and now professionals have also joined this group. New techniques and methods, such as reusable software development and obfuscations [Linn and Debray, 2003], are being used to create a copy (variant) of the original malware. Due to these incentives and efforts, malware programs have grown in sophistication, from a simple file infection virus to programs that can propagate through networks, can change there shape and structure (polymorphic and metamorphic) with a variety of complex modules to execute malicious activities. These stealthy malware programs have also infiltrated the new platforms, such as smartphones and IoTs (Internet of things), etc.

The total number of new malware programs are on the rise. Malware growth in the last ten years (2011 − 2020) reported by AV-TEST, an independent IT security institute, is shown in Figure 4.1. The report shows a significant growth, almost 16 times, in the number of malware programs in these ten years. Most of the cyberattacks are carried out by installing malware that performs different malicious activities. The numbers in Figure 4.1 indicate and confirm that malware poses a major threat to cybersecurity.

Several techniques and methods have been developed to mitigate and reduce the threat posed by malware. There are two main purposes of these methods, finding the structure or behavior of the malware program, using static, dynamic, or hybrid analysis techniques. If behavior of the program is suspicious/malicious or it is a variant of some previous malware, then appropriate actions are applied on the malware program. These actions include, deleting, repairing, or quarantining the malware and isolating the effected computers and networks, etc.

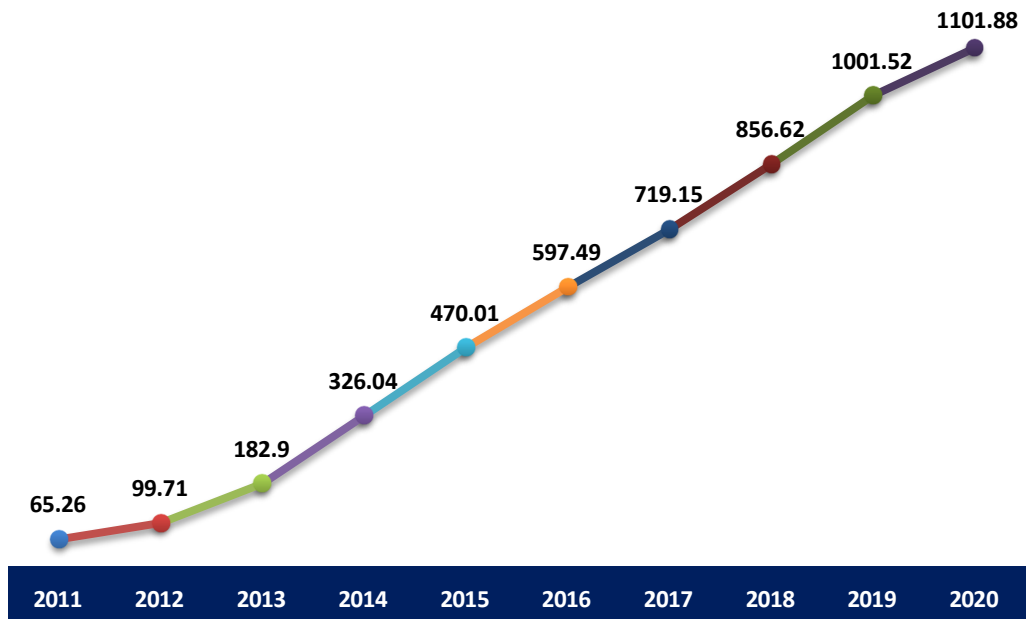

Figure 4.1: Malware growth in millions from 2011 − 2020

### 4.1.1 Classification

The following are some of the traditional classifications of malware.

- □ **Virus**: A computer program that replicates itself. It does so by modifying and inserting its own code to other programs. After this replication the areas affected become infected with a virus.

- □ **Worm**: It is similar to a virus. The only difference is that a virus becomes a worm when it starts spreading to other computers through networks.

- □ **Trojan**: It is a virus that hides from or misleads the users of its true intents. The name is driven from the deceptive Trojan horse mentioned in the ancient Greek story.

- □ **Spyware**: A malicious software that collects and gather information about a person or organization and delivers it to another entity to be used for harmful purposes.

□ **Ransomware**: A malicious software that threatens the victim to publish or block access to victim's data unless a ransom (money) is paid.

□ **Rootkit**: A computer program that enables another program to get access to a computer or its software that is otherwise not accessible. Some malware programs use a rootkit to hide.

## 4.2 Analysis and Detection

There are basically two main approaches for analysing a malware program, static and dynamic analysis. To maximize the benefits these two approaches are combined into one called hybrid analysis. Here we briefly discuss the first two.

**Static analysis**: This analysis is performed without executing the program. It is used mostly by antimalware software for automatic malware analysis and detection. A simple static analysis is performed using string or instruction sequence scanning and matching [Szor, 2005]. A more sophisticated analysis is performed using complex static analysis techniques, such as control flow analysis [Aho et al., 2006], value set analysis [Balakrishnan and Reps, 2010], opcode-based analysis [Bilar, 2007], and model checking [Clarke Jr et al., 2018]. Some of the advantages and disadvantages of static analysis are as follows.

Advantages:

□ The main advantage is that there is no need to run the program. Therefore, no possibility of infecting the system on which the program is running.

□ It captures all the paths taken by an executable program and is easier to automate. The possibility of hiding an executable (malicious) path by a malware is close to impossible.

□ Depending on the method used, static analysis has the potential of being used in a real-time malware detection.

□ These advantages makes static analysis suitable to be used in an industrial antimalware software product.

Disadvantages:

□ When conducted manually, static analysis can be time consuming.

□ It is non-trivial for a static analysis tool to support different platforms.

□ When a malware cannot be unpacked, then the instructions on the disk will be different than the instructions at runtime. This makes the analysis incorrect.

□ This also makes it non-trivial for a static analysis tool to detect malware that introdcues changes during runtime, such as polymorphic and metamorphic malware.

**Dynamic analysis**: This analysis is preformed when the program is executing. It is used by antimalware software for manual analysis to either, find an anomaly, or get a specific signature of a malware to be used latter for malware detection. A simple dynamic analysis is carried out by running the program in an emulator or simulator, and logging important or all the tasks performed by the program. A more complex analysis is carried out by first instrumenting (implement code instructions that monitor specific components in a system) and then running the program to collect as much information as possible. Some of the advantages and disadvantages of dynamic analysis are as follows.

Advantages:

☐ The main advantage of dynamic analysis is the it is very useful when a program cannot be unpacked during static analysis

☐ The program automatically unpacks itself during execution. This makes it easy to detect malware that make changes during runtime, such as polymorphic and meta-morphic malware.

☐ With the availability of tools dynamic analysis can be performed on any platform.

☐ A high malware detection rate is possible with dynamic analysis.

Disadvantages:

☐ The main disadvantage of dynamic analysis is that there is a possibility of infecting the host system.

☐ Dynamic analysis only captures the paths that are executed during runtime, and may miss a malicious path. One possible solution is to force a conditional branch to take multiple paths. This is a non-trivial problem and is time consuming, and may make the analysis impractical.

☐ Running a program in a controlled environment for automatic analysis may take more time than performing an automatic static analysis on the same program.

☐ Anti-reverse engineering (if malware can tell that it is being analysed) techniques may make the analysis unreliable for malware detection.

There are basically two approaches for malware detection, signature-based and anomaly-based. The difference between these two is the way they gather information for detecting the malware. Both these can utilize either of the three analysis approaches, static, dynamic, or hybrid.

**Signature-Based**: In this detection approach, a signature also called pattern is built. This pattern is stored in a database and used latter for comparison in the detection process. Pattern matching techniques are used, that looks for a particular pattern of the malware to perform the detection. These signatures or patterns are usually preconfigured or prede-termined by domain experts and need an update whenever new malware or their variants appear. One of the drawbacks of such a system is that it will not be able to detect zero-day (whose signature is not in the database) or previously unknown malware. One of its advantage is that it consumes less resources and is faster than other systems and can be used in real-time. This is one of the main reasons this approach is used by most of the antimalware systems.

**Anomaly-Based**: In this detection approach, an anomaly is detected by differentiating between the normal and abnormal behavior of a system/program. For this system to work correctly it is necessary to build a profile of a model of normal system/program behavior. These profiles are then used to detect new behavior that significantly deviates from them. One of the advantages of this system is that it may be able to detect zero-day malware attacks. One of the drawbacks of this system is that it consumes more resources and can be very time-consuming. Compared to signature-based technique anomaly-based have a high percentage of false positives. Similar to signature-based new behaviors/profiles could be added to enhance and further improve malware detection.

## 4.3 Reverse Engineering

Malware reverse engineering is a process of gaining insight into a malware program. The purpose is to determine the functionality, origin, and potential impact on the system by the malware program. This knowledge enable security engineers to neutralize and mitigate the effects of malware. A reverse engineer utilizes a variety of tools to investigate these details. The popular WannaCry malware was reverse engineered to discover the mechanisms (a kill switch in this case) to stop its spread. Here we are going to discuss some of the main tools used for reverse engineering a malware program.

**Debuggers** (e.g., OllyDbg, WinDbg, GDB) – One of the basic tools used by reverse engineers to get insight into a malware program. Debuggers are use to manipulate the execution of a program and let the engineer control part of the program while it is running. For example, a portion of the current program's memory can be inspected to find possible exploits present in the malware. Debuggers also help when dealing with obfuscated (hidden) code.

**Disassemblers** (e.g., IDA Pro) – A disassembler takes apart an executable (binary) program and produces its assembly code. Sometimes a decompiler is used to produce the source code. This process converts binary instructions into higher level code which is easier to understand and analyse by a human. Most of the malware programs are present in binary form. Disassemblers are used to produce higher level code to analyse these programs.

**Network Analyzers** (e.g., Wireshark, Network Miner) – It is very crucial to know how the malware is interacting with other machines on the network. For example, what connections are being made and what data is being sent. Network analyzers help security engineers to perform malware traffic analysis, and investigate and find all these details. This helps engineers to render malware harmless before it can perform damage and any infection can be thwarted.

**Sandboxing** (e.g., Cuckoo Sandbox) – This a technique where a malware program is run in a protective software box without causing any damage to the real machine. Sandbox tools help reverse engineers observe the program's behavior and output activity in a proactive layer of security. It also allows dumping of memory to have a better look at what is happening in the memory. It is very useful in quarantining and eliminating zero-day threats.

**Patching** (e.g., *x*64Dbg) – As malware is evolving, some of the complex and modern-day malware contain defense mechanisms to detect if they are being reverse engineered. In such a case, it becomes impossible to reverse engineer such a malware. To get rid of these defense mechanisms patching is used, i.e., a reverse engineer manually modify the malcious code. After an engineer identifies the defense mechanisms in a malware program, he/she will remove such a code and save the updated malware program as a new executable to be analyzed without the impact of the defense mechanisms.

## 4.4 Cyberattacks

A definition of cyberattack by Merriam Webster dictionary is: *an attempt to gain illegal access to a computer or computer system for the purpose of causing damage or harm*. An attack can be active or passive. Active attack alter or affect system resources, whereas passive attack does not affect system resources but only learn or make use of information from the system.

With coming of the cyberspace, most of the enterprises and organizations are adapting to this new pace of technology. At the same time they are facing a new wave of cyberattacks. Stuxnet malware targeting industrial control systems caused substantial damage and infected 200,000 computers. Shamoon, another similar malware was used for cyberwarfare against some of the national oil companies in the middle east. Recently Twitter got hacked where hackers were able to steal US high profile accounts, and Megellan Health, a Fortune 500 company, faced a sophisticated ransomware attack affecting thousands of patients. Cyberattacks because of malware are on the rise and are a serious threat to an organization's financial and other resources. A chronological timeline of such and other high-profile cyberattacks on different companies are shown in Figure 4.2. The number of breaches range from 134 million accounts in the year 2008 to 538 million accounts in the year 2020. The average cost of a malware attack on a company is 2.4 million USD. These attacks accentuate the vulnerabilities of the current cyberinfrastructure and also underline the significance of integrating cybersecurity as part of the complete system.

## 4.5 Evolution

A chronological timeline of evolution of malware in the last 50 years is shown in Figure 4.3. Because of the space constraints, only notable malware programs are listed in the Figure. The following paragraphs discuss some of these malware.

The world's first virus was written in 1971 for mainframe computer running the TENEX operating system. It was a harmless program named *Creeper*, which was able to spread (creep) to other computers using local connections. The program would display I'M THE CREEPER: CATCH ME IF YOU CAN. The first PC virus was written for MS-DOS in 1986 called *Brain*. Like the first virus it did no harm and even came complete with the name, addresses and phone numbers of the authors. It was written to protect the proprietary software application on a floppy disk from illegal copying. *AIDS Trojan* was the first ransomware written in 1989. Before this the viruses were mostly harmless. The virus would render all the files on computer inaccessible and demand money to be sent to a specific address to restore the access. textitMelissa virus was responsible for carrying out the first social engineering attack in 1999. This led to one of the biggest malware attacks of all the time by the virus textitILOVEYOU in 2000. It used email to spread, and would overwrite files, steal usernames, passwords, etc, and lock the victim out of her/his email address. It compromised 45 million computers (10% of all connected computers at that time) and caused over 8 billion US dollars in damages.

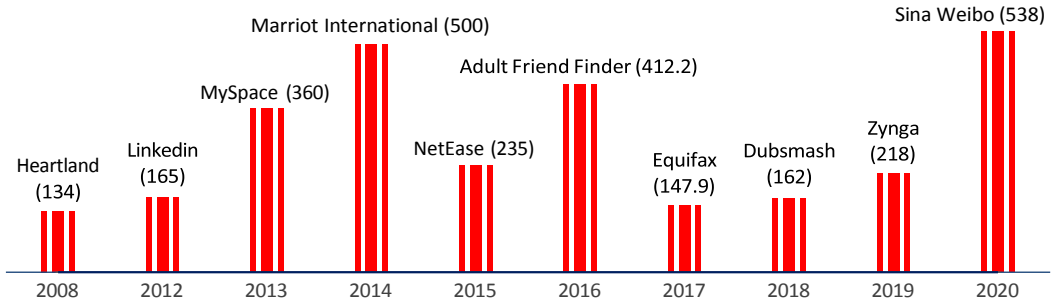

Figure 4.2: A chronological timeline of high-profile cyberattacks from 2008 − 2020 with user accounts effected in millions

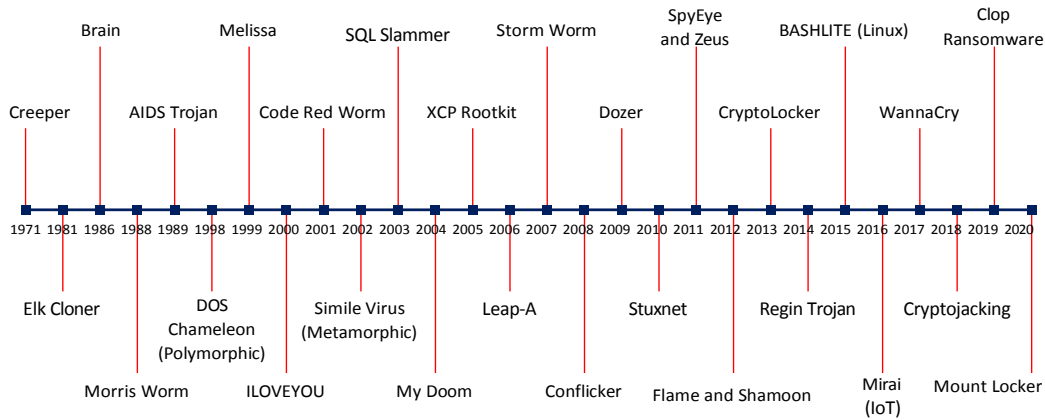

Figure 4.3: A chronological timeline of notable malware programs from 1970 − 2020

Till now malware was a threat to only businesses, personal finances, etc. In the year 2010, malware evolution took a major leap, and *Stuxnet*, a self-replicating malware was discovered for attacking nuclear facilities in many different countries. To prevent detection

Stuxnet used rootkit to hide in the infected machine. Various variants of this malware was latter used for similar purposes. Symantec, a leading antimalware company, reported that Stuxnet was a complex piece of code. Different news agencies including BBC, The Guardian, and The New York Times, reported that the complexity of the code indicates only a nation-state would have the ability to produce it. The reports that followed the discovery of Stuxnet indicated a start of a cyberwarfare, i.e., now it is possible to launch a cyberattack instead of military strike on the nuclear or other sensitive facilities of a country.

Recently, as we are stepping into the use of cryptocurrency, malware attacks have focused more on mining and stealing the cryptocurrency. In this regard, some of the recent emerging forms of malware are Mount Locker, Cryptojacking, and Clop Ransomware. *Cryptojacking* found in 2018, hides and uses the computer or mobile device resources to steal and mine cryptocurrency. *Clop Ransomware* found in 2019, encrypts all files in an enterprise and then demands money for a decryptor to decrypt all the affected files. *Mount Locker* found in 2020, is also a ransomware. It steals the victims' unencrypted data and threatens to publish the data, and demands that the victims pay multi-million dollar ransom payments to recover their data.

### 4.5.1 Hidden Malware

To counter the simple signature-based detection techniques malware uses stealthy techniques to hide themselves. As we are moving into cyberspace, these techniques have evolved in sophistication and diversity. Now on one of the major challenges plaguing malware detection is obfuscation. The problem is that even legitimate developers obfuscate their code to protect from reverse engineering attacks. Similar techniques are used by malware writers to prevent analysis and detection. Obfuscation can be used to create variants of the same malware in-order to escape detection. Obfuscation obscure the code to make it difficult to understand, analyse and detect malware embedded in the code. There are basic three types of techniques to obfuscate the code and are discussed below.

**Packing** is a technique where a malware program is packed (compressed) to escape detection. There are various packing algorithms used for this purpose. The malware program needs to be unpacked before it can be analyzed. In general entropy analysis is used to detect packing, but to unpack the packing algorithm used to pack the malware program must be known. Legitimate software also use packing to distribute and deploy itself. Therefore, the program needs to be unpacked before it can be labeled as malware or benign (legitimate).

**Polymorphism** is a technique where different encryption algorithms are used to morph the static binary code. A different (new version of) morphing is performed with each run of the program. During each run, the malware is first decrypted and then written to memory for execution. The malware program carries a different signature with each run of the code, i.e., the signature of the program changes with each execution of the program. This makes it difficult for an antimalware program that employs simple signature-based detection technique to detect such malware. The morphing does not change the functionality of the malware, i.e., the code is semantically the same for each instance of its execution. It is possible for the current advanced signature-based techniques to perform semantic analysis to detect such malware during runtime.

**Metamorphism** is a technique where different obfuscations are used to morph the dynamic binary code to escape detection. This technique does not use any encryption or decryption and changes the program instructions with each run of the infected program. This phenomenon is also called dynamic code obfuscation. One peculiar property of this malware is that it carries its own morphing engine, which generates the newly morphing code. There are two types of metamorphic malware based on the channel of communication. *Open-world malware* – where there is communication with other sites on the Internet and the malware updates itself with new features through this communication. *Closed-world malware* – where there is no communication to the outside world and the malware generates the newly morphing code using a binary transformer or a metalanguage.

### 4.5.2 Obfuscations

As mentioned above, malware employs different obfuscations to keep them in stealth mode to escape detection. The main purpose of these obfuscations is not to change the behavior of the program but to change its shape/structure which makes difficult its analysis and detection. There are several obfuscation techniques [Nagra and Collberg, 2009] that are and can be employed by malware programs. Here only few, including some basic and as well as advance obfuscation techniques are presented.

**Dead Code Insertion**: In this basic technique, some code is added that either does not execute or has no effect on the results of a program. An example of this technique is shown in Table 4.1.

Table 4.1: An example of dead code insertion with the corresponding original and modified assembly code

| Original assembly code | Modified assembly code |
|---|---|
| mov ebx, [ebp+4] | mov ebx, [ebp+4] |
| | add ebx, 0x0 ; dead code |
| | nop ; dead code |
| | nop ; dead code |
| jmp ebx | |

**Register Renaming**: In this basic technique, registers are reassigned in a snippet of a binary code. This process changes the byte sequences and hence the signature of the binary code. In this case, a simple signature-based detector will not be able to match the signature. An example of this technique is shown in Table 4.2, where register *edx* is replaced by *eax*.

**Instruction Reordering**: In this basic technique, the order of instructions is changed. For this purpose generally either commutative or associative operator are used. This change produces a different structure of a program but keeps the behavior intact. A simple example of instruction reordering is shown in Table 4.3.

Table 4.2: An example of register renaming with the corresponding original and modified assembly code

| Original assembly code | Modified assembly code |
| --- | --- |
| `lea edx, [RIP+0x203768]` | `lea eax, [RIP+0x203768]` |
| `add edx, 0x10` | `add eax, 0x10` |
| `jmp edx` | `jmp eax` |

Table 4.3: An example of instruction reordering with the corresponding original and modified C and assembly codes

| Original C code | Modified C code |
| --- | --- |
| `a = 10; b = 20;` | `a = 10; b = 20;` |
| `c = a * b;` | `c = b * a;` |
| Original assembly code | Modified assembly code |
| `movl [rbp-0xc], 0xa0 ;a=10` | `movl [rbp-0xc], 0xa0 ;a=10` |
| `movl [rbp-0x8], 0x14 ;b=20` | `movl [rbp-0x8], 0x14 ;b=20` |
| `eax, [rbp-0xc] ;` | `eax, [rbp-0x8] ; (reordered)` |
| `eax, [rbp-0x8] ;a*b` | `eax, [rbp-0xc] ;b*a (reordered)` |
| `[rbp-0x4], eax ;c=a*b` | `[rbp-0x4], eax ;c = b*a` |

**Branch Functions**: In this advance technique, the flow of control in a program is obscured using a branch function. The target of all or some of the unconditional branches in the program is changed by the address of the branch function. The main objective is to only change the structure and not the behavior. THerefore, to keep the behavior of the program intact, the branch function makes sure that the branch is correctly transferred to the right target for each affected branch.

**Jump Tables**: Operating system uses jump tables to implement function and system calls. Compilers use jump tables to implement switch-case statements in a language. In this advance technique, jump tables are used to modify the flow of control of a program. A malware writer uses either one or a combination of the following approaches to accomplish this. (1) An artificial jump table is created and artificial jumps are added to the existing jump table. (2) The target of a jump in the table is changed to point to malicious code.

**Self-Modifying Code**: One of the basic properties of polymorphic and metamorphic malware is that they modify their own code. That means part of their code is self-modifying. Not only malware, benign programs also modify their code. For example, an optimizing program may change its instructions to improve performance. A malware program may change its instructions to hide code to prevent reverse engineering or to escape detection from antimalware programs. Table 4.4 depicts a snippet of a self-modifying original and obfuscated (order of instructions and flow of control changed) code.

In the above paragraphs, we have discussed a few of the possible obfuscations.

Table 4.4: An example of instruction reordering with the corresponding original and modified C and assembly codes

| Original assembly code | Modified assembly code |
|---|---|
| | `mov ebx, 0x402364` |
| | `jmp j2` |
| `mov ebx, 0x402364` | `loop:mov edx, [ebx]` |
| `add ebx, 0x100` | `mov [ecx], edx` |
| `push edx` | `jmp j3` |
| `loop:mov edx, [ebx]` | `j1:jmp j4` |
| `mov [ecx], edx` | `j2:add ebx, 0x100` |
| `dec ebx` | `push edx` |
| `inc ecx` | `jmp loop` |
| `cmp ebx, (0x402364+0x100)` | `j3:dec ebx` |
| `jne loop` | `inc ecx` |
| `pop edx` | `jmp j1` |
| | `j4:cmp ebx, (0x402364+0x100)` |
| | `jne loop` |
| | `pop edx` |

For a detailed discussion of various obfuscation techniques the reader is referred to [Nagra and Collberg, 2009]. One of the main purpose of these obfuscations is to create variants of the same malware to escape detection from the antimalware tools. To evaluate the resilience of current commercial antimalware tools against variants of known malware, we carried out an experimental study. For this experimental study, 12 malware variants were generated from 30 different families (class/type) of known Android malware. 16 commercial antimalware tools were tested with these variants. To generate these variants, following 7 obfuscations were applied: NOP = no-operation insertion; CND = call indirection; FND = function indirection; RDI = removing debug information; RVD = reversing order; RNF = renaming fields; RNM = renaming methods. The results of this experimental study are shown in Table 4.5. Only five of the tools achieved a detection rate of more than 80%. The majority, i.e., 11 out of 16, performed very poorly with overall detection rate ranging from 8.33% − 72.91%. These results expose weaknesses of current antimalware tools, and indicate that by just applying some basic obfuscations to a known malware program it is possible to deceive an antimalware.

## 4.6  Issues and Challenges

**Anti-Reverse Engineering**: Modern malware programs use anti-reverse engineering techniques to escape detection. Some of the techniques are sandbox detection, debugger de-tection, and binary instrumentation detection. Such malware programs are non-trivial or sometimes impossible to detect even after using hybrid analysis. These type of malware may require an expert malware analyst to carry out a deep analysis of the malware. This may be very time-consuming process, and by the time the analysis is complete, the malware may have inflicted the damage.

Table 4.5: Detection results of the 16 commercial antimalware tools tested with 192 variants of 30 malware families

| Antimalware | Detection Rate (%) | | | | | | | |
|---|---|---|---|---|---|---|---|---|
| | NOP | CND | FND | RDI | RVD | RNF | RNM | Overall |
| Kaspersky | 92.85 | 92.58 | 92.00 | 92.59 | 96.43 | 93.10 | 92.59 | 93.22 |
| Sophos | 92.85 | 92.58 | 92.00 | 85.18 | 96.43 | 93.10 | 92.59 | 92.18 |
| AVG | 78.57 | 85.71 | 92.00 | 92.59 | 78.57 | 93.10 | 92.59 | 87.50 |
| BitDefender | 85.71 | 92.58 | 92.00 | 85.18 | 85.71 | 86.20 | 85.18 | 86.45 |
| DrWeb | 85.71 | 53.57 | 84.00 | 85.18 | 89.28 | 79.31 | 85.18 | 80.31 |
| ESET | 89.28 | 53.57 | 84.00 | 88.88 | 17.85 | 89.65 | 85.18 | 72.91 |
| Microsoft | 39.28 | 75.00 | 88.00 | 33.33 | 32.14 | 34.48 | 33.33 | 40.10 |
| VIPRE | 17.85 | 57.14 | 32.00 | 14.81 | 14.28 | 13.79 | 14.81 | 21.35 |
| Symantec | 14.28 | 53.57 | 8.00 | 3.70 | 7.14 | 10.34 | 7.40 | 15.10 |
| Qihoo-360 | 10.71 | 53.57 | 8.00 | 7.40 | 7.14 | 6.89 | 7.40 | 14.58 |
| Fortinet | 10.71 | 53.57 | 8.00 | 7.40 | 7.14 | 6.89 | 7.40 | 14.58 |
| TrendMicro | 3.57 | 53.57 | 4.00 | 7.40 | 3.57 | 3.57 | 3.70 | 11.45 |
| McAfee | 7.14 | 53.57 | 4.00 | 3.70 | 3.57 | 3.57 | 3.70 | 11.45 |
| TotalDefense | 7.14 | 53.57 | 0.00 | 0.00 | 0.00 | 0.00 | 0.00 | 8.85 |
| Malwarebytes | 3.57 | 53.57 | 0.00 | 0.00 | 0.00 | 0.00 | 0.00 | 8.33 |
| Panda | 3.57 | 53.57 | 0.00 | 0.00 | 0.00 | 0.00 | 0.00 | 8.33 |

**Automation**: The number of new malware programs including their variants are increasing significantly as shown in Figure 4.1. This is one of the major challenges posed by the current malware programs. Manual analysis and detection of malware is time-consuming and prone to errors, and moreover it does not scale well to the exponential rise in the rate of new unique malware generated. We have moved into the new digital age, but malware analysis is still a manual task. We need to apply the new technologies, such as artificial intelligence, cloud computing, and deep learning, and integrate them with other classic techniques, such as program analysis, programming languages, information flow analysis to successfully automate the process of malware analysis and detection.

**Morphing**: Mutating or transforming the structure (shape) of a malware program during runtime using different techniques, such as polymorphism and metamorphism, makes it non-trivial or impossible to detect such malware. Some of these malware use dynamic code loading, i.e., initially they do not contain any malicious code and when they feel safe (e.g., when they are not being analyzed) download the malicious code from the Internet. Dynamic analysis is usually performed to detect such complex malware, but it has its own drawbacks. Moreover, these malware use anti-reverse engineering to escape such detection.

**Native Code**: This is any machine code that is directly executed by a processor. The two most popular operating systems are Windows and Android. Majority of the malware programs on Windows are native-code programs. 86% of the most popular Android applications contain native code [Sun and Tan, 2014]. Almost all the malware detection using static analysis for Android operates on bytecode level. Native code for each processor (e.g., Intel and ARM, etc.) is different. Handling these differences while analyzing native code is non-trivial, i.e., a separate binary (native code) analyser is to be developed for each processor. Therefore, designing and developing a cross-architecture malware analyser is also a challenge for analysing native code.

**Obfuscation**: Malware writers use obfuscations to make their code difficult to understand and comprehend for analysis, or to create variants of the same malware to escape detection. Even legitimate developers obfuscate their code to protect their software. This makes it non-trivial to make a difference between legitimate and malicious software. As shown in Table 4.5, majority of the commercial antimalware tools were not able to successfully detect the variants.

## 4.7 Solutions

In the previous section we discussed about some of the issues and challenges plaguing malware analysis and detection. In this section we present and discuss some of the practical solutions to these issues and challenges. We describe two systems that makes an attempt to solve some of the issues and challenges described above. The first one called MAIL (malware analysis intermediate language) utilizes static analysis and semantic signature-based techniques, and the second system utilizes dynamic analysis and anomaly-based techniques.

### 4.7.1 Static Malware Analysis

MAIL, an intermediate language, tries to solve two of the challenges *automation* and *native code*, by providing a step towards automating and optimizing malware analysis and detection. Intermediate languages are used in compilers to provide portability and ease of optimization. Here we discuss some of the advantages of MAIL.

1. The first step during static analysis of a native code malware program is to disassemble the program into its assembly code. A straight translation is not good enough because of the large number (200+ to 500+) of assembly instructions present in popular instruction set architectures (e.g., Intel, ARM and IBM PowerPC). For a successful analysis equivalent instructions are grouped in one MAIL instruction.

2. MAIL abstracts the complexity of these instructions and makes the language more transparent to static analysis.

3. MAIL can be used with different platforms, such as Intel and ARM the two most popular architectures use in Servers, Desktops and smartphones. This avoids a separate static analysis for each of the platform.

4. MAIL can be easily translated into a string, tree, or graph and hence can be optimized for various analysis that are required for malware analysis, such as pattern matching and data mining. Moreover, each MAIL statement is annotated with patterns that can be used by a tool to optimize malware analysis and detection.

There are other intermediate languages available and are being used in research and industry for malware analysis and detection. The reason for choosing and discussing MAIL here is its well-defined formal model and detailed explanation of its application and use in automating and optimizing malware analysis and detection. Here we present a short introduction to MAIL and its use. For more information about MAIL the reader is referred to [Alam et al., 2013].

**Design of MAIL**

MAIL is designed as small, simple, and extensible language, and contains 8 basic statements, such as *assignment*, *control*, *libcall*, and *jump*. For example, the control statement represents the following control instructions.

```
control ::= ("if" condition (jump | assignment))
            ("else" (jump | assignment))? ;
```

Assembly language instructions can be mapped to one of the 8 statements of MAIL. An example of the translation from binary (hex dump) to MAIL of a small piece of code from a real malware program is shown in Table 4.6.

Table 4.6: Translating the small piece of code of a real malware from binary (hex dump) to MAIL

| Hex dump | x86 assembly | | MAIL Statement |
|---|---|---|---|
| 83c001 | ADD EAX, 0x1 | —→ | EAX = EAX + 0x1; (assign) |
| 83f800 | CMP EAX, 0x0 | —→ | compare(EAX, 0x0); (libcall) |
| 7511 | JNZ 0x401236 | —→ | if(ZF==0) jmp 0x401236; (control) |

There are three MAIL statements assignment, libcall, and control in the above code. First the hex dump is converted to assembly (Intel x86) and then translated to MAIL statements. The first statement increments EAX, the second compares the values in EAX to 0, and the third jumps to an address a location in the program) if the zero-flag (ZF) is 0. Compare is one of the libs of MAIL and sets the zero-flag based on the results of the comparison. As we can see from the above example, it is much easier to understand the MAIL statements and hence meaning of the program, than the assembly code or hex dump.

**MAIL Patterns for Annotation**

Each MAIL statement is annotated using a pattern. This pattern describes the type and class of the statement. There are 21 patterns [Alam et al., 2013] in MAIL for this purpose. These annotations can be sed latter for pattern matching during malware analysis and detection. A MAIL annotated program is shown in Table 4.7.

Table 4.7: Snippet of a MAIL program annotated with patterns

| MAIL Statement | Pattern |
|---|---|
| `RAX = RAX + 0xf;` | (ASSIGN) |
| `[sp=sp+1] = 0x132; call (0x4b8);` | (CALL_C) |
| `jmp (0xed6);` | (JMP_C) |
| `jmp (0x068);` | (JMP_C) |
| `EDI = EDI;` | (ASSIGN) |
| `[sp=sp+0x1] = EBP;` | (STACK) |
| `EBP = ESP;` | (ASSIGN) |
| `EAX = [EBP+0x8];` | (ASSIGN) |
| `EAX = [EAX];` | (ASSIGN) |
| `compare([EAX], 0xe06d7363);` | (LIBCALL) |
| `if (ZF == 0) jmp 0x17a;` | (CONTROL_C) |

The snippet of a MAIL program shown in Table 4.7 consists of eleven statements annotated with five patterns. Five of these patterns are ASSIGN, one CALL, two JUMP_CONSTANT (jump to a known or constant address), one STACK, one LIBCALL, and one CONTROL_CONSTANT (jump to a known or constant address). These annotations and patterns can be utilized by a machine learning algorithm to extract and select important features and characteristics of a program for malware analysis and detection.

**Application of MAIL**

Here we present and discuss two of the previous researches applying MAIL to analyze and detect malware.

DroidClone [Alam and Sogukpinar, 2021] exposes code clones to help detect malware. It uses MAIL to find code clones in Android applications. To reduce the effects of obfuscations, MAIL helps DroidClone to use specific control flow patterns and detect clones that are syntactically different but semantically similar upto a threshold. Features extraction is performed by serializing each function in a MAIL program. Features selection is performed using TF-IDF (term frequency and inverse document frequency). A SimScore is computed for each sample in the dataset. The samples with their SimScore values are used for training a classifier for malware detection. A new sample is tagged as malware if SimScore of the sample is greater than or equal to a certain threshold. DoridClone when evaluated against various obfuscations was able to successfully provide resistance against all the trivial and some non-trivial obfuscations. The results of these evaluation is shown in Table 4.8.

Table 4.8: Detection rates of different Android bytecode obfuscations implemented to test the resistance of DroidClone against various obfuscations

| Obfuscation | Description | Detection Rate |
|---|---|---|
| ICI | Manipulating call graph of the application. | 31/31 = 100% |
| IFI | Hiding function calls through indirection. | 30/30 = 100% |
| JNK | Inserting non-trivial junk code, including sophisticated sequences and branches that change the control flow of a program. | 15/28 = 53.6% |
| NOP | Inserting No operation instruction. | 32/32 = 100% |
| RDI | Removing debug information, such as source file names, local and parameter variable names, etc. | 31/31 = 100% |
| REO | Reordering the instructions and inserts non-trivial goto statements to preserve the execution sequence of the program. Inserting goto statements changes the control flow of a program. | 17/29 = 58.6% |
| REV | Reverse ordering the instructions and inserting trivial *goto* statements to preserve the execution sequence of the program. Hence changing the control flow of a program. | 29/30 = 96.7% |
| RNF | Renaming fields, such as packages, variables and parameters, etc. | 31/31 = 100% |
| RNM | Renaming methods. | 31/31 = 100% |

DroidNative [Alam et al., 2017] is a tool that operates at the Android native code level and implements two different techniques to optimize malware analysis and detection. The first technique builds an annotated control flow graph of MAIL program. The second technique performs statistical analysis of MAIL patterns' distributions to develop a set of heuristics that help in selecting the appropriate number of features and also reduces the runtime cost. When tested with traditional malware variants, DroidNative achieved a detection rate of 99.48%, compared to the commercial tools that range from 8.33% to 93.22% as shown in Table 4.5.

### 4.7.2 Dynamic Malware Analysis

First we give an overview of how to perform dynamic malware analysis and then discuss specific techniques for such analysis. If the malware program is packed, then the first step in dynamic malware analysis is to unpack the program. Once the program is unpacked, it is executed in a controlled environment such as a sandbox, etc. In addition to keep the execution of the malware program safe, this gives the ability to stop execution at any point for inspection of the program. Some other techniques that are also used to control malware execution are debugging and binary instrumentation. The duration of execution depends on various factors, such as the time available and the extent of the analysis.

A good dynamic analysis always include some mechanisms to explore the full behavior of the malware program, such as forward symbolic execution to explore multiple execution paths of the program, etc. When the execution and the analysis is complete, a report is generated. For example, if the malware stole the password, the analysis report will indicate which methods or parts of the code were responsible, and how it was stolen e.e., from memory or through phishing attack, etc. One of the main advantages of dynamic analysis is that it makes it easy and possible to analyse packed, polymorphic, and metamorphic malware. To escape detection by dynamic analysis malware programs use anti-reverse engineering techniques. There are different techniques used to defeat anti-reverse engineering. One of them is patching, in which the malware analyst will identify the defense mechanisms in a malware program, remove them, and save the updated program as a new executable to be analysed without the effects of the defense mechanisms. This is a non-trivial process and may require an expert and experienced malware analyst to successfully perform this process.

Most of the dynamic analysis platforms execute the malware for a limited time before terminating the analysis. Therefore, one of the techniques to evade dynamic analysis employed by the malware is to use delay mechanisms, such as sleep commands. CIASandbox [Lin et al., 2015] utilizes a virtual machine (VM) to execute the malware and collects information about the system calls through the traces left in the VM's memory and stack. VTCSandbox [Lin et al., 2018] extends CIASandbox to utilize the VM time controller to manipulate time inside the VM and accelerate analysis of the malware. The tick counter inside the VM is manipulated to speed up the execution inside the VM. This helps skip the sleep time or other delays employed by malware to suspend their operations.

MEGDroid [Hasan et al., 2021] is a model-driven approach recently proposed for dynamic Android malware analysis. MEGDroid enhances the model-driven event generation process by generating appropriate events for malware analysis using model-to-model and model-to-code transformations. Another advantage that MEGDroid provides is the inclusion of human in the loop to take his/her experience and knowledge into account to further improve malware analysis.

DeepAMD [Imtiaz et al., 2021], a recent proposed technique utilizes deep artificial neural networks (ANN) to analyze and detect Android malware. DeepAMD performs both static and dynamic analysis. The static layer utilizes a binary classifier and classify the samples either as malware or benign. The dynamic layer is then used to further classify the malware (which is detected by the static layer as malware) into 4 different categories. The static layer achieves an accuracy of 93.4% for malware classification, whereas the dynamic layer achieves an accuracy of 80.3% for malware category classification. This shows that deep ANN are promising for analyzing and detecting malware but still there is a lot of room for improvement.

**Memory Forensic Analysis**

There are certain situations in which it is not possible to perform dynamic analysis of a malware program in a controlled environment. Such situations can be, when the program is not able to finish its execution and crashes or some advance attacks or malicious behaviors leave no track on the hard drive. In these cases, analysis is performed on the memory dumps called memory forensic analysis. In this analysis the first step is to acquire memory. Then the memory dump is analysed to extract data relating to malware and associated information that can provide additional context about the behavior of the malware program. Various tool, such as Volatility Suite, are available for such purpose.

## 4.8 Future Work

Malware programs and attacks carried out by them are increasing in sophistication and complexity. To keep pace with the new emerging threats and attacks and stimulate further research in this thriving area, this section discusses some of the future research directions including the problems and challenges faced.

### 4.8.1 Automation

Currently there is an exponential growth in the number of new malware. To successfully analyse and detect this shear number of malware we need to completely automate this process. But a complete automation of malware analysis and detection is non-trivial. There are lot of efforts, some of them discussed in this chapter, that have tried to fully automate this process, achieving only partial success. After a failed static and dynamic analysis the suspicious program goes through a manual analysis by a malware analysis expert. There are some caveats during this whole process. (1) What if the static or dynamic analysis falsely detects the program as malware. That means more of the benign programs will be rejected. This may not be a problem where security of the system has a higher priority, otherwise, this may significantly affect the business. (2) What if more of the programs are detected as benign. If security is a higher priority than these benign programs will also analyzed manually, which will require a greater number of manual analysis. If the program is positively detected as benign, then this will waste the time and money spent on the manual process. This caveats highlight the limitations to fully automate the process of malware analysis and detection. A successful partial automation reduces false and increases true positives.

### 4.8.2 Embedded Systems Malware

With the proliferation of intelligent embedded devices, including Internet of things and smartphones, comes the new wave of malware attacks. These devices with more CPU power and memory are prone to more sophisticated malware attacks. Because of their limited energy resources running a complete malware detector on these devices is quite challenging. To combat this challenge, a full malware analysis and detection can be provided as in-cloud service.

### 4.8.3 Adversarial Machine Learning and Malware Analysis

Artificial intelligence powered technologies such as machine learning, deep learning, reinforcement learning, game theory, and others can provide solutions to some of the main challenges faced by malware analysis and detection in the new digital transformation age. A recent proposed artificial intelligence technology Adversarial Machine Learning [Huang et al., 2011] is a technique that generates a class of attacks that deteriorate the performance of classifiers on specific tasks. This helps develop approaches to counter these attacks and strengthen malware detection system by improving the classifiers. A recent study [Hu and Tan, 2017] proposed a Generative Adversarial Networks [Goodfellow et al., 2014] based approach, that takes original samples and produce adversarial examples to defeat machine learning malware detectors. Another recent research [Martins et al., 2020] presents a systematic review of the application of adversarial machine learning to intrusion and malware detection. 20 recent research works were studied, and the authors concluded that adversarial attacks could retrograde the performance of the intrusion and malware classifiers. All the classifiers studied show similar results on normal data, but most of them got effected except neural networks and random forests on the manipulated data.

The best defense against any cybersecurity system is to know the current and future risks to the system. These and other such research, that are still in their infancy, inspire studies about adversarial systems and potential risks to malware analysis and detection systems and provide guidance and direction to reduce these risks. There is a need to do further research to develop new techniques and methods to apply adversarial machine learning to improve malware analysis and detection systems.

### 4.8.4 In-Cloud Malware Analysis

A cloud is a computing environment where a program or an application runs on several connected computers. These computers can be physical or virtual. The virtual servers running on physical servers are not bound to the physical server. The virtual servers can move around and scale up and down without affecting the client. In-cloud refers to the services provided by the cloud to its users. Malware analysis and detection can be one of the services that can be provided as in in-cloud service. There are some advantages and disadvantages of providing an in-cloud malware analysis and detection service.

**Advantages**: (1) Multiple antimalware engines can be combined to improve the detection rate. (2) It can be easily extended, by adding other antimalware engines. (3) Very useful for resource constrained devices, as it can provide compute intensive and deep malware analysis. (4) Enhances the malware detection by providing correlation of information among different antimalware engines, such as sharing the behavior of a malware program.

**Disadvantages**: (1) It is highly dependent on the trust and privacy provided by the cloud. (2) Running a file on the client and then replicating it on the cloud can expose the client to possible malware. (3) More antimalware engines can produce more false positives. (4) A full replication of the client is required to detect the new generation of malware such as polymorphic and metamorphic malware.

A complete malware analysis and detection system comprises of both static and dynamic analysis. Such a complete system consumes lot of computing resources and is more suitable for in-cloud based malware analysis and detection. Figure 4.4 presents an overview of such a system. The client can run either an LWA (lightweight agent) or an LWE (lightweight antimalware engine) depending on the resources available, that can scan/detect files and send only a suspicious file to the cloud for further analysis and detection. The cloud runs multiple antimalware engines, shown as $e_1$, $e_2$, $e_3$, ..., $e_m$. The signatures of new and known malware programs are stored in databases, shown as $s_1$, $s_2$, $s_3$, ..., $s_k$. To shield the client from the malware, the in-cloud system has three layers, LWA, LWE, and antimalware engines. These layers separate execution of the file for dynamic analysis, and also reduces the bandwidth required for transferring the files from the client to the cloud. To get more details about this system readers are referred to [Alam et al., 2014].

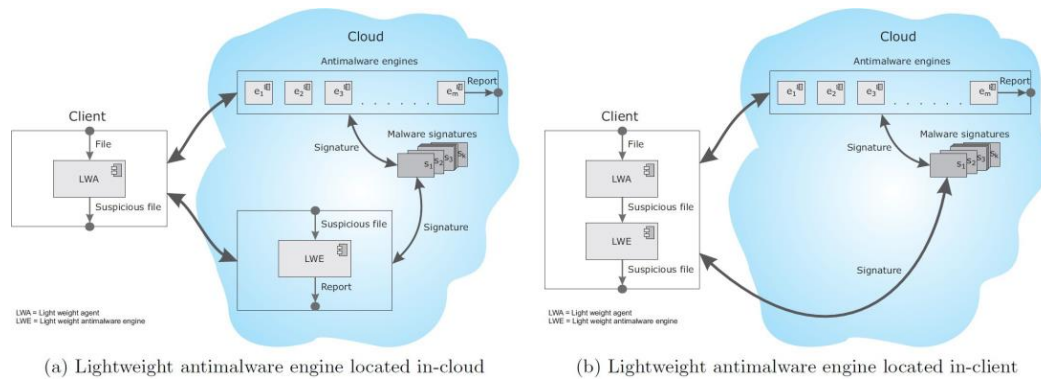

(a) Lightweight antimalware engine located in-cloud          (b) Lightweight antimalware engine located in-client

Figure 4.4: An overview of the hybrid in-cloud layered malware analysis and detection system [Alam et al., 2014]

# Chapter 5

# Biometrics

*All warfare is based on deception.*
(Sun Tzu, The Art of War, 500 BC)

## 5.1    Basics

In cybersecurity there are three common factors used for authentication: *Something you know* e.g., passwords, *Something you have* e.g., credit cards, and *Something you are* e.g., fingerprints. Biometrics falls into the third factor. They are costlier but some of them are highly effective and accurate compared to the other two factors when used properly. The most practiced and popular authentication method is the use of passwords, because they are free and easy to maintain. But, passwords have some inherent limitations. Simple passwords can be easily guessed especially based on social engineering methods or broken by dictionary attacks [Klein, 1990], complex passwords are difficult to remember. Most of the user employ the same passwords across different applications, and hence with a broken password an adversary can gain access to multiple machines/accounts. These inherent limitations in the use of passwords can be ameliorated by employing better techniques for authentication. The new biometric security paradigm, *Forget About Passwords And Cards You Are Your Own Key* says it all.

Biometrics is the combination of two words bio = life and metrics = measurements, and when we combine them it means measurements (of attributes such as physical and behavioral) of a living being. Biometrics are used for identification (authentication) of human beings based on their biological or behavioral characteristics. Some of these biometrics can be fingerprints, face, eye iris, gait, odor, keyboard dynamics, and signature dynamics, etc. Biometrics offer the following advantages over other authentication factors.

☐ They provide distinctive information for each person that can be utilized as a technique for individual identification and authentication.

☐ They are much easier to use, because they cannot easily be forgotten.

□ They offer better security, privacy, and assurance, because they cannot easily be
shared, borrowed, or observed.

□ They are much more difficult to spoof (fake or steal), because of the same reasons
mentioned above.

To present an efficient use of a biometric, we share a real-life example [Cathy, 2002] of
finding a person (an Afghan girl) after 17 years by the National Geographic with the help
of John Daugman, the inventor of automatic iris recognition [Daugman, 1994]. The picture
of a young Afghan refugee girl on the cover of National Geographic in June 1985 became a
mystery for 17 years. Finally in 2002 the crew from National Geographic TV & Film found
the Afghan girl in a Pakistani village. After 17 years of hardships her youth was erased. So
to make sure she is the same Afghan girl, the National Geographic took a new picture of
the woman found and turned to John Daugman for the evidence. After applying biometric
techniques on her iris patterns (from the old to new picture) and mathematical calculations,
the evidence (numbers) that Daugman got left no doubt in his mind that the eyes of the
young Afghan girl and the woman found belong to the same person. The two pictures of
the Afghan refugee used by Daugman to find the evidence are shown in Figure 5.1.

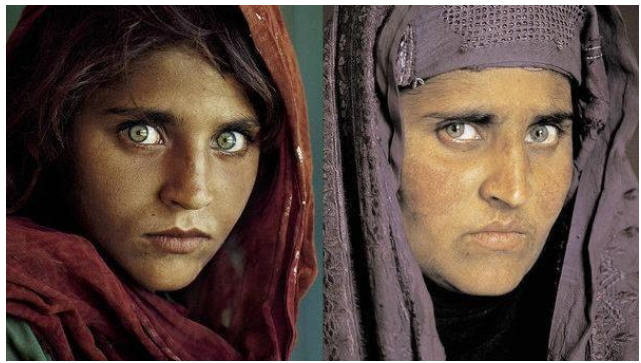

Figure 5.1: The two pictures (left – taken in 1984, right – taken in 2002) of the Afghan
refugee used by Daugman to find the evidence, using his automatic iris recognition tech-
nique, that they belong to the same person. (images from [Daugman, 2002])

The above real-life example shows an efficient use of a biometric the eye iris,
which is considered as the best currently available biometric. So, what are those
properties/requirements that make a biometric successful? There are seven proper-
ties/requirements that are used to assess the suitability of a biological or behavioral charac-
teristic to be used successfully in a biometric identification system [Jain et al., 2006].

1. Universality – every person should have the characteristic.

2. Uniqueness – different from every other with 100% certainty.

3. Permanence – characteristic being measured should never change.

4. Collectability – characteristic can be measured quantitatively.

5. Performance – achieving acceptable accuracy, speed and robustness by the system.

6. Acceptability – how well people are willing to accept the system.

7. Circumvention – how easy it is to deceive the system by deceptive techniques.

No biometrics is expected to effectively satisfy all the properties listed above. Each biometrics has its strengths and weaknesses, and its use is very application dependent. Privacy and security of biometrics are a major concern when providing biometrics. Here is a paragraph from Apple's website [Apple, 2020] that can be taken as a guidance about what kind of privacy and security one should be looking for when providing biometrics.

"Your fingerprint data is encrypted, stored on device, and protected with a key available only to the Secure Enclave. Your fingerprint data is used only by the Secure Enclave to verify that your fingerprint matches the enrolled fingerprint data. It can't be accessed by the OS on your device or by any applications running on it. It's never stored on Apple servers, it's never backed up to iCloud or anywhere else, and it can't be used to match against other fingerprint databases."

### 5.1.1 Classification

Biometric methods are classified into two classes/types based on the user's and biometrics viewpoint.

**Static** – In static method a user is identified by a fixed unchangeable characteristic of the user. This characteristic is always present and does not require any special action of the user. Some of the examples of static biometric methods are fingerprint, face recognition, iris scan, and vein recognition, etc.

**Dynamic** – In dynamic method a user is identified by a behavioral characteristic of the user. Different actions of the user lead to different biometric data. Some of the examples of dynamic biometric methods are keystroke dynamics, voice recognition, and signature dynamics, etc.

### 5.1.2 Phases

Biometric identification is the process of establishing a person's identity based on the person's biometric data. A biometric identification system has two phases.

**Enrollment** – Where the data about a person is collected and stored into a database. This phase can be slow, as it is done once, and may require multiple measurements.

**Recognition** – Detection by the biometric system. A person's newly acquired biometric data is compared to the enrolled biometric data, and a similarity score is generated. This similarity score is then used to identify the person. This phase must be quick, simple, and accurate.

Figure 5.2 shows the enrollment and recognition phases of a typical biometric system. *Enrollment* – During preprocessing the raw biometric data is cleaned up. Then most important features are extracted that help in recognition. These features are enrolled/stored in the biometric database. *Recognition* – The new feature is preprocessed and then the corresponding features are extracted to be matched with the enrolled features from the database. A similarity score is generated that is verified and a match or a no match decision is made.

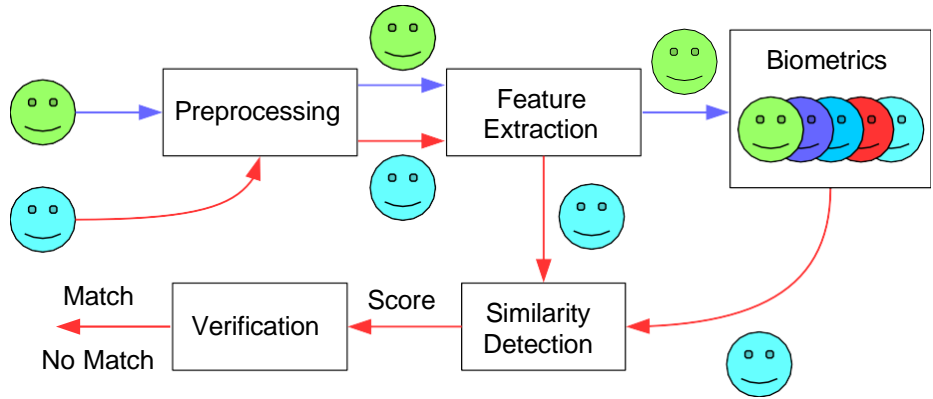

Figure 5.2: An overview of the enrollment and recognition phases of a biometric system

### 5.1.3   Testing

There are basically three error rates (metrics) that are measured to test/compare different biometric systems. **False Acceptance Rate** (FAR) – When a person is mistakenly authenticated as someone else. **False Rejection Rate** (FRR) – When the system fails to authenticate a genuine person. Based on these errors an **Equal Error Rate** (EER) is computed, which is the rate when the fraud and insult rates become equal. That is, the parameters of the system are adjusted until the fraud rate and insult rate are precisely in balance. EER is a useful measure for comparing different biometric systems. A biometric system with lower EER is considered to be more accurate. The other two metrics which are also measured and used to test a biometric systems are: **Failure to Enroll** – The number of people unable to use the system, and **Failure to Acquire** – The number of people unable to enroll with the system.

An identification rate is determined by computing the FAR and FRR. This measures the utility of the biometric system for a particular application. The enrollment phase plays a significant role in enhancing the identification rate. Therefore, care and special efforts are required to obtain a good enrollment template, and may be repeated many times to enhance the template.

### 5.1.4 Spoofing

Biometric **spoofing** is a technique where an attacker forges and counterfeits biometric identifiers to impersonate a legitimate user and gain access to a biometric identification system. Biometric **anti-spoofing** is a process where a biometric identification system employs mitigation techniques to defeat spoofing.

Biometric spoofing is technically possible, and there will always be someone up to the challenge to hack the latest technology. But, biometric spoofing is harder than spoofing or defeating other access control approaches, such as passwords and credit cards. The most common approach used for authentication, the PIN and passwords, are more easier to break. The attackers and criminals usually take the least resistance path. Modern biometric systems, especially those using more than one biometric (combining fingerprints with eye iris or face), require more time, skill, and sometimes luck to successfully spoof them.

## 5.2 Methods

A number of biometrics have been proposed, such as fingerprint, voice/speech, face, eye iris, gait, odor, keyboard dynamics, and signature dynamics, etc. Here we are going to present the methods using some of these biometrics. We will cover only those methods that are widely used and satisfy most of the properties discussed in section 5.1. We also list some of the spoofing and anti-spoofing techniques used for each method.

### 5.2.1 Fingerprinting

Fingerprinting is the oldest method in use for identifying a person. The widespread use (in forensics and criminal justice) of fingerprinting started when Galton developed a classification system based on minutia [Galton, 1892, Galton, 1893]. This system enabled efficient searching, and verifying that fingerprints do not change over time. Some of Galton's classifications (minutia – loop, whorl and arch) are shown in Figure 5.3.

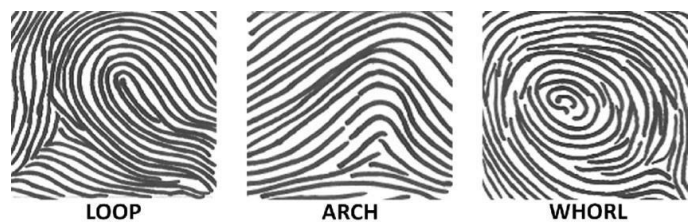

Figure 5.3: Example of three common finger patterns (minutia) found in Galton's classification system (image from [Walsh et al., 2016])

In this method after capturing the image, the image is enhanced by removing the noise and using other enhancing techniques, such as binarization and thinning. Then, various points (minutiae – features) are extracted from the enhanced image. These extracted points are compared with the stored information (enrollment phase). A statistical match is made with a predefined threshold. The accuracy of a fingerprint system can be enhanced by using not just one, but more than one fingers. However, this also increases the identification time of the system.

*Spoofing*: Applying artificial fingerprints, or dead fingers.

*Anti-Spoofing*: Measuring skin temperature, resistance, or capacitance.

### 5.2.2 Face Recognition

Now a days, a camera has become an integral part of a computer system (desktop, laptop, and smartphones, etc). Therefore, face recognition has also become an established and the most common method of human recognition. The environments plays an important role in face recognition. For example, face recognition in a controlled environment is much easier than in an uncontrolled environment. For example, a uniform background and lighting, and an identical pose constitute a controlled environment. Whereas, the background or lighting may not be uniform, and the face orientation, positions may be different. Some of the advantages and disadvantages of face recognition are: Availability of cameras, it is fast and discreet, and is mostly acceptable by public. Variations in face due to aging, makeup, hair, and glasses, and uncontrolled environments cause problems.

There are two approached used for face recognition: (1) Transform Approach – Eigenspace vector is built from an image of a face. Two faces are considered to be equal if their eigenspace vectors are sufficiently close. (2) Attribute-Based Approach – Where facial attributes like nose, eyes, etc are extracted from an image of a face, and then geometric properties among these features are used for matching faces.

*Spoofing*: Photos, and facial disguise, such as makeup, hyper-realistic (e.g., a 3D model) masks.

*Anti-Spoofing*: Challenge-response method (e.g., move the face before recognizing), and image quality assessment, and deep feature fusion.

### 5.2.3 Iris Scanning

The human iris has shown extraordinary variations in the texture and is therefore, very distinguishable and stable. Traditionally, texture filters, such as Gabor filter, are applied to an image of the iris to extract the iris code (features) [Daugman, 2009]. The following properties of a human iris makes it a prime biometric for authentication: (1) An extremely rich physical data structure. (2) It is genetically independent, i.e., no two eyes are the same, not even for the same person or identical twins. (3) The pattern of an iris is stable throughout the lifespan. (4) It is physically protected by cornea of the eye that does not inhibit external viewability. (5) The pattern of an iris can be recorded/encoded from a much larger distance (up to one meter) compared to some other biometrics.

In this method, first the iris is located using a scanner. Then a black and white photo (image) is taken and is transformed into an iris code of 2 bytes. Two iris codes , $x$ and $y$, are compared using their hamming distance: $d(x, y) =$ number of non-match bits / number of bits compared. A lower value of hamming distance (e.g., 0.08) indicates that the two irises are the same and vice versa.

*Spoofing*: Photos, contact-lenses, artificial eye.

*Anti-Spoofing*: Challenge-response method (e.g., blink the eyes before scanning), and image quality assessment.

### 5.2.4 Keystroke Dynamics

Keystroke dynamics represent the use of the keyboard by a person. Every person type on a keyboard in a distinguishable way. This biometric is not very unique to a person but is distinct enough to enable identity authentication [Leggett and Williams, 1988]. Unlike other biometrics, the temporal information of the keystrokes can be used to identify a user using only software with no additional hardware. Keystroke dynamics are cost effective, user friendly, and provide continuous user authentication with a potential of high accuracy. Beside these advantages, as a behavioral biometric Keystroke dynamics have also some limitations. Emotions, stress, and drowsiness, etc, may change the typing style of a person. The typing characteristics can also change if a different keyboard is used with a different layout of keys.

There are two types of text typing used during authentication: *static text* e.g., a short text like a password, and *free text* e.g., a longer free text. For more secure applications a free text should be used for continuous authentication. Keystroke dynamics are based on time durations between the keystrokes, inter-key delays, and dwell times. Inter-key delays can be digraphs (time delays between two keys), trigraphs (time delays between three keys) and so on. Dwell time is the time a person holds down a key. Keystroke dynamics is still in its infancy and not yet ready to be used in a high security environment.

*Spoofing*: Keystroke dynamics biometric is hard to spoof, but is still possible by obtaining key-press timings using a keylogger or timing attack.

*Anti-Spoofing*: Sensor-enhanced (sensors on mobile devices, such as accelerometer, gyroscope, etc.) keystroke dynamics.

### 5.2.5 Voice Recognition

Voice is a behavioral biometric, is not sufficiently distinctive and may not be appropriate to be used for identification of an individual from a large database of identities. The quality of a voice signal depends on the microphone, communication channel, and digitization techniques used for constructing the signal. There are two types of voice recognition systems, one is text-dependent and the other is text-independent. The first one uses a text/phrase known to the system and the second one uses an arbitrary text/phrase. The latter is more difficult to implement. Features are extracted from several normalized and decomposed frequency channels. These features can be either in time-domain or frequency-domain.

Some of the advantages and disadvantages of a voice recognition system are: (1) Unobtrusiveness. (2) Voice in general is an acceptable biometric in all societies. (3) Part of voice/speech of a person changes with age, medical conditions, emotional state, etc. (4)

Misread prompted phrase. (5) Channel mismatch, such as use of different microphones for enrollment and recognition. (6) Poor room acoustics, such as background noise.
*Spoofing*: Recorded voice, and mimicking.
*Anti-Spoofing*: Prompting, i.e., each time when authenticating prompt a different phrase.

### 5.2.6 Comparison

Table 5.1 compares the five biometric methods based on the seven characteristics discussed in section 5.1. Out of the five biometric methods, iris scanning is given the highest score in all the characteristics except one, the acceptability, because of the close proximity during the scanning some users may experience discomfort/unease in their eyes. The discomfort/unease can be caused by the psychological effect of the eyes being close to the scanning device. We gave a score of H = High instead of M = Medium to collectability of iris scanning, because of the current technical advancements in the scanning make this process much easier than it used to be. Keystroke dynamics is given the lowest score because of a low score in the five characteristics, including universality, uniqueness, permanence, and performance.

Table 5.1: Comparison of the five biometric methods based on the seven characteristics discussed in section 5.1, where H = High, M = Medium, and L = Low

| Biometric Method | Characteristic | | | | | | |
|---|---|---|---|---|---|---|---|
| | Universality | Uniqueness | Permanence | Collectability | Performance | Acceptability | Circumvention |
| Iris Scanning | H | H | H | H | H | L | L |
| Face Recognition | H | H | M | H | L | H | H |
| Fingerprinting | M | H | H | M | H | L | M |
| Voice Recognition | M | L | L | M | L | H | H |
| Keystroke Dynamics | L | L | L | M | L | M | M |

## 5.3 Multi-Biometric Systems

To overcome the challenges faced by a single biometric system a multiple biometric systems [Ross et al., 2006], such as face recognition and fingerprinting, is used. Such systems increase the reliability by combining more than one biometric for authentication. It also becomes difficult to simultaneously spoof more than one biometric systems. Moreover, the use of more than one biometric increases the universality of the system. They should be designed such that they should not increase the cost considerably, and are easy to use e.g., no complex enrollment procedures. Beside these advantage there are some issues when using multi-biometric systems. We discuss some of these issues in the following sections.

### 5.3.1 Normalization

Combining scores from different biometric systems for a successful authentication is challenging. These scores are the result of different algorithms and are therefore heterogeneous. Score normalization is necessary to compute a comparable score in the same domain. Different techniques are used to normalize these scores and we discuss few of them here. For a detailed list readers are referred to [Jain et al., 2005]. (1) *Min-max normalization* – This is based on the minimum and maximum score values produced by a biometric system. The min and max scores are shifted to (normalized between) 0 and 1. If there are no min and max scores (i.e., the values are not bounded) produced a set of max and min values are estimated and then apply min-max normalization. This method is not reliable when the values are not bounded, because of the use of the estimated min and max values. (2) *Z-score* – This is the most commonly used normalization. A z-score is computed using the arithmetic mean and the standard deviation of the given data. Similar to min-max normalization this also suffers from outliers. (3) *Tanh-estimators* – This normalization is both reliable and highly efficient. The only problem is, that the three different intervals of the distribution of data must be defined in an improvised manner, and are difficult to determine experimentally. These parameters limit the effectiveness of this technique if chosen incorrectly.

### 5.3.2 Fusion

Fusion is the process of combining different systems into one system. The purpose is to increase the global response of the system. To answer the question *when – at which level – to fuse multi-biometric systems* there are three fusion policies currently in use. (1) *Fusion at feature extraction level* – This fusion occurs when the features are extracted. These features are combined into a unique feature vector, to be compared with the stored template to obtain a global score. This process can become complex and less flexible if the process contains large feature vectors. (2) *Fusion at matching level* – This fusion occurs when the matching takes place. A feature vector from each biometric system is compared with the template associated with the corresponding biometric system. The matching scores are then fused to obtain a unique global score. This process is simple and flexible. (3) *Fusion at decision level* – This fusion occurs when the decision takes place, i.e., at end of the matchings from all the corresponding biometric systems. Each biometric performs its own decision and all the decisions are combined into a unique global decision using boolean binary rules. This makes the process simpler and more flexible, but lot of information is lost during the process.

Out of the three fusion types, fusion at matching level is commonly used, because of the easy availability of matching scores and they contain enough information for authentication.

We have not discussed what fusion techniques can be used at each of the fusion level. Readers interested in these fusion techniques are referred to [Singh et al., 2019].

## 5.4 Soft Biometrics

So far we have discussed primary biometrics, such as fingerprints, face, iris, voice, and keystroke dynamics, etc. These biometrics include characteristics that are specifically used for authenticating a person. Soft biometrics contain additional characteristics of a person, and does not have enough information for identification but can be used for other purposes, such as augmenting primary biometrics. Soft biometrics are collected by extracting ancillary information from the primary biometrics, such as facial measurements, certain permanent marks on the face, color of the skin, hair, and eyes, etc. They give semantics (meanings) to a person, e.g., when describing a person *young white male*. In this section we present a brief introduction to soft biometrics. For a detailed study and survey on soft biometrics readers are referred to [Dantcheva et al., 2015, Nixon et al., 2015].

### 5.4.1 Benefits

Some of the advantages/benefits of soft biometrics are as follows:

- ☐ They can be used to augment primary biometrics by improving their recognition accuracy and speed. They improve speed by filtering out unnecessary subjects from a large biometric database.

- ☐ They can be extracted and stored without much privacy concerns, because they only provide partial information (such as *female*, *tall*, *young* etc) about an individual.

- ☐ Their collection is not intrusive, because they can be collected during the primary biometrics collection and also their collection can be performed from a distance, such as height of a person.

- ☐ They can be used in a multi-biometric system when it is difficult to collect one of the primary biometrics.

### 5.4.2 Application to Cybersecurity

Soft biometrics has a number of applications. Here we discuss some of their applications in the domain of cybersecurity. The first cybersecurity application where they can be used is *Forensics* or *Surveillance*. For example, locating a person of interest based on a set of soft biometrics. A person of interest in a busy airport can be defined using soft biometrics as, age: 25-35, gender: female, hair color: blond, clothes: blue T-shirt, and accessories: wearing sun glasses and carrying a brown handbag. A video surveillance camera can be used to identify such a person even with low resolution images when faces are occluded or not visible. A second application can be *Age Specific Access Control*, where children should be prevented from accessing certain websites, or movies.

### 5.4.3   Spoofing

Because of the nature of soft biometrics they are much easier to spoof than primary biometrics. Cosmetics is a major threat towards spoofing soft biometrics. Using cosmetics eye color can be spoofed with color lenses, hair can be dyed, and the overall facial appearance can be modified by makeup. For example, female subjects attempt (by applying facial makeup) to look like males and vice versa as shown in Figure 5.4.

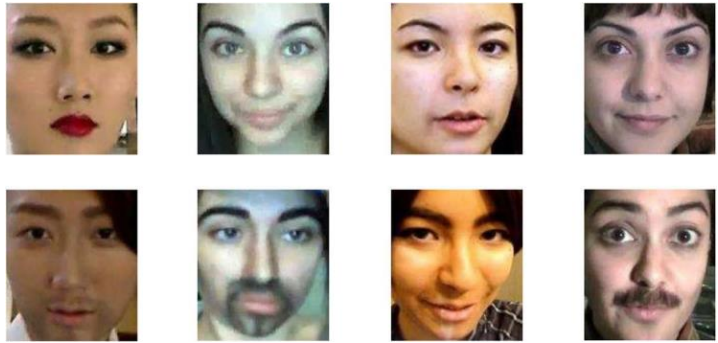

Figure 5.4: Example of gender, female to male, spoofing (image from [Chen et al., 2014]). The upper row are the images before the application of makeup, and the bottom row are the images after the makeup.

## 5.5   Privacy

Privacy is always a concern when dealing with biometrics. There is also a problem of false security. In general biometrics are considered easy to use and safer than passwords or PINs. But, when a password is stolen it can be changed, and this can be done repeatedly until the password is strong enough. Unfortunately, this is not the case with biometrics. If a biometric is stolen it cannot be altered, i.e., the victim remains vulnerable as long as the attacker has the victim's biometric information. There are other privacy concerns, such as the biometric information could be used for other purposes. For example, face recognition might be used for surveillance purposes and fingerprint information checked against forensic database. These are some of the privacy issues that are at the heart of many objections to the use of biometrics.

### 5.5.1   Data Protection

Biometric data contains personal information and is used to identify individuals. Therefore, this data needs to be protected for the privacy of the individuals. Privacy concerns may be higher, if this data is stored in a central database than in a personal/local device retained by the individual.

To protect privacy of information, data protection legislation are enacted to ensure that personal data, including biometric data, is not processed without the knowledge and consent (except in special cases) of the subject. Currently there are no laws in the world that protect consumers' biometric data [Nguyen, 2018]. However, the General Data Protection Regulation (GDPR) [E.U., 2019, Nguyen, 2018] for European Union (E.U) States does address biometric data. This act has made some real international impact, such as some States, including California, Texas, and New York, in USA have now passed a biometric privacy law. GDPR has a clear focus on biometric data privacy, and some of the provisions related to biometrics in GDPR are:

- □ E.U. residents are gaining more control over their biometric data.

- □ There is only one set of rules directly applicable to all the E.U. States

- □ The right to be forgotten – The consent must be explicit, and the subject has the right to withdraw the consent at any time.

- □ Data breach must be notified within 72 hours otherwise biometric companies could be hit with massive penalties.

- □ GDPR is a global law, because it not only affects E.U. but also Non-E.U. companies.

- □ Now, the companies and organizations only have to deal with the supervisory authority of the country where they are established. The supervisory authority acts as a central lead authority and deals with the processing activities of the company/organization across all the E.U. States.

Over two years have passed since the GDPR was enacted in 2018. GDPR is a great example and work in progress of setting up privacy laws that are good both for consumers and the businesses. We hope other countries and corporations will take some concrete actions to legalise data privacy especially biometrics data (which is more sensitive than other data) globally.

## 5.6 Biometric Attacks

As any other cybersecurity system, biometric systems are also vulnerable and susceptible to various attacks and threats. Hackers were able to steal 5.6 million fingerprints in a US government (Office of Personal Management) hack [Peterson, 2015]. Due to this incident, the director of the Office of Personal Management resigned. Recently, in August 2019, a huge data breach was discovered in the security platform of BioStar 2, a biometric security company [VPNMentor, 2019]. BioStar 2 uses fingerprinting and face recognition to identify and authenticate users. Over 1 million fingerprint records and facial recognition information were leaked.

### 5.6.1 Types

Various type of attacks [Ratha et al., 2001] can be launched against biometric systems. Some of these are:

- **Fake Biometric**: In this attack, a possible reproduction of the biometric is used to deceive the system. Examples include, a fake fingerprint made from silicon, a fake face mask, contact lenses, etc.

- **Replay Attack**: In this attack, the old biometric information is injected between the sensor and the processing system. First the sensor transmission is intercepted and copied, then the information is modified and finally replaying the attack. Examples include, an old copy of fingerprint or recorded ausio signal, etc.

- **Overriding Feature Extract**: In this attack, either the feature extractor or the features in the communication channel are overridden or replaced to produce the features desired by the attacker, and then stored as part of the existing features in the database.

- **Overriding Matcher**: In this attack, the similarity detector (matcher) is manipulated to produce an artificially high or low score to deceive the authentication system.

- **Tampering Template**: In this attack, the templates (either stored in the database or when in the communication channel) are manipulated and modified to either accept or refute the authentication of a person.

- **Overriding Final Decision**: In this attack, the final decision (accept or reject) can be overriden by the result desired by the attacker. This attack separates a biometric system with its application, i.e., makes a biometric system useless, and open up the system to potential dangerous attacks.

### 5.6.2 Protections

Here we discuss some of the techniques used to resist and protect against attacks on a biometric system.

**Liveness Detection**

Liveness Detection is a technique which detects whether the biometric is provided by a live person or not. In this technique different measurements are taken, such as temperature for fingerprints, movement of face or eye for face recognition, and blinking eye for iris recognition, etc. Challenge response is another technique that can be used for liveness detection. In this case, a challenge is presented to the user, and a successful response from the user indicates that the user is a live person. The challenge can be, an expression for face recognition, such as smiling etc., and repeating sequence of digits for voice recognition, etc. Liveness detection requires extra hardware and software, which increases cost of the overall system.

**Biometric Cryptosystem**

A biometric cryptosystem is developed by combining biometrics with cryptography [Rathgeb and Uhl, 2011]. This provides a high degree of security while eliminating the need to remember any password or carry a token. In a biometric cryptosystem instead of using some other authentication scheme (e.g., passwords) the key is bound to a biometric or generated a biometric. It is significantly difficult to copy, share, and forge biometrics compared to passwords. In addition to replacing password-based key-release, it also provides biometric template protection. Biometric cryptosystems are classified as key-binding and key-generation systems. In key-binding a chosen key is bound to the biometric template, and in key-generation key is generated from the biometric template and a given biometric sample. For more information and a detailed survey on biometric cryptosystems the reader is referred to [Rathgeb and Uhl, 2011].

**Hiding Biometrics**

Cryptography is used to hide information/message and requires a key to decode the encrypted message. In this case, the user knows that the message is encrypted. There are other techniques that just hides the message from the user without employing any encryption. One such technique is steganography. In this technique the original data (message) is hidden within data. For example, hiding a secret text message inside an image. Integrating biometrics with steganography augments both access control and secure transmission of sensitive biometric data. Generally images are used to hide the message. There are two approaches to image steganography, based on the embedding (hiding) of data either into frequency or spatial domain. Hiding data into the frequency domain is more preferred because of less distortion to the image and resilience to attacks. Based on different studies conducted, embedding biometric data into the frequency domain is a recommended approach [Douglas et al., 2018].

**Cancellable Biometrics**

Cancellable biometrics are produced by intentional and repeatable distortion of the biometric signals. These distortions are based on different transformation of the biometric signals. Some of these transformation methods are techniques based on cryptography – e.g., image hashing, bio hashing, etc., non-invertible transformations – e.g., wavelet and polar transformations etc., filter based – e.g., random and Gabor filtering etc., and huffman encoding, etc. The cancellable biometric should have the following properties: *Irreversibility* – The original biometric cannot be recovered if the generated one is compromised. *Revocability* – The ability to reissue new protected template to replace the compromised one. *Unlinkability* – Unability to distinguish if two protected templates, from different applications, are derived from an identical subject or different subjects. *Performance Preservation* – The performance of the biometric system should not decline because of the new formulation. Cancellable biometrics is an emerging research area and is open to lot of improvements. Most of the cancellable biometrics are not able to preserve the performance, so there is a need to develop novel techniques to improve this specific area. For more information and a detailed survey on cancellable biometrics the reader is referred to [Rathgeb and Uhl, 2011, Manisha and Kumar, 2019].

## 5.7 Future Work

Biometrics provide us with positive, reliable and irrefutable identification. It gives you comfort to know that your health care system does not only rely on your social security number as proof of your identity to get access to your medical records. In the past few years, biometrics technology have significantly improved, but still there are lot of challenges to resolve before it becomes the mainstream identification technology. To encourage further research in protecting the hardware, this section discusses some future research directions including the problems and challenges faced.

### 5.7.1 Wearable Biometrics

Wearable devices, such as smartphones, smart watches and activity trackers, have become ubiquitous, and presents an extraordinary opportunity for biometrics researchers [Blasco et al., 2016]. These devices store and transmit personal information, such as health and financial data about its user. Moreover, the sensors on these devices record different attributes of the user, such as physical (walking distance etc.) and biological (heart rate etc.). It is common now days for a person to carry more than one such device. Therefore, there is a need to develop and improve portable authentication signatures across devices belonging to the same user. These devices are resource constrained, with limited space and power. There is also a need to develop and improve simple user authentication mechanisms suitable for these resource constrained devices.

### 5.7.2 DeepFakes

DeepFakes are synthesized realistically looking images and videos. Recently, these images are generated using Generative Adversarial Networks (GANs) [Goodfellow et al., 2014]. Also, the raw biometric data of a person can be modified for malicious purposes. For example, digital images of a person's face can be edited and modified using Adobe Photoshop for creating a false match. Some examples of these fake images and videos are shown in Figure 5.5. These fake images and videos can be shared on social media and other platforms with the intention of creating a proliferation of fake images and videos with the originals. This makes it difficult to identify the original and unmodified image/video. Therefore, it is extremely necessary for a biometric system to validate the integrity of the biometric data and its source [Farid, 2016]. In this direction, there is a need to develop and improve the algorithms for detecting DeepFakes as well as maliciously modified images and videos. Also necessary is to infer the trail of modifications done to produce the fake images and videos.

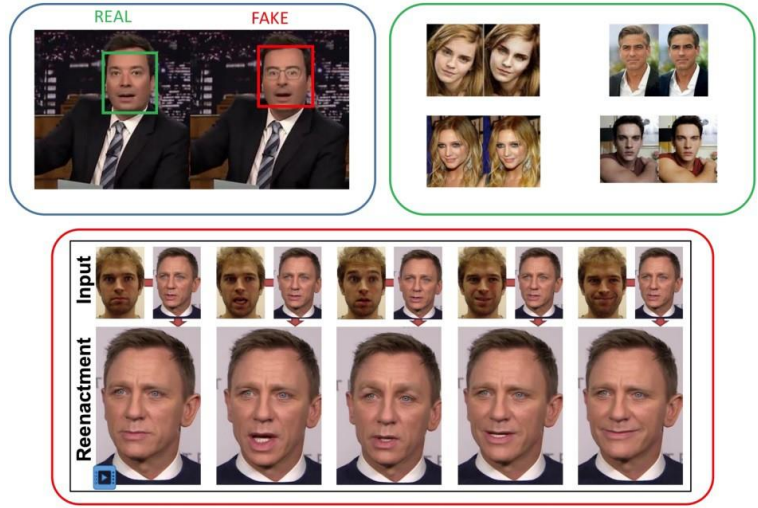

Figure 5.5: Examples of digitally modified images and videos. Top left is a GAN generated image. Top right is a near duplicate images of four subjects using Adobe Photoshop. Bottom is a fake video rendered by transferring the facial expressions of one person to another person: source [Ross et al., 2019]

### 5.7.3 Usability Testing

Biometrics are commonly used to authenticate a person's identity by processing the person's physical and behavioral attributes. Therefore, the human factor usability, plays a big role in successful development, implementation and deployment of a biometric system. There is a need to test the usability of a biometric system, to evaluate how well the system handles a variety of user interface expectations. For example, testing the cognitive or physical difficulties of using the system can help the designers and developers incorporate this information, and fine tune the interfaces that would increase usability of the system. Ease of participation is another usability factor that should be considered while designing the overall biometric system. For example, physical differences, such as height and weight. Is there any assistance provided to adjust the system accordingly? Are there any accommodations made for users who are not familiar with the system?

### 5.7.4 Personal Privacy

With the advancement in machine learning algorithms, it is possible to extract private information from the biometric data of a person [Dantcheva et al., 2015]. When such data is extracted without the user's consent, then it represent a privacy breach. Similarly, biometric data of a person can be used to link information from different sources and reveal the identity of the person. For example, matching biometric data about the face of an unknown person to a face on a social media platform can reveal sensitive information about the person through data accretion [Acquisti et al., 2014]. There is a need to develop methods and techniques to instill privacy into biometric data without compromising accuracy of the system. Researchers have started working on controllable privacy where artificial intelligence techniques (such as adversarial networks [Goodfellow et al., 2014]) are used to suppress the private cues in the raw image [Mirjalili et al., 2018]. GDPR [E.U., 2019, Nguyen, 2018] for European Union (E.U) States has reinforced to develop privacy preserving methods in biometric systems. This is a great example and work in progress of setting up privacy laws. We hope other countries and corporations will follow and legalise biometrics data privacy, which is more sensitive than other data.

# Chapter 6

# Cyber Intelligence

*The opportunity of defeating the enemy is provided by the enemy himself.*
(Sun Tzu, The Art of War, 500 BC)

## 6.1 Introduction

Artificial intelligence (AI) [Russell and Norvig, 2020] is a type of intelligence demonstrated by machines. Some of the major differences between machine and human intelligence are: *Emotions/Feelings* — Humans are social creatures. They have emotions and feelings. Through these feelings they can understand and interact with other humans better than machines. *Consciousness* — Humans have better awareness of the surroundings, especially self-awareness about situations and how to act in different situations. *Unused/Unseen Changes/Events* — Machine takes much more time than humans to adjust to unused changes or sometimes may not be able to adapt to unseen changes or events. *Decision Making* — Machines are not biased when making decisions. Whereas, humans have the ability of learning to make decisions based on experienced scenarios. *Speed of Execution* — Machines can process huge amounts of data at a much faster rate than humans. *Accuracy* — Machines can achieve perfection and are more accurate than humans at performing tasks, especially monotonous tasks. Moreover, machines perform tasks without any *human error*.

The fundamental question that AI is trying to answer is: *How to make machines think and act like humans*? What are all those steps that an AI capable machine has to go through to mimic like humans? Merriam-Webster [Merriam-Webster, 2021a] dictionary defines AI as:

*The capability of a machine to imitate intelligent human behavior.*

Another formal but a more elaborative definition given in [Kaplan and Haenlein, 2019] is:

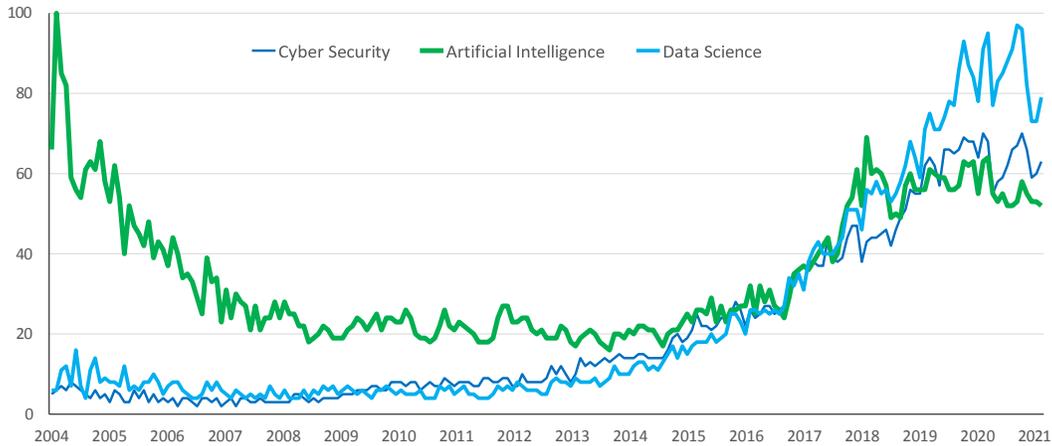

Figure 6.1: Google search trends [Google, 2021] by month, from January 2004 to January 2021, for the terms *Cyber Security*, *Artificial Intelligence*, and *Data Science*. The y-axis depicts the relative search frequency for the term. A value of 100 is the peak popularity for the term. A value of 50 means that the term is half as popular.

*The ability of a system to correctly interpret external data, to learn from such data, and to use those learnings to achieve specific goals and tasks through flexible adaptation.*

From the second definition, we can deduce some of the steps of an AI machine. (1) The first step is the presence of some kind of data. This data comes from an external source of the machine, and contains enough information about specific tasks so that an AI machine can learn and complete the tasks. (2) The second step is to learn from such data. (3) The third step is to utilize the learnings to achieve and complete specific tasks. During this process an AI machine should be flexible enough to adapt to different situations to complete the tasks.

The graphs shown in Figure 6.1 depict the popularity by month of the two terms, *Cyber Security*, *Artificial Intelligence*, and *Data Science*, when used during Google searches performed from January 2004 to January 2021. This development in the use of terminology based on the current most popular search engine is useful and of value to identify trends. These three terms start gaining popularity (i.e., a value > 40%) from the year 2018. AI was already a popular term but then its popularity decreased during the years 2007 − 2017 i.e., for almost 10 years, and stayed between 20% − 25%.

Currently, with the advancement in the computing power and the presence of large amount of data AI is experiencing a resurgence, as shown in Figure 6.1 the popularity of the two terms *Artificial Intelligence* and *Data Science* started increasing from the year 2018. AI has become an essential part of the new technology and is helping solve many challenging problems in computer science, software engineering and cybersecurity. The four major approaches and methodologies used for processing and solving problems using AI are as follows:

1. *Formal Logic*: Logic is a science that studies the principles of reasoning. For example, two truths lead to a third truth. In formal logic the reasoning/conclusion is based on clearly stated claims and evidences. An example would be, *John plays tennis on Thursdays when it is not raining. Today is thursday and it is not raining, therefore John will play tennis.*. Formal logic is used to develop an objective reasoning system to process and solve problems. Logical techniques can be used to formalize the reasoning problems. This helps understand the problem and make it easy to implement the solutions.

2. *Probability*: AI develops predictive models from uncertain data or imperfect and incomplete information. Probability is used to express the chance of occurring a particular event. The value generally lies between 0 (will never occur) and 1 (will certainly occur). This property of working with uncertain data makes probability a strong tool for predicting outcomes and making decisions in AI. Probability distribution helps AI to make statements and reason in the presence of uncertainty. Different techniques from information theory [Cover, 1999] are used to measure the uncertainty in a probability distribution. For example, Bayesian inference is used to update the probability of a hypothesis as more information and evidence become available.

3. *Analogy*: Analogy (also, comparison, inference, or correspondence) is one of the main inference methods in human cognition to learn from the past. Analogical reasoning is a powerful tool and extremely popular in AI applications to learn from the past experiences and adapt to different situations. In practice analogy is specified by indicating the most significant similarities. An analogical reasoning has the following form [Bartha, 2019]: (1) *A* is similar to *B* in some known features $\{f_1, f_2, f_3, ...\}$. (2) *A* has some other feature $f_n$. (3) Therefore, *B* also has the feature $f_n$, or some other feature similar to $f_n$. For example, *based on these known similarities between earth (A) and mars (B): orbit the sun ($f_1$); have moons ($f_2$); revolve on axis ($f_3$); subject to gravity ($f_4$), we can analogically infer that since earth supports life ($f_n$) mars may support life ($f_n$).*

4. *Artificial Neural Network* (ANN): An ANN is a collection of connected artificial neurons (just like neurons in a human brain) and is inspired by the workings of a human brain. In this approach neurons learn from processing specific tasks. After each learning the processed output is compared with the target output. The difference between the processed and target outputs is the error. Neurons are aggregated into layers, and these layers perform different adjustments. ANN adjusts itself according to a learning rule and the error value. After a sufficient number of adjustments ANN produce output which is almost similar to the target output. This is when the training is terminated. One of the major limitation of ANNs is that for achieving a higher accuracy they require a large set of data to train.

## 6.2 Cyber Intelligence

The number and variety of cyberattacks have significantly increased. It is becoming increasingly challenging to protect against these attacks. We need to find solutions to provide a holistic protection against a wide range of cyberattacks on a variety of target systems. AI techniques and systems have flexible and adaptable behavior and can help overcome these challenges by providing a holistic protection against cyberattacks. *Cyber Intelligence* is the collection of data and information related to the security of an organization, business, or state, and then utilizing and processing this information to develop and implement strategies to protect their operation, assets and interests. Figure 6.2 shows cyber intelligence as an ongoing process.

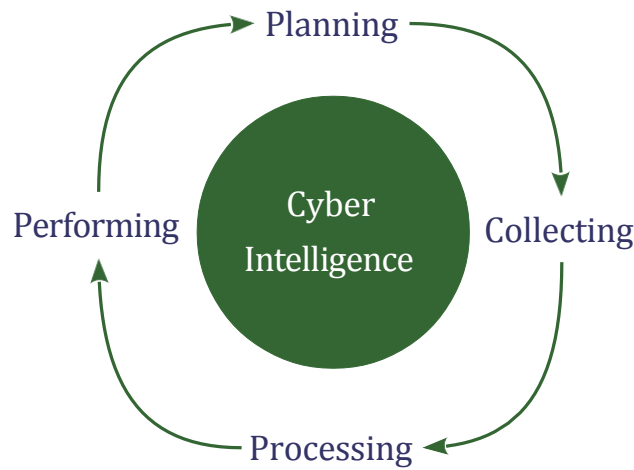

Figure 6.2: Cyber intelligence

### 6.2.1 Planning

The first process is to set some achievable goals and objectives of the cyber intelligence operations. A key part of this step is to identify the assets and then assess the risks and threats to them. Understanding the vulnerabilities, and what risks and threats do they pose is critical in setting the goals and objectives. Other tasks may include establishing the priorities, organizing roles and responsibilities.

### 6.2.2 Collecting

The next important process in cyber intelligence is to collect and gather the information. This commonly entails the collection requirements, i.e., what data needs to be collected and where to find that data. For example, every information about past and present risks and threats are collected. Data can be collected from the end points, networks, and databases, etc. Appropriate methods and tools are selected for collecting the data.

### 6.2.3 Processing

Once the information/data is collected, it is standardized or transformed to a usable and readable format to perform any analysis on the data. In the analysis stage the data is turned into usable and actionable intelligence, so that effective counter measures are developed to either stop or mitigate the risks and threats. The processed intelligence helps to understand the realities and risks involved and devise strategies to protect the information and data.

### 6.2.4 Performing

After the raw data is transformed to intelligent data, different actions are implemented to be performed and a comprehensive defense strategy is developed against the cyberthreats and cyberattacks. Different techniques and metrics are developed to analyse the performance of the defensive strategy. This analysis helps the system learns and in turn achieve specific goals and objectives through flexible adaptation.

## 6.3 Challenges and how AI can help

A cybersecurity system requires gathering and analyzing a range of data and information for gaining cyber intelligence. However, there are several challenges in collecting, processing and performing cyber intelligence. Some of these main challenges and how AI can help fully or partially solve them are discussed below.

### 6.3.1 Automation

Generally in a traditional cybersecurity system, file hashes, signatures based on rules, or manually defined heuristics are used to detect cyber threats. These techniques have their own merits but require lot of manual work to be successful and keep up with the changing cyber threat landscape. Some of these processes are automated, such as signature matching, but not all of them. This manual intervention is error prone and also financially infeasible. Therefore, there is a compelling need to completely automate these and other such tasks. AI powered technologies have already automated a number of different tasks in retail, wholesale, and business services. Similarly, the above mentioned, and the other long and repetitive tasks in the workflows of cybersecurity system, such as testing, basic threat analysis, and data deception can be fully automated. This automation also eliminates the problem of skills shortage in cybersecurity industry [Smith, 2018].

Unlike manual interventions, automation does not have limitations for processing information, and can therefore, continuously scan and monitor important assets. Humans, can only scan and identify a small fractions of security threats and definitely can not work round the clock. Whereas, machines do not have these limitations and can scan, process and identify threats accurately from a large set of data generated by today's information systems. This leaves the experts with more time and resources to manually investigate and manage more serious security threats.

Different types (machine, deep, reinforcement, and others) of learnings are branches of AI that reduces the need for manual programming and enables a system learns from previous experiences. This enables a cybersecurity system to automatically learn from previous cyber attacks and thus allowing the system to prevent any future attack. Suh c a learning can successfully defend against known attacks. What about unknown or zero-day attacks. To defend against such attacks, certain rules for a normal working system can be defined. The AI powered system can learn from such rules, and any anomaly can be detected as a threat to the system. This is where some human intervention is required. To fully automate this process is where most of the cybersecurity research into AI is advancing, such as learning from the adversary – adversarial machine learning [Martins et al., 2020]. AI can make the fully automated and also the human-in-the-loop systems more trustworthy.

### 6.3.2 Big Data

To implement an effective cybersecurity system, data and information from all systems across the entire organization need to be considered. The use of electronic devices have become ubiquitous and the amount of data they process has increased exponentially. This data is produced and processed at a high rate, and is not always structured, which increases the complexity of storing and processing this data. Heterogeneity of data and its sources makes identification and collection of data difficult.

The ability of AI to work well with data analytics makes it the prime approach to deal with big data. This ability of AI comes from the fact that AI is not just computing but actually learning. AI needs to work and learn from (a lot of – big) data to make successful and accurate predictions that help improve the performance of a cyber intelligence system. AI learns from big data and generates new rules for future cyber analytics. Good data is the lifeblood of AI. Therefore, it is essential to develop methodologies for collecting (mining) and structuring (formatting) data before applying any AI techniques.

Cybersecurity data science (also known as data mining) [Sarker et al., 2020] is the study of collecting, maintaining, processing, analyzing and communicating structured, semi-structured and unstructured cybersecurity data. It prepares and helps AI to apply different learning algorithms to discover patterns and making predictions from this data. This process can effectively learn from the training data to get insight into the cybersecurity incident patterns, and detect the current and predict the future cyber threats.

### 6.3.3 Adaptation

Cyber space is dynamic, i.e., it changes and expands constantly. The old traditional methods were good to secure static systems but are not suitable for a dynamic environment. In order to adapt and secure a highly dynamic environment we need to use and develop new cybersecurity dynamic technologies that can cater for these changes and expansions.

AI powered technologies can help us develop such a secure system that can self-adapt and adjust in the face of ongoing attacks. AI predictive analytics [Fong et al., 2021] can provide insights into a future attack, that can help protect sources of the system and improve resilience over time. This can also help identify adversarial operations early in the attack. Such AI analytics can be used to monitor and track current and future cyber attacks and in turn adapt the system to counter them.

Game theory [Morgenstern and Von Neumann, 1953] when used with AI can be applied to simulate a possible future attack or model different cybersecurity scenarios. Game theory models can be of dual nature, and can be used both for cyber offense and defense. These models can be used to assess the security of a system by analyzing the capabilities and incentives of an attacker. In turn, these assessments can help adapt the system to improve it's security.

## 6.4 Practical Applications of AI to Cybersecurity

As mentioned above, AI powered technologies such as machine learning, deep learning, reinforcement learning, adversarial learning, data science, game theory, and others can provide solutions to some of the main challenges, and help develop and implement successful cybersecurity systems. Here we discuss and present some of the current research and work that highlights practical applications of some of these technologies to cybersecurity.

### 6.4.1 Adversarial Machine Learning

Adversarial machine learning [Huang et al., 2011] is a technique that learns from previous attacks by adversaries and helps generate a class of attacks that deteriorate the performance of classifiers on specific tasks. These attacks manipulate the data either during the training (poisoning attacks) or deployment (evasion attacks). To counter these attacks different approaches discussed in [Martins et al., 2020] have been proposed to adapt the classifiers to improve their classification.

Martins et al. [Martins et al., 2020] presents a study and systematic review of the application of adversarial machine learning to intrusion and malware detection. The study based on 20 recent research works concluded that adversarial attacks can retrograde the performance of intrusion and malware classifiers. All the classifiers studied showed similar results on normal data, but under the manipulated (attacked) data, decision trees, naive bayes, and support vector machines were the most effected, while neural networks and random forests remained volatile.

One of the real threats in many cybersecurity problems, a zero-day attack, remains a primary open issue for adversarial machine learning. There is a need to improve and develop new approaches to enhance the adversarial training of the classifiers and counter these and other attacks. Machine learning algorithms should be able to deal with zero-day attacks using anomaly detection but may require human intervention when required. We also need to develop practical methods to analyse machine learning systems, that can explain, visualize, and interpret these systems. Such analysis will help investigate the behavior of such systems and decide whether to trust their results. The future research in robust artificial intelligence [Dietterich, 2017] and interpretability of machine learning [Lipton, 2018] can give us more understanding about the potentials and limits of machine learning applications in the field of cybersecurity and other fields.

### 6.4.2 Game Theory

Game theory [Morgenstern and Von Neumann, 1953] is the science of decision making. It provides a theoretical framework and models the interactions between two or more competing players to help make decisions that are interdependent. The common assumption made is that the players are rational, whereas in real life human behavior diverges from this model. Evolutionary game theory [Smith and Price, 1973] resolve this issue, in which the model presumes no or bounded rationality on part of the players.

Game theory plays a significant important role in logic and computer science. In AI game theory helps in making decisions. Recently the concept of **Generative Adversarial Networks** (GANs) [Goodfellow et al., 2014] have been proposed that consists of two models, a discriminative model and a generative model. Figure 6.3 presents a general architecture of GANs.

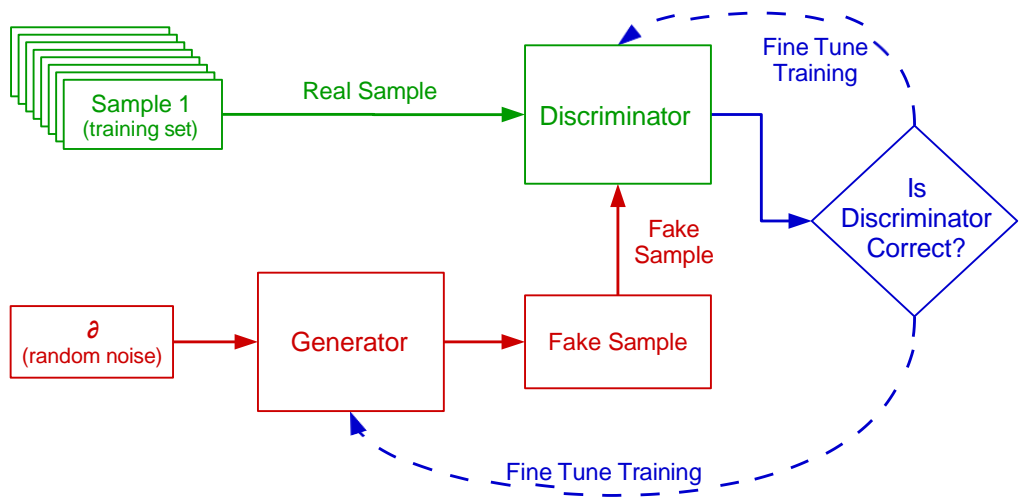

Figure 6.3: A general architecture of generative adversarial networks

The generative model generates fake samples that are suppose to have the same distribution as the original data samples. The discriminative model discriminates among the real and the generated fake samples. This is like a game, where one model tries to be better than the other model. The game continues until the two models become experts on what they are doing. The discriminative model becomes an expert in discriminating the real from the generated fake samples, whereas the generative model increases its ability to generate samples that can successfully deceive the discriminative model. Back propagation (Fine tune training) is applied to increase the accuracy of both the models. Back propagation is a commonly used technique to train machine learning algorithms. The method is used to optimize an error function, which is the difference between the predicted/actual and the desired output. A prediction error is calculated and the initial weight assigned is adjusted. After a number of iterations the learning process stabilizes and the models under training starts producing a value equal or very close to the desired output.

### 6.4.3 Improving Cybersecurity Using Generative Adversarial Networks

To improve the detection of adversarial attacks one should know about the type of attack, whether poisoning or evasion. Then the machine learning algorithm is trained and prepared for the specific attack. Recently, this is achieved by including poison samples, generated using GANs from the original training samples [Yin et al., 2018], in the training dataset [Anderson et al., 2016]. GANs generated adversarial samples have also been used in improving malware detection [Kim et al., 2018] and steganography encoding [Shi et al., 2017].

GANs have also been used to improve adversarial systems to trick and defeat cybersecurity systems. The research work carried out in [Hitaj et al., 2019] presented a password generating GAN called PassGAN that was able to generate fake passwords. PassGAN was trained on over 1 million passwords, and was able to match $51\% - 73\%$ passwords from the testing dataset.

Due to the increase in computing power, now a days machine learning algorithms have gained popularity and are being used to detect malware. These algorithms extract features from programs and use a classifier to classify between a benign and a malware program. On the other hand, to defeat these machine learning malware detectors, malware authors have gained motivation to attack these detectors. For this purpose adversarial examples are used to defeat a machine learning based malware detection tool. Most of the time such a tool is a black box to the attackers. The attackers do not have any information about the architecture and parameters/features used by the detectors. Recently the research work presented in [Hu and Tan, 2017] proposed a GAN based approach, called malGAN shown in Figure 6.4.

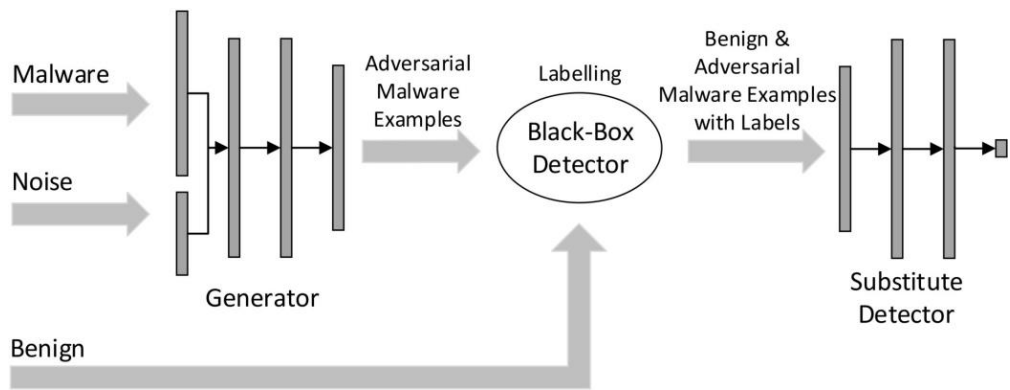

Figure 6.4: Architecture of malGAN: source [Hu and Tan, 2017]

malGAN takes original samples and produce adversarial malware examples to defeat machine learning malware detectors. The experimental results show that GANs can be used to successfully defeat and bypass malware detectors by generating similar and realistic features, and hence fooling the system in believing that such features are not adversarial.

These and other such research works inspire studies about adversarial systems and discussions about the potential cybersecurity risks to systems. These inspirations in turn give insights into specific behaviors of adversaries and also provide guidance and direction to reduce the cybersecurity risks to a system. The best cybersecurity defense for a system is to know and reduce the current and future cybersecurity risks to the system.

## 6.5 Future Work

The above sections have presented discussion about some of the challenges faced by cybersecurity and how AI can help fully or partially solve these challenges. This section discusses some major future research directions including the problems and challenges faced in applying AI to cybersecurity.

### 6.5.1 Hybrid Augmented Intelligence

Despite the fact that recently AI has achieved significant improvements in cybersecurity, it is still not possible to fully and automatically adjust a system to the changes in the environment; assess and learn all the risks, threats and attacks; autonomously choose and apply appropriate countermeasures to reduce the risks, and protect against these threats and attacks. To augment cybersecurity, a strong interdependence and collaboration between AI systems and human factors is required. An expert cybersecurity analyst should be part of the AI analytic process to help facilitate and enhance decision making processes. This is referred to as hybrid augmented intelligence [Zheng et al., 2017] or human-AI collaboration [Cichocki and Kuleshov, 2021]. This human-AI collaboration can significantly increase the performance of a system over a system that only utilizes either AI processes or human cybersecurity experts. The development and study of hybrid augmented intelligence is critically needed to improve the role of AI in cybersecurity.

### 6.5.2 Explainable Artificial Intelligence

AI has definitely improved cybersecurity but this success is often been achieved through increased model complexity. It becomes very difficult to know their operation and how they come to decisions. This ambiguity about AI has made it problematic to adopt AI in sensitive critical domains such as cybersecurity. Recently there is a surge in research to develop methods that interpret and explain machine learning models [Linardatos et al., 2021]. This surge in research has led to the revival of the field *Explainable AI*. This field of study focus on the understanding and explaining the operations of the AI algorithms rather focusing on the predictive power of these algorithms. The more interpretable an AI system is the easier it is to identify the cause and effect relationship between the inputs and outputs of the system. A detailed survey and study is presented in [Linardatos et al., 2021] on the current AI interpretability methods. A classification of these methods is shown in Figure 6.5.

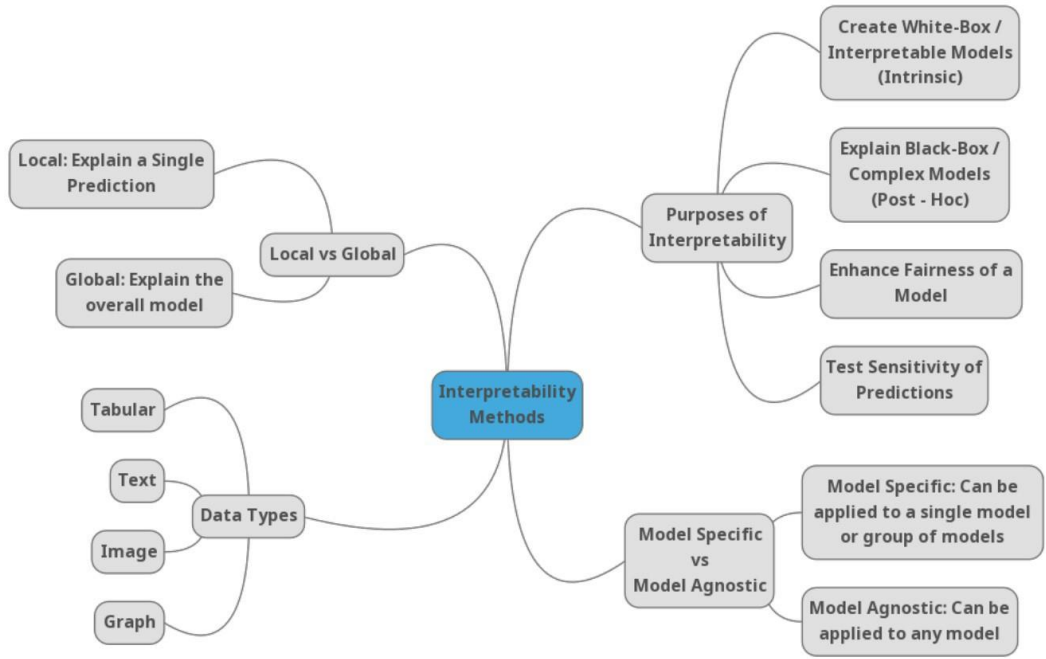

Figure 6.5: A mind-map of the classification of AI interpretability methods: source [Linardatos et al., 2021]

If an AI interpretability method can be restricted to only a specific family of algorithms, then it is called model specific otherwise model agnostic. The classification depends on various factors, such as the scale (local or global) of interpretation, and the type of data on which these methods could be applied.

Explainable AI is an emerging field of study. There are a great number of AI interpretability methods developed [Linardatos et al., 2021] but still they have not become part of the standard AI workflows and pipelines. Explainable AI has a lot of potential in improving the role of AI in cybersecurity and still has a lot of unexplored and uninvestigated features.

# Chapter 7

# Cyber Forensics

*He who only sees the obvious, wins his battles with difficulty; he who looks below the surface of things, wins with ease.*
(Sun Tzu, The Art of War, 500 BC)

## 7.1 Introduction

Every contact by a perpetrator leaves behind traces [Chisum and Turvey, 2000]. To find and make a case against the perpetrator these traces or evidences need to be found, collected, secured, studied and analyzed. Forensic science is the use of scientific methods and expertise to gather and analyze evidences that can be used in criminal or other investigations in a court of law. Cyber forensics is a branch of forensic science that processes and deals with the evidences found in digital devices.

The evidences collected from digital devices can be used for several different purposes. Some of them are: (1) *Electronic Discovery* — This is the process of searching, finding, locating, and securing any electronic data latter to be used in a civil or criminal forensic case. (2) *Intelligence* — The data gathered and collected from digital devices can provide actionable intelligence. This intelligence contain information that can lead to different type of missions, such as securing national interests, decreasing or eliminating crimes like kidnapping, child exploitation and human trafficking etc. (3) *Administrative* — Digital evidences collected can help with the administrative investigation, such as a potential misuse of computers of a company or organization. For example, playing games instead of inspecting and scrutinizing images from security cameras.

### 7.1.1 Basics

To perform cyber forensic analysis a person should be familiar with the inner workings of a computer. Some parts of the computer, such as the memory and storage plays a more important role in forensic analysis than others, such as the processor. This section explains some of the technical details that are required to collect evidences and conduct a thorough examination to give an accurate opinion.

To understand computers we need to understand their language. The language of a computer is called **binary**. In binary, which is a base 2 numbering system, there are only two numbers (options) 0 or 1. Each 0 or 1 is called a bit. Computers generally work with a chunk of bits called bytes. One byte is made up of 8 bits. These bytes relate to letters and numbers. For example using the ASCII character set the letter `A` is represented by `01000001`. We can also write this as 65 in decimal (a base 10 numbering system) or 41 in hexadecimal (a base 16 numbering system). To make computers perform certain tasks we provide a set of instructions, that are a collection of bits that the computer understands and obeys. These instructions have specific meanings and are encoded using bits and bytes. For example, the following sequence of bits tell one computer to add two numbers:

```
1000110010100000 [ symbolic notation —— add A,B ]
```

This translation from a symbolic notation (easy to understand by a human) to the language of the computer (also called binary/machine language) is performed by compilers. A *compiler* is a software program that translates programs written in a high-level language to instructions that the hardware can execute. As a cyber forensic analyst one should understand the importance and use of these bits and bytes as a form of information. There are tools available to perform analysis and make sense and interpret these bits and bytes.

**Processor**

A processor in a computer is a hardware that processes and take actions depending on the information present in the bits and bytes passed to the processor as instructions. A processor is where action takes place. i.e., where different tasks are executed. It is important to understand functioning of some basic elements of a processor. These elements may contain some important evidences that need collection. The contents present in the CPU (central processing unit) registers and caches of a processor should be given a higher priority for collection. A CPU is the brain of a processor that retrieves and executes instructions. CPU registers are the smallest but fastest memory available to a CPU for storage. Caches, another faster memory available to a CPU, are bigger in sizes but slower than registers. Because of their proximity to the CPU and design, caches are faster than other storage devices. CPU registers and caches are volatile (do not keep state when powered OFF) memory elements and are very crucial to inspect and analyze during live forensic analysis and provide context for non-volatile (keep state when powered OFF) data [Wang et al., 2011].

**File Carving**

Information in a computer is generally stored in fies. A computer file is a combination of instructions, i.e., code and data written as bits or bytes. During cyber forensics one must look at these bits and bytes to extract and interpret evidences from a file. *File carving* is the process of mining and extracting files from a storage device. The files are usually present in the form of raw bytes, i.e., there is no metadata information available about the files and the file system that originally created the files is damaged. A file is usually identified by the header and the footer. A fragmented file is much more difficult to extract than a continuous file.

### File Signature

The next step is to identify the file. The most common way to identify a file is by its extension. This method is not very reliable. During cyber forensic analysis a file is identified based on the header and not the extension. Generally the files whose headers do not match their extensions are tagged as suspicious. This comparison is called *file signature* analysis. A hex browser can be used to examine the file header. Different type of files have different formats and type of headers. There are several forensic tools available to perform file signature analysis and determine the exact type of the file.

### Storage Devices

A storage device is a piece of hardware that is used to store data. All types of data, whether code or data associated with the code, any file created, operating system, and applications, etc., is saved on a storage device. Every computer, whether desktop, laptop, tablet, smartphone and embedded systems have some kind of storage device within it. Standalone external storage devices that can be connected across single or multiple computers are also available. There are basically two types of storage devices, primary and secondary storage.

**Primary storage devices** generally are volatile and act as a short term memory. Due to its proximity to the processor and design, data can be read and written to a primary storage device extremely quickly. ROM, RAM, caches and registers are some of the examples of primary storage devices. ROM (read only memory) is non-volatile and contains data that cannot be changed easily, i.e., the data cannot be changed using a normal write or delete operation. RAM (random access memory) is a volatile memory and is constantly being written to and read from. Caches and registers are volatile and gives the processor a much faster (than ROM and RAM) access to data and instructions.

**Secondary storage devices** generally are non-volatile and act as a long term memory. They are much slower to access than primary storage devices. Hard disk drive (HDD), solid state drive (SSD) and optical storage devices are some of the examples of secondary storage devices. HDD is a magnetic disk with tiny particles that can be magnetized to represent bits. A `1` is stored if the particle is magnetized else a `0`. SSD is a type of flash memory where electronic circuit is used to store bits. A `1` is stored if the circuit is charged else a `0`. Unlike HDDs there are no moving parts in SSDs. SSDs are faster, smoother and last longer than HDDs. Example of optical storage devices are CDs, DVDs, and Blu-ray Discs. A laser is used to read and write to these devices. Change in reflectivity of the laser beam is read as binary.

**ARP Cache**

Address resolution protocol (ARP) translates an Internet protocol (IP) address to a media access control (MAC) address and vice versa. MAC address is a unique identifier embedded into every network card (ethernet or WiFi card) by the card manufacturer and cannot be changed. ARP cache is a table that maps IP addresses to MAC addresses to connect to destinations on the network. An example of a simple ARP cache table is as follows:

```
Internet Address        Physical (MAC) Address

192.168.3.1             00-14-22-01-23-45
192.168.3.201           40-d5-48-br 55-b8
192.168.3.202           00-12-23-01-22-35
```

An entry in the ARP cache can be poisoned (changed) by an attacker to divert the future communications to the attacker. So, now the attacker is secretly in the middle of all the communications. Therefore, this attack is also called a *Man in the Middle* attack. A poisoned ARP cache, can be detected during cyber forensic analysis, is an indication that an attack is taking place.

**Kernel Data Structures**

A kernel is the core of an operating system that controls everything and performs all the critical operations in the system. Sifting through the kernel memory images to find important data structures that reveal the inner state of the system is therefore an important function during any cyber forensic analysis [Case et al., 2010]. There are several structures used by a kernel to perform certain operations, here we discuss two of them that plays an important role in keeping the current state of the system. (1) **Process Table**: It is a data structure that contains the list of all the processes in a system. It also contains a PCB – process control block – for each process. A PCB contains all the information about a process, such as the process ID, state – running or waiting –, program counter, opened files, and scheduling algorithms, etc. (2) **File Table**: It is a data structure that contains information about all the files in the system, such as the file status, offset, and other file attributes including pointers to other tables. Sockets, special file types used during interprocess and network communication, are also listed in this table.

**Routing Table**

A network routing table contains information about the routes to different network destinations. The table helps in routing packets from one node in a network to another (destination node indicated by the IP address in the packet) in the same or other network. To reach the destination a packet may have to pass through different routing tables, each sending the packet to the next node/router in the path. A routing table has many attributes that are critical and can be used in a cyber forensic investigation. Traceback mechanisms can use these attributes, such as addresses of destinations next in line to the router, to trace the attacking path and the attacker [Davidoff and Ham, 2012].

**Data Types**

Cyber forensic analysis mostly deals with some kind of data. Here we discuss three different types of data that are mostly encountered by a cyber forensic analyst. (1) **Active Data**: This is the data that is in use everyday, and resides in the allocated space of the storage drive. (2) **Latent Data**: Data that is physically present on a storage drive but logically unavailable. Deleted or partially overwritten is an example of latent data. Cyber forensic tools may be required to find and collect this type of data. (3) **Archival Data**: This is the data that is backed up (archived) on a storage drive. If it is too old, i.e., legacy data, it may be difficult to retrieve.

**File Systems**

A file system separates the data into pieces. Each piece of data is called a file. There are several file systems but we are going to discuss only the basic and important file systems currently in use and encountered most during cyber forensic analysis. (1) **FAT**: File allocation table is the oldest of the common file systems. FAT has a limit 8 characters for the filename and 3 characters for the extension. (2) **NTFS**: NT file system developed by Microsoft is an extension to FAT. NTFS improves performance, reliability, and disk space usage. It also allows ACL (access control list) based permission control over the files. (3) **APFS**: Apple file system is developed by Apple for macOS and iOS. It is cross platform and supports international friendly file names. It supports full disk encryption.

Here we also discuss some other terminologies used with file systems. **Allocated and Unallocated Space** — Allocated means space being used by data, and unallocated means space not being used by data. Unallocated does not mean the space is empty, it may not be empty but essentially invisible. **Data Persistence** — Data residing on a hard drive is persistent until it is overwritten wih some other data. **Slack Space** — If a file is partially overwritten, then slack is the space (part of the file) that is not overwritten and can be recovered. **Swap Space** — An operating system writes data into a page file, called swap space, to free up room in the memory. This space may contain files and fragments (part of the memory) that no longer exist anywhere else on the drive. The swap space may have some evidences overlooked by the perpetrators.

## 7.2 Collecting Evidence

Collecting evidence is the foremost piece of work during cyber forensic analysis. Here we provide a general checklist of the important tasks involved during collecting evidence from a forensic scene.

- ☐ List the kind of and number of devices present

☐ Check if the devices are ON or OFF

☐ What tools are required at the scene

☐ Prioritize the evidence collection, such as follow the following order when collecting evidences

    1. Contents of CPU registers and cache

    2. Contents of the routing and process table, ARP cache, and kernel statistics (structures)

    3. Memory

    4. Swap space

    5. Storage devices

☐ Secure the evidence

    1. Computers and wireless devices must be made inaccessible (e.g., removing the ethernet cable or the modem) after making sure that no volatile data would be lost

    2. Isolate the wireless devices (e.g., put them inside an electronic shield) such as cell phones from the network signals

☐ Write everything, i.e., document the scene

☐ Take pictures as evidences and part of the documentation

☐ Take notes

## 7.2.1 Chains of Custody

Whenever an evidence is recovered and collected, it moves to different locations for storage, analysis and examination before it is presented in the court. For example, a computer is first collected, then logged in at the lab and stored. It may be taken out from the storage for analysis and checked back in many times, and so on. To keep track of these changes (custody, transfer, analysis and disposal), all these chains of custodies of an evidence must be documented. Without this detailed trailing and accounting the evidence may loose its trustworthiness. This chain of custody establishes that the evidence is in fact related to the alleged crime.

## 7.2.2 Cloning

*Forensic cloning* is an exact, i.e., bit for bit, copy of a piece of cyber evidence, and is also known as the forensic image. A forensic image contains both active and latent data. Cloning a device can be a time consuming process and may be performed at a lab which, provides a much more stable and safe environment, and not at the scene. Sometimes (mostly in civil cases) it is not possible to take the device to the lab, such as a business machine that generates revenues while on the scene and the hard drive cannot be replaced. In this case cloning is performed on the scene.

The main reason of cloning a device is that digital evidences are volatile. Therefore, examination on the original device is hardly performed, and the device is cloned. There may be exigent or urgent circumstances, such as a missing child, or no tools or techniques available at hand, that can lead to examining the original device. Cloning is also performed to make a backup of the evidence, so that one can be used for examining and the other to fall back on.

Before starting the cloning process it is always recommended to have some kind of write blocking in place. Write blocking preserves the original evidence during the cloning process by preventing any data being written to the original evidence device. The destination device/drive must also be forensically cleaned prior to cloning an evidence device/drive. A *forensically cleaned* drive is one that can be proven to be free from any data at the time the clone is created. Cloning is complete and successful when the hash values (also called *digital fingerprint*) of the source and the clone match. The final result of a cloning process is a forensic image of the source device/drive. After finishing, a forensic cloning tool leaves a paper trail that should be kept as part of the documentation.

### 7.2.3 Risks and Challenges

Some of the risks and challenges of collecting the evidence are as follows.

1. Bad sectors and damaged or malfunctioning drives.

2. A corrupt boot sector or a failing motor can also create complications.

3. Interacting with a running can cause changes to the system, which in turn can compromise the integrity of the evidence.

4. Turning the computer OFF (accidentally or for other reasons) makes the data in memory under threat of destruction.

5. Sudden power loss could damage the data.

### 7.2.4 Digital Fingerprints

Hashing helps produce digital fingerprints of the evidence or the forensic image. Any change to the device/drive, even by a single bit, results in a radically different digital fingerprint (hash value). This means any tampering or manipulation of the evidence is readily detectable. Cryptographic hash functions are used for producing these kind of digital fingerprints. Some of the key properties of cryptographic hash functions are: One-way — There is no feasible and practical way to invert the hash. Deterministic — The hash function produces the same hash, no matter how many times it is used. Collision Resistant —

No two inputs hash to the same value. Following is an example of a cryptographic hash function:

```
SHA256(The Grey Wolf) =
1101001110110111101000100111110111100010100101001001001100110011001
0111010111001110000001100000010011010001100011001111100100011
1010110001110100010000111010010001010011111001111001000011000
0101010011100011011111011110011000011101011010101101001111001
00100100111010010100

SHA256(The Gray Wolf) =
0111111001111111010100000010011010011101011100111101000101010
0111000011011111000011110000000100100010100001001101100001
0110111011111000111001001111111001101000001011100011011101110
0011010001001000111111001110101011001000011000010111010110
0001000111110100000000
```

In the example above just by changing one character (`e ⟶ a`) of the input text, the cryptographic hash function `SHA256()` has generated a very different hash value.

### 7.2.5 Final Report

The examination should be documented with sufficient details so that it can be duplicated by another examiner. All the major forensic tools have a very good reporting features that help generate useful reports. However, these standard reports are very difficult to read and understand by a non-technical reader. It is necessary that the report not only includes the standard report generated by the tool, but should also contain all the details written in plain English. This way, a cyber forensic report becomes more effective and useful even for a non-technical reader.

## 7.3 Antiforensics

Antiforensic is any approach, technique and method that makes forensic analysis difficult, practically impossible, or time consuming. During this process, the forensic data is either erased, so that the data is not available for analysis, or obfuscated, so that the data is practically impossible to find and extract. For example, hiding or destroying e-mails, financial and health records, and so on. The use of wiping applications, e.g., the use of defragmentation application to destroy the data. The use of different obfuscation techniques, such as packing, polymorphism and metamorphism [Alam et al., 2015], etc., to hide malicious contents of an application (e.g., a malware). The same techniques can be used for legitimate purposes. For example, software companies hide their proprietary application using some of these techniques. Therefore, proving the intent is critical and foremost when dealing with any antiforensic data.

Here we are going to discuss some of the antiforensic techniques used in practice for hiding and erasing data.

### 7.3.1 Hiding Data

For a discussion on the three obfuscation techniques **packing**, **polymorphism** and **metamorphism** the reader is referred to Section 4.5.1 of Chapter 4. Here we are going to discuss two of the other popular techniques for hiding data.

#### Encryption

One of the most common method used for hiding the evidence data is encryption. This is the process of converting also called encoding a plain text (i.e., the original data) into a cipher text (i.e., an alternative form of the original data). A cipher text contains information about the original data but is unreadable by a human or a computer without the cipher (the encoding algorithm). Classical ciphers [Menezes et al., 2018] used were substitution, and transposition ciphers. These ciphers are quite easy to crack using brute force attack. Modern ciphers are more secure than classical ciphers and can survive a large range of attacks. The two popular modern ciphers used in practice are symmetric and asymmetric keys algorithms [Menezes et al., 2018].

*Symmetric key algorithm* (e.g., DES and AES) — In this algorithm the same key is used for encryption and decryption. The sender and the receiver needs to share a secret key in advance. The sender uses this key to encrypt the message (data), and on the other end the receiver uses the same key to decrypyt the message.

*Asymmetric key algorithm* (e.g., RSA) — In this algorithm two different keys (private and public) are used for encryption and decryption. The sender uses a public key (available to everyone) to encrypt the message, and the receiver uses her/his secret private key (only available to the receiver) to decrypt the message.

As methods of encryption are very common and frequently encountered during forensic analysis, most of the forensic tools provide software for decryption of the encrypted forensic data.

#### Steganography

This is also known as hidden writing, and is the attempt to hide the fact that data is being transmitted or present. The first known attempt of steganography was in 440 BC when a Greek general shaved the head of a slave and wrote a message (of a Persian invasion) on the slave's head [Stamp, 2011]. After his hair had grown back and covered the message, he was sent through the enemy lines to deliver the message to another Greek general. The modern version of steganography involves hiding data in media, such as image files, audio data, or even software. Different methods are used for this purpose. Some of the popular methods used in practice are least significant bit and discrete cosine transform embedding [Cox et al., 2007, Provos and Honeyman, 2003]. Steganography can be used by malware to avoid detection, and is called *Stegware*. To countermeasure such a threat, the data needs to be transformed in a way that destroys the payload (malware), and is called content threat removal [Wiseman, 2017].

It is difficult to detect steganography especially when high compression rates are used. The best approach is to compare the modified files with the original unmodified files. When detected it is easy to extract the data with forensic tools.

There are some limitations of steganography: (1) *Capacity* — The size of the data to be hidden in a file can increase the size of the file. Therefore, a general assumption is that only a small amount of data is hidden using steganography. (2) *Visual Distortion* — When encrypting data into an image, the quality of the image can be distorted. This can be easily detected visually by the investigator.

In order to effectively hide the evidence data and provide limited access to an investigator, the attackers normally combine steganography with other encryption methods.

### 7.3.2 Erasing Data

When a disk or file is deleted, an investigator can still perform recovery of the data or files. Permanent data deletion is not achieved by simple file deletion commands. Several different specific methods are used to permanently delete the data. Here we are going to discuss few of these methods.

#### Wiping

Data wiping or cleaning is the process of overwriting the data, to render the data unrecoverable, and hence achieving data sanitization. This technique allows to write zeros, ones or random data onto either a selected portion or across all sectors of a drive. The data is removed once the overwriting process is complete.

#### Secure Erase

For modern drives a single overwrite is enough to permanently erase the data. But, for older magnetic drives this is not the case [Gutmann, 1996]. Therefore, either they are overwritten multiple times or some other techniques are used to permanently delete the data, such as disk degaussing.

There are different options for securely erasing the data, and some of them are: (1) Overwriting once, either zeros, ones or random data and then erasing the data. Similar to data wiping. (2) Overwriting 7 times, either zeros, ones or random data and then erasing the data. (3) Overwriting 35 times, either zeros, ones or random data and then erasing the data. As we can see the last two are the most secure erasing techniques.

#### Disk Degaussing

Disk degaussing is a technique, where the data is removed using a magnetic field. It is simply a demagnetizing process to erase a hard drive, and achieves a 100% secure removal of the data. It is a hardware process and different type of degaussers are available for such purpose. Therefore, may not be very convenient and cost effective for an attacker.

**Defragmentation**

It is the process of physically moving the fragmented parts/pieces of a file together. This way it frees more larger free contiguous spaces and also speeds up file access. Since this process moves data from one location to another on the drive, the data can be overwritten in the process. The overwritten data may have some evidentiary significance.

## 7.4  Future Work

Due to the increase in cyberattacks against systems, states, and corporations, these entities have started taking cybersecurity more seriously. Therefore, without a doubt cyber forensics has become a critical part of cybersecurity. With the new digital transformation, such as IoTs, Cloud computing, big data, cyber forensics faces lot of challenges to be resolved. To encourage and stimulate further research in cyber forensics, this section discusses some future research directions including the problems and challenges faced.

### 7.4.1  Tools

Software and hardware tools are at the forefront of cyber forensics technology. The cyber forensics technology is evolving at a rapid pace, but the tools are not adapting with this evolution. The tools currently available are not practical. There is a strong variability between their outputs. There is a lack of shared datasets; existing tools can not get validated; and hence can not be used in practice. There is a lack of information and standardization of tools, so it is difficult to use them when merging cyber forensics with other disciplines (non-technical areas). There is a need for collaboration between experts in different fields and disciplines to create standardized benchmarks for testing and streamline the process of tool development. There are no tools available in certain areas of cyber forensics, such as steganography, quantum computing, and IoTs, vehicle and software defined networks forensics [Luciano et al., 2018].

### 7.4.2  IoT Forensics

Internet of Things (IoTs) are very vulnerable to cyberattacks and criminal activities. When IoTs security fails and there is an attack it becomes necessary to perform thorough forensics analysis. Current IoT forensic techniques and methods are still in research and evaluation stages [Jones et al., 2016]. Tools and techniques addressing IoT forensics have not yet been introduced. Due to the resource constrained nature of IoTs, the current traditional approaches and methods need to be enhanced and adapted to suite IoTs. Some areas that need further work and improvement are following. (1) *Collecting Evidence* − In IoTs it is not easy to collect the evidence. IoTs components are not always in one location and may not be connected or in use. Therefore, to collect the evidence requires knowledge of the IoTs architecture, physical and digital characteristics, and locations of their components. In this regard, there is a need to develop new tools to help in collecting the evidence. For example a visualization tool that shows in real-time the operation and location of IoTs, which can help identify the areas and locate the components for collection. (2) *Analysis* – Analysis of the evidence is a critical part in cyber forensics. In IoTs, virtualization of the crime scene can provide a good investigation platform. (3) *Testbeds* – To explore different attack scenarios testbeds [Poudel et al., 2017] need to be developed and implemented. These testbeds will also allow for forecasting and detecting security attacks before they can cause damage to the IoTs. The information collected can also help in reversing or minimizing the damages caused by these attacks. (4) Complexity – Complexity and diversity of IoT devices pose a significant challenge to forensic analysis of IoTs. For example, different file formats and systems that are proprietary and closed source, and diverse communication protocols of IoTs make it difficult or unable to acquire and examine IoTs evidentiary data.

### 7.4.3 Cloud Forensics

With the proliferation of cloud computing in our digital lives, attackers and adversaries have started targeting clouds for their malicious and criminal activities [Choo et al., 2017]. These activities can range from extracting and stealing personal information, to storing and hiding incriminating and illegal materials. Cloud service providers are continuously making efforts to prevent their services from being criminally exploited. For example Dropbox and Microsoft have implemented a child abuse detection software on their clouds storage product. There are other traditional solutions that have been proposed and are being used to protect and secure the cloud. Still there are times, e.g., crime investigation involving cloud, when cloud forensic is needed to gather evidence of a crime or incident. In cloud forensics data is commonly analyzed at three main stages, data-at-rest on the clients, data-in-transit, and data-at-rest on the servers. Traditional techniques may not be applicable to cloud forensics. Some of the challenges of cloud forensics are: (1) Investigators may not have physical access to the evidence. (2) Data may be distributed between countries and within datacenters, which may result in jurisdictional and operational challenges for collecting the evidence. (3) Trust on the cloud service for maintaining and providing the correct log data and reports about the activities of users of interest, e.g., a court order [Zawoad et al., 2015].

As cloud and related technologies (e.g., federated cloud, fog and edge computing) advance there is a need to develop and implement new forensic methods to keep pace with the advancement. Cloud computing uses advance and sophisticated techniques to secure data-at-rest and data-in-transit. Therefore, it is crucial to develop novel data collection techniques to obtain evidential data from cloud [Choo et al., 2017]. These novel techniques should circumvent advance encryption and anti-forensic features without compromising the integrity of the evidence. While developing these forensic techniques one should keep a balance between the security needs of the system and rights of the individuals to privacy. Another major future research area is the development of effective visualization of evidential data for forensic analysis [Tassone et al., 2017].